\documentclass[11pt]{article}
\usepackage{amssymb,amsmath,amsfonts,amssymb,mathrsfs,bm,color,graphicx,subfigure}
\usepackage{CJK}
\usepackage{epstopdf}
\usepackage{epsfig}
\usepackage{color}
\usepackage{hyperref}
\hypersetup{
	colorlinks=true,
	linkcolor=blue,}

\numberwithin{equation}{section}

\begin{document}

	\title{Traveling waves of a generalized Rotation-Camassa-Holm equation with the Coriolis effect}

	\author
	{
		N'Gbo N'Gbo $^1$\footnote{npaulrene2@outlook.com} \,\,\,\,\,Yonghui Xia$^2$\footnote{Corresponding author. Yonghui Xia, yhxia@zjnu.cn, xiadoc@163.com}\,\,\,\,\,Tonghua Zhang$^3$\footnote{tonghuazhang@swin.edu.au}\,\,\,\,\,
\\
		{ \small \textit{1. Department of Mathematics, Shanghai University, Shanghai, 200444, P.R. China}}\\
		{ \small \textit{2. Department of Mathematics, Zhejiang Normal University,	Jinhua, Zhejiang, 321004, P. R. China}}\\
		{ \small \textit{3. Department of Mathematics, Swinburne University of Technology, Hawthorn, Victoria, 3122, Australia}}\\
		{ }\\
	}
	
	\maketitle
	\begin{abstract}
		In this paper, we analyze the dynamics of a  generalized Rotation-Camassa-Holm equation, which is the $\theta$-equation augmented with the Coriolis effect, induced by the earth rotation. The generalized Rotation-Camassa-Holm equation (named as Rotation-$\theta$ equation) is a generalization of a family of models (including the Rotation-Camassa-Holm equation for $\theta=\frac{1}{3}$, asymptotic Rotation-Camassa-Holm equation for $\theta=1$ and the Rotation-Degasperis-Procesi (DP) equation for $\theta=\frac{1}{4}$). Our study is conducted via the bifurcation method and qualitative theory of dynamical systems. The existence of not only smooth solitary wave solutions, periodic wave solutions, but also peakons and periodic peakon solutions is shown. The chosen values $\theta$, allow us to assess the difference in behavior between the classical $\theta$-equation and the Rotation-$\theta$ equation. We conclude that the Coriolis effect does affect the traveling wave solutions. We summarize the bifurcations and explicit expressions of waves solutions in three theorems in Section 4. A conclusion ends the paper.

	\end{abstract}
	
\textbf{Keywords:} Rotation-Camassa-Holm equation; solitary wave solution; periodic wave solution; peakon solution; periodic peakon solution.
	
	\section{Introduction}
	\noindent
	
	The Coriolis force is an inertial force acting on objects that are in motion within a frame of reference. The deflection of an object due to the Coriolis force is called the Coriolis effect. The Coriolis effect is caused by the rotation of the Earth and is known to perturb wave propagation.
	In 1993, Camassa and Holm \cite{CamassaH}, derived the famous classical Camassa-Holm (CH) equation, as a model for shallow-water waves. The model is formulated as follow:
	\begin{equation}\label{CH-Equation}
		u_{t}+2\lambda u_{x}-u_{xxt}+3uu_{x}=2u_{x}u_{xx}+uu_{xxx},
	\end{equation}	
where	$x\in \mathbb{R}, \textit{t}>0$, and $u$, the velocity field, is a function of the position $x$ at time $t$ (see \cite{LiuH}). We write $u(x,t)$, depicting the water's free surface over a flat bottom. $u_x$ and $u_t$ are the partial derivatives of $u$, with respect of $x$ and $t$.
	
	The extensive literature on the CH-equation contains investigations of every-other aspects of the models via countless methods. For examples, Constantin and Bressan \cite{constantinCH1} utilized a transformation method to obtain global conservative solutions for the CH equation. The significance of the CH and DP equations as shallow water waves models was proven in \cite{constantinCH2}. The scattering problem of the CH equation was considered in \cite{constantinCH3}. Wei et al \cite{WeiMJ} studied the travelling wave solutions for a generalized CH equation using the bifurcation method of dynamical system. Further, Du et al \cite{DuZ}  proved the existence of solitary wave solutions for a delayed CH equation by geometrical approach. Furthermre, Ge and De \cite{DuZ-AML} studied the shallow water wave equation by singular perturbation approach.  Most recently, Chu et al \cite{CHUjf2} considered the spectral problem for a modified CH equation. Sun et al \cite{SunXB1} proved coexistence of the solitary and periodic waves in convecting shallow water fluid.
	
	 The importance of the CH equation in fluid dynamics is due to its remarkable properties. Among which, conservation laws, integrability, peakons, soliton solutions \cite{CamassaH,constantinCH1}. Gui et al. and Luo et al. in \cite{GuiGL, Gui2, Gui3,Luo} derived a series of shallow-water waves equations with the Coriolis effect, namely, the Rotation-Green-Naghdi (R-GN), the Rotation-Korteweg-De Vries (R-KdV) and Rotation-Camassa-Holm (R-CH) equations. The R-CH equation reads,
	\begin{equation}\label{R-CH Equation}
		\begin{cases}
			m_{t}+um_{x}+2u_{x}m+ku_{x}-\frac{\beta_0}{\beta}u_{xxx}+\frac{\omega_1}{\alpha^2}u^2u_{x}+\frac{\omega_2}{\alpha^3}u^3u_{x}=0,~~~x\in \mathbb{R},\textit{t}>0, \\
			m=u-u_{xx},\\
		\end{cases}
	\end{equation}
	where the constants appearing in the equation are defined as:
	$k=\sqrt{1+\Omega^2}-\Omega,  \alpha=\frac{k}{1+k^2},$
	$\beta_0=\frac{k(k^4+6k^2-1)}{6(k^2+1)},  \beta=\frac{3k^4+8k^2-1}{6(k^2+1)},$
	$\omega_1=\frac{-3k(k^2-1)(k^2-2)}{2(1+k^2)^3}, \omega_2=\frac{(k^2-2)(k^2-1)^2(8k^2-1)}{2(1+k^2)^5}$,
	and $\Omega$ is the Coriolis frequency caused by the Earth rotation (\cite{NiLD}\cite{CHUjf1}). System (\ref{R-CH Equation}) is analogous to the CH-equation, but describing the motion of the fluid with the Coriolis effect from the incompressible and irrotational two-dimensional shallow-water
	in the equatorial region. For more details on the equatorial waves modeling and dynamics, one can refer to \cite{CHUjf-SPAM,Constantin,Constantin2,TsunamiNumerical,Constantinexactsol,Constantinexactsol1,Tsu exactsol}.
	We notice that, without the Coriolis effect ( i.e. $\Omega=0$), we have the classical Camassa-Holm with $\lambda$ equal to one. In \cite{TuXY}, the authors proved the global existence and uniqueness of the energy
	conservative weak solutions in the energy space $H^1$ of system (\ref{R-CH Equation}). Moreover, Fan et al \cite{Fan} obtained peakon weak solutions for the two component R-CH-equation formulated in \cite{FanL}. Recently, Liang et al \cite{Liang} analyzed the bifurcations and exact solutions of an asymptotic R-CH equation considered in \cite{Constantin4}. 
	
	The study of the CH-equation brought forward many generalizations of the model. Following these ideas, Liu \cite{LiuH} derived a type of nonlocal dispersive models, namely, the $\theta$-equation:
	\begin{equation}\label{theta-Equation}
		\begin{cases}
			m_{t}+\theta um_{x}+(1-\theta)u_{x}m=0,\\
			m=u-u_{xx}.\\
		\end{cases}
	\end{equation}
	It is easy to see that for $\theta=\frac{1}{3}$ and $\theta=\frac{1}{4}$, the $\theta$-equation, respectively, becomes the CH and DP \cite{Degasperis} equations. Wen \cite{WenZS2} will later study the bifurcations of traveling wave solutions of Eq. (\ref{theta-Equation}).
	
	Motivated by the aforementioned researches, we aim to investigate the behavior of the $\theta$-equation, at the equator neighborhood by mean of the transformation technique of dynamical
	system (\cite{Li-Book,Lijibin1,zhang,LiLiu,DUzj,SongYL2,SunXB2}).
	In this paper, we study the 
	Rotation-$\theta$ equation ($\theta$-equation augmented with the Coriolis effect) taking the form of:
	\begin{equation}\label{Rotation-Theta}
		\begin{cases}
			m_{t}+\theta um_{x}+(1-\theta)u_{x}m+ku_{x}-\frac{\beta_0}{\beta}u_{xxx}+\frac{\omega_1}{\alpha^2}u^2u_{x}+\frac{\omega_2}{\alpha^3}u^3u_{x}=0,~~~x\in \mathbb{R},\textit{t}>0, \\
			m=u-u_{xx},~~~x\in\mathbb{R}.\\
		\end{cases}
	\end{equation}
	Notice that, for $\theta=\frac{1}{3}$, system (\ref{Rotation-Theta}) is exactly the Rotation-CH considered in \cite{GuiGL}.  For $\theta=1$,  system (\ref{Rotation-Theta}) reduces to the asymptotic Rotation-CH considered in \cite{Liang}. For $\theta=\frac{1}{4}$,  system (\ref{Rotation-Theta}) reduces to the Rotation-Degasperis-Procesi equation. To study the travelling wave of (\ref{Rotation-Theta}),
	we transform the partial differential equation (PDE) to an ordinary differential equation (ODE). We introduce the new variable $\xi$ such that $\xi=x-ct$ with  $c$ denoting the wave speed.
	Next, we let $u(x,t)=\phi(\xi)$. Substituting $u(x,t)$ in system (\ref{Rotation-Theta}), we obtain
	\begin{equation}\label{Eq.5}
		-c\phi_{\xi}+c\phi_{\xi\xi\xi}+\phi\phi_{\xi}+k\phi_{\xi}-\frac{\beta_0}{\beta}\phi_{\xi\xi\xi}+\frac{\omega_1}{\alpha^2}\phi^2\phi_{\xi}+\frac{\omega_2}{\alpha^3}\phi^3\phi_{\xi}=(1-\theta)\phi_\xi\phi_{\xi\xi}+\theta\phi\phi_{\xi\xi\xi}.
	\end{equation}
	Integrating  Eq. (\ref{Eq.5}) once, taking the integration constant null, leads to
	\begin{equation}\label{Eq.6}
		-c\phi+c\phi_{\xi\xi}+\frac{1}{2}\phi^2+k\phi-\frac{\beta_0}{\beta}\phi_{\xi\xi}+\frac{\omega_1}{3\alpha^2}\phi^3+\frac{\omega_2}{4\alpha^3}\phi^4=\frac{1}{2}(1-2\theta)(\phi_{\xi})^2+\theta\phi\phi_{\xi\xi}.
	\end{equation}
	For convenience, we set  $C_1=c-\frac{\beta_0}{\beta}$, $C_2=\frac{\omega_1}{3\alpha^2}$, $C_3=\frac{\omega_2}{4\alpha^3}$,and $K=-c+k$.
	
	Then, we pose $\frac{d\phi}{d\xi}=y$ and obtain the planar system
	\begin{equation}\label{planar system}
		\begin{cases}
			\frac{d\phi}{d\xi}=y,\\
			\frac{dy}{d\xi}=\frac{(\theta-\frac{1}{2})y^2+C_3\phi^4+C_2\phi^3+\frac{1}{2}\phi^2+K\phi}{\theta\phi-C_1}.\\
		\end{cases}
	\end{equation}
	Subsequently, we derive the first integral of system (\ref{planar system}). From system (\ref{planar system}), we write
	\begin{equation}\label{Eq.8}
		[(\theta-\frac{1}{2})y^2+C_3\phi^4+C_2\phi^3+\frac{1}{2}\phi^2+K\phi]d\phi-(\theta\phi-C_1)ydy=0.
	\end{equation}
	Multiplying Eq.(\ref{Eq.8}) by the integrating factor $\mu(\phi)=(\phi-\frac{C_1}{\theta})^{\big(\frac{1-3\theta}{\theta}\big)}$, we obtain the exact ordinary differential equation
	\begin{equation}\label{Eq.9}
		[(\theta-\frac{1}{2})y^2+C_3\phi^4+C_2\phi^3+\frac{1}{2}\phi^2+K\phi](\phi-\frac{C_1}{\theta})^{(\frac{1-3\theta}{\theta})}d\phi-\theta(\phi-\frac{C_1}{\theta})^{(\frac{1-2\theta}{\theta})}ydy=0.
	\end{equation}
	Therefore, we have
	\begin{align}\label{Eq.10,11}
		H(\phi,y)&=\int-\theta(\phi-\frac{C_1}{\theta})^{(\frac{1-2\theta}{\theta})}ydy \\ &=-\frac{1}{2}\theta(\phi-\frac{C_1}{\theta})^{(\frac{1-2\theta}{\theta})}y^2 + p(\phi).
	\end{align}
	We easily get, $p'(\phi)=(C_3\phi^4+C_2\phi^3+\frac{1}{2}\phi^2+K\phi)(\phi-\frac{C_1}{\theta})^{(\frac{1-3\theta}{\theta})}$. It follows,
	\begin{equation}\label{Eq.12}
		p(\phi)=\int (C_3\phi^4+C_2\phi^3+\frac{1}{2}\phi^2+K\phi)(\phi-\frac{C_1}{\theta})^{(\frac{1-3\theta}{\theta})}d\phi.
	\end{equation}
	In the above equation, set $m=\frac{1-3\theta}{\theta}$ with $m\in\mathbb{Z}$. Finally, we arrive at the general first integral of system (\ref{planar system}):
	\begin{equation}\label{general first integral}
		\begin{aligned}
			H(\phi,y)=-\frac{1}{2}\theta y^2(\phi-\frac{C_1}{\theta})^{m+1}+C_3\sum_{n=0}^{m}\frac{\binom{m}{n}\phi^{m-n+5}(\frac{C_1}{\theta})^n}{m-n+5}\\
			+C_2\sum_{n=0}^{m}\frac{\binom{m}{n}\phi^{m-n+4}(\frac{C_1}{\theta})^n}{m-n+4}+\frac{1}{2}\sum_{n=0}^{m}\frac{\binom{m}{n}\phi^{m-n+3}(\frac{C_1}{\theta})^n}{m-n+3}\\
			+K\sum_{n=0}^{m}\frac{\binom{m}{n}\phi^{m-n+2}(\frac{C_1}{\theta})^n}{m-n+2},\\
		\end{aligned}
	\end{equation}
	where $\binom{m}{n}=\frac{m!}{(m-n)!}$ is the binomial coefficient. One can verify that for $m=0$, i.e. $\theta=\frac{1}{3}$.  Eq.(\ref{general first integral}) only differs from the first integral in \cite{Liang} by the value of certain parameters. Thereafter, we  consider the  cases of $\theta=\frac{1}{4}$ , $\theta=\frac{1}{2}$, $\theta=1$ in this paper. 
	
	The rest of our work is conducted as follows. In section 2, we present the bifurcations of phase portraits of system (\ref{planar system}) and classification of equilibria based on the theory of dynamical systems. In section 3, dynamics of traveling wave solutions are discussed. The main results are given in Section 4. Lastly, a conclusion ends the paper.
	
	\section{Bifurcations of Phase Portraits of system (\ref{planar system})}
	\subsection{Case of \textbf{$\theta=\frac{1}{4}$}}

	If $\theta=\frac{1}{4}$, then system (\ref{planar system}) takes the form
	\begin{equation}\label{planar system22}
		\begin{cases}
			\frac{d\phi}{d\xi}=y,\\
			\frac{dy}{d\xi}=\frac{-\frac{1}{4}y^2+C_3\phi^4+C_2\phi^3+\frac{1}{2}\phi^2+K\phi}{\frac{1}{4}\phi-C_1},\\
		\end{cases}
	\end{equation}
	with the corresponding Hamiltonian:
	\begin{equation}\label{Hamiltinian2}
		\begin{aligned}
			H(\phi,y)=-\frac{1}{8}(\phi-4C_1)^2y^2+\frac{C_3}{6}\phi^6+\frac{1}{5}(C_2-4C_1C_3)\phi^5\\
			+(\frac{1}{8}-C_1C_2)\phi^4+\frac{1}{3}(K-2C_1)\phi^3-2C_1K\phi^2=h.\\
		\end{aligned}
	\end{equation}
	Imposing the transformation $d\xi=(\theta\phi-C_1)d\tau$, the system (\ref{planar system22}) becomes the regular system
	\begin{equation}\label{planar system23}
		\begin{cases}
			\frac{d\phi}{d\tau}=y(\frac{1}{4}\phi-C_1),\\
			\frac{dy}{d\tau}=-\frac{1}{4}y^2+C_3\phi^4+C_2\phi^3+\frac{1}{2}\phi^2+K\phi.\\
		\end{cases}
	\end{equation}
	The difference between the systems (\ref{planar system22}) and (\ref{planar system23}) is in the dynamical behavior at the singular line $\phi=\frac{C_1}{\theta}$.
	
	\subsubsection{ Qualitative analysis of singular points of system (\ref{planar system23})}
	We investigate the equilibrium points of the system (\ref{planar system23}).
	
	The point (0,0) is an obvious equilibrium of the system (\ref{planar system23}). On the line $\phi=4C_1$, we have two equilibrium points $y_\pm=\pm2\sqrt{f(4C_1)}$ for $f(4C_1)>0$.
	To identify the other equilibrium points of system (\ref{planar system23}), we write
	\begin{equation}\label{polynomial1}
		f(\phi)=C_3\phi^4+C_2\phi^3+\frac{1}{2}\phi^2+K\phi=\phi*g(\phi),
	\end{equation}
	where
	\begin{equation}\label{polynomial2}
		g(\phi)=C_3\phi^3+C_2\phi^2+\frac{1}{2}\phi+K.
	\end{equation}
	$f(\phi)=0$ for $\phi=0$ or $g(\phi)=0$.
	
	We obtain,
	\begin{equation}\label{g'}
		g^{'}(\phi)=3C_3\phi^2+2C_2\phi+\frac{1}{2}
	\end{equation}
	and
	\begin{equation}
		g^{''}(\phi)=6C_3\phi+2C_2.
	\end{equation}

	Notice that, $g(0)= K$, $g^{'}(0)=\frac{1}{2}$. Additionally, $\Delta_{g^{'}}=4C_2^2-6C_3$. Assuming $C_3<0$ and $\Delta_{g^{'}}>0$.
	$g^{'}(\phi)$ admits two zeros denoted $\tilde\phi_\pm=\frac{-2C_2\pm\sqrt{\Delta}}{6C_3}$, with  $g^{''}(\tilde\phi_-)>0$ and $g^{''}(\tilde\phi_+)<0$.
	\newline
	So $g(\phi)$ has at most 3 zeros denoted $\phi_{n}, n=1,2,3$.
	
	Thus, we have the following conclusions:

	1. When $K\neq0$, and $\Delta_{g^{'}}>0$,
	
	(i)  if $g(\tilde\phi_{+})<0$, then $g(\phi)$ has only one zero, and system (\ref{planar system23}) admits four equilibrium points, $E_0(0,0)$, $E(\phi^*,0)$ and $S_\pm(4C_1,y\pm)$.
	
	(ii) if $g(\tilde\phi_{+})=0$, then $g(\phi)$ has a simple zero and a double zero. System (\ref{planar system23}) admits five equilibrium points $E_0(0,0)$, $E_i(\phi_i,0)$, $n=1,2$ and $S\pm(4C_1,y_\pm)$.
	
	(iii) if $g(\tilde\phi_{-})<0$ and $g(\tilde\phi_{+})>0$, then $g(\phi)$ has three simple zeros and the system (\ref{planar system23}) admits six equilibrium points, $E_0(0,0)$, $E_i(\phi_i,0)$, $n=1,2,3$ and $S\pm(4C_1,y_\pm)$.

	(iv) if $g(\tilde\phi_{-})=0$, then $g(\phi)$ has a simple zero and a double zero. System (\ref{planar system23}) admits five equilibrium points, $E_0(0,0)$, $E(\phi^*,0)$ and $S_\pm(4C_1,y_\pm)$.
	
	(v) if $g(\tilde\phi_{-})>0$, then $g(\phi)$ has only one zero, and system (\ref{planar system23}) admits four equilibrium points, $E_0(0,0)$, $E(\phi^*,0)$ and $S_\pm(4C_1,y\pm)$.
	
	2. When $K\neq0$, for $\Delta_{g^{'}}=0$ and $\Delta_{g^{'}}<0$,
	$g(\phi)$ admits only one zero, the conclusion  are similar to (1-i) and (1-v).
	
	3. When $K=0$, if
	
	(i) $C^2_2>2C_3$, then the system (\ref{planar system23}) admits five equilibrium points, $E_0(0,0)$, $E_{1,2}(\phi_{1,2},0)$ and $S_\pm(4C_1,y_\pm)$.
	
	(ii) $C^2_2=2C_3$, then the system (\ref{planar system23}) admits four equilibrium points, $E_0(0,0)$, $E(\phi,0)$ and $S_\pm(4C_1,y_\pm)$.
	
	(iii) $C^2_2<2C_3$, then the system (\ref{planar system23}) admits three equilibrium points, $E_0(0,0)$ and $S_\pm(4C_1,y_\pm)$.

	We only consider the cases 1(i)-(v) and 3(i), then Fig.1 directly follows
	
	\begin{center}
		\begin{tabular}{ccc}
			\epsfxsize=5cm \epsfysize=5cm \epsffile{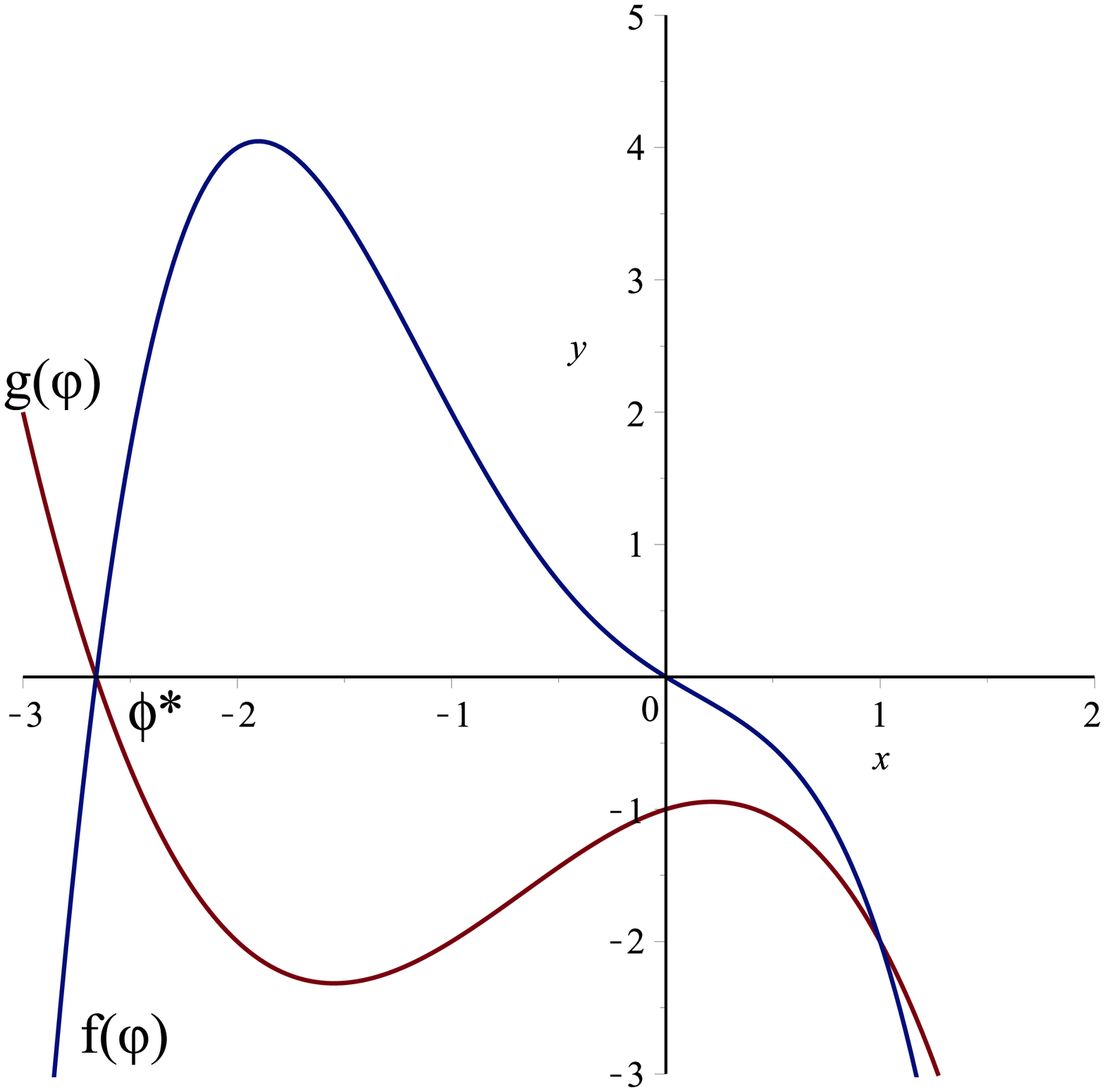}&
			\epsfxsize=5cm \epsfysize=5cm \epsffile{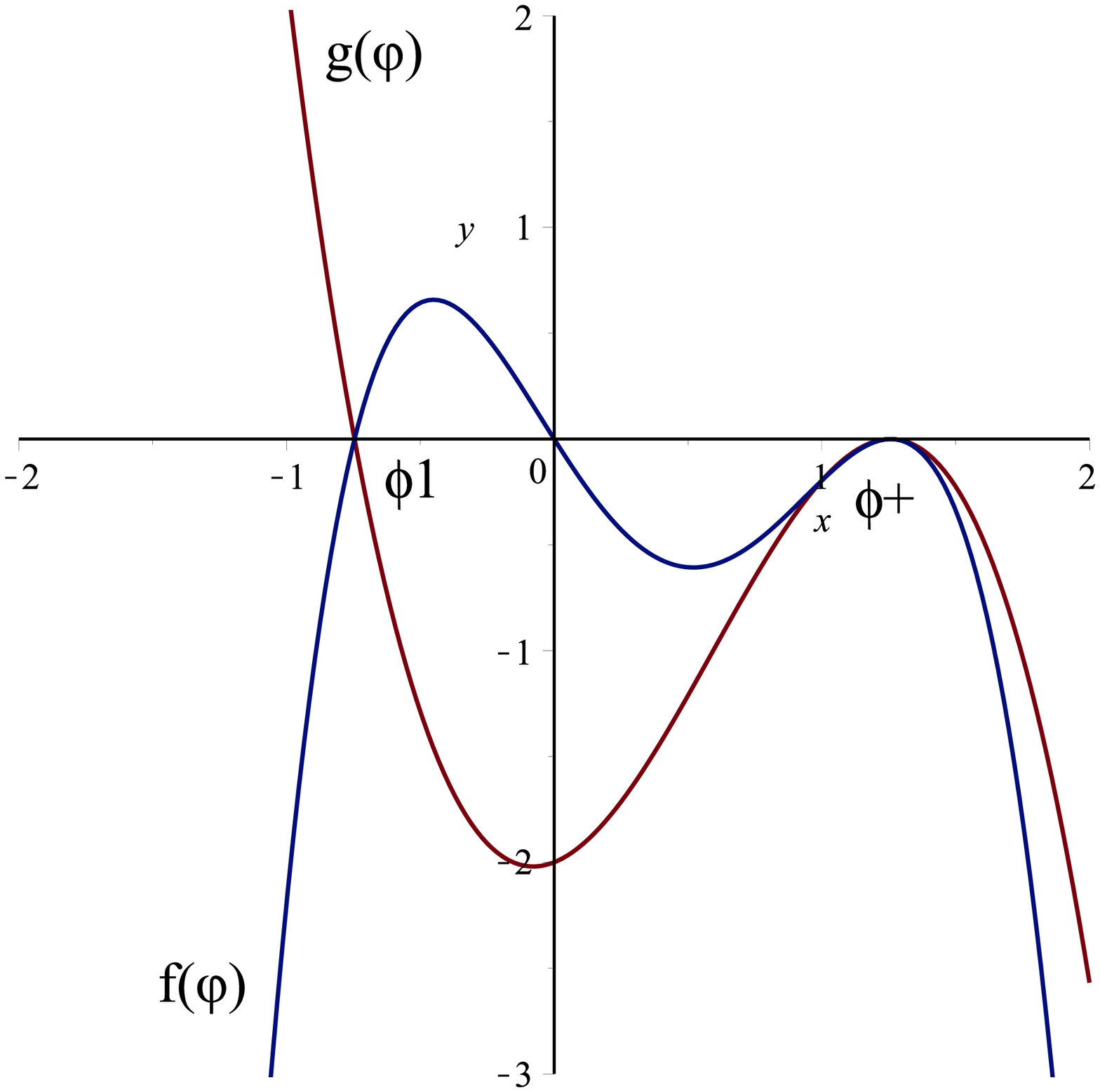}&
			\epsfxsize=5cm \epsfysize=5cm \epsffile{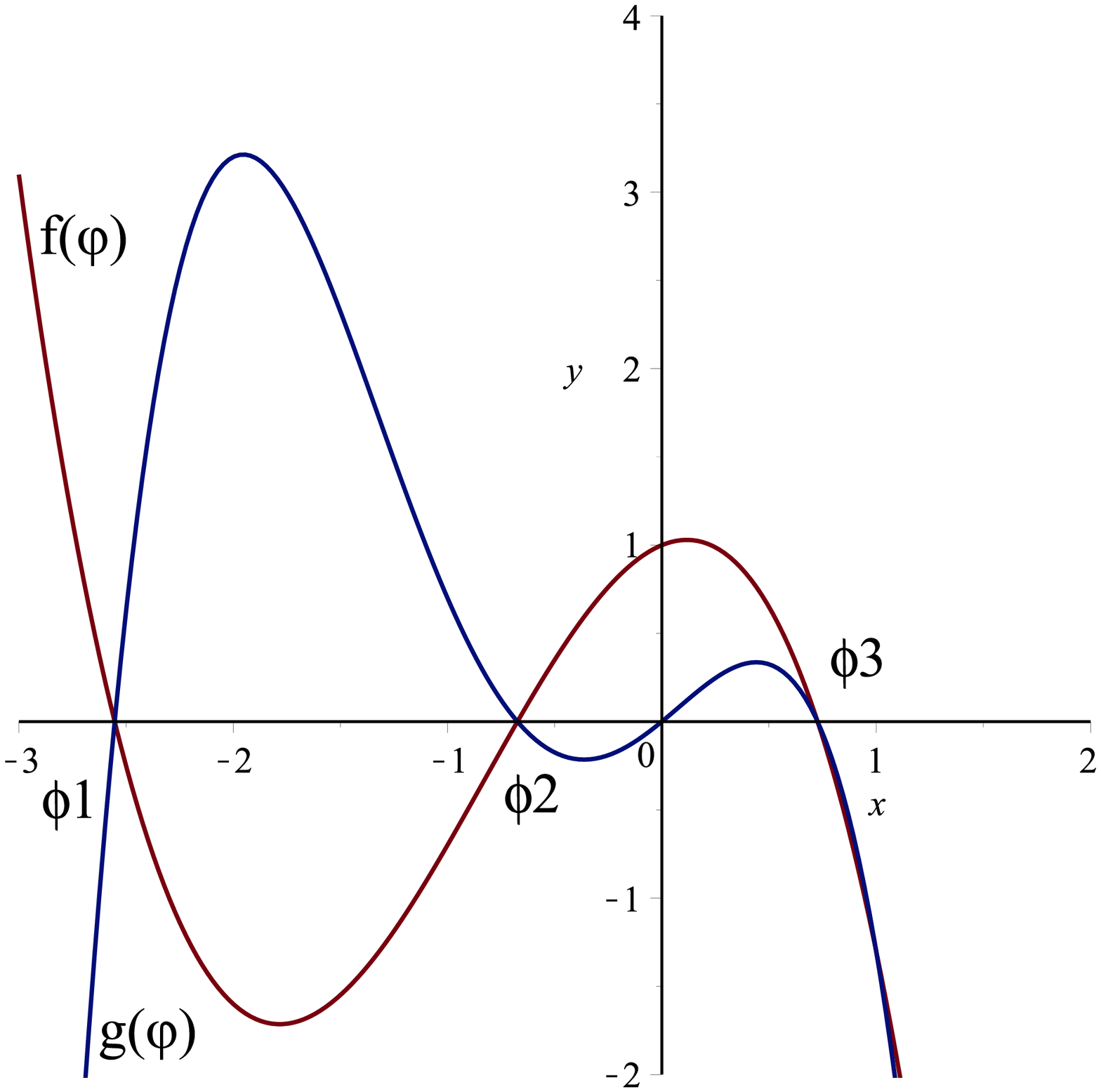}\\
			\footnotesize{ (a) $K\neq0, \Delta_{g'}>0, g(\tilde\phi_{+})<0$.} & \footnotesize{ (b) $K\neq0, \Delta_{g^{'}}>0, g(\tilde\phi_{+})=0$.}&
			\footnotesize{(c) $K\neq0, \Delta_{g'}>0, g(\tilde\phi_{+})>0$.}
		\end{tabular}
	\end{center}

\begin{center}
	\begin{tabular}{ccc}
		\epsfxsize=5cm \epsfysize=5cm \epsffile{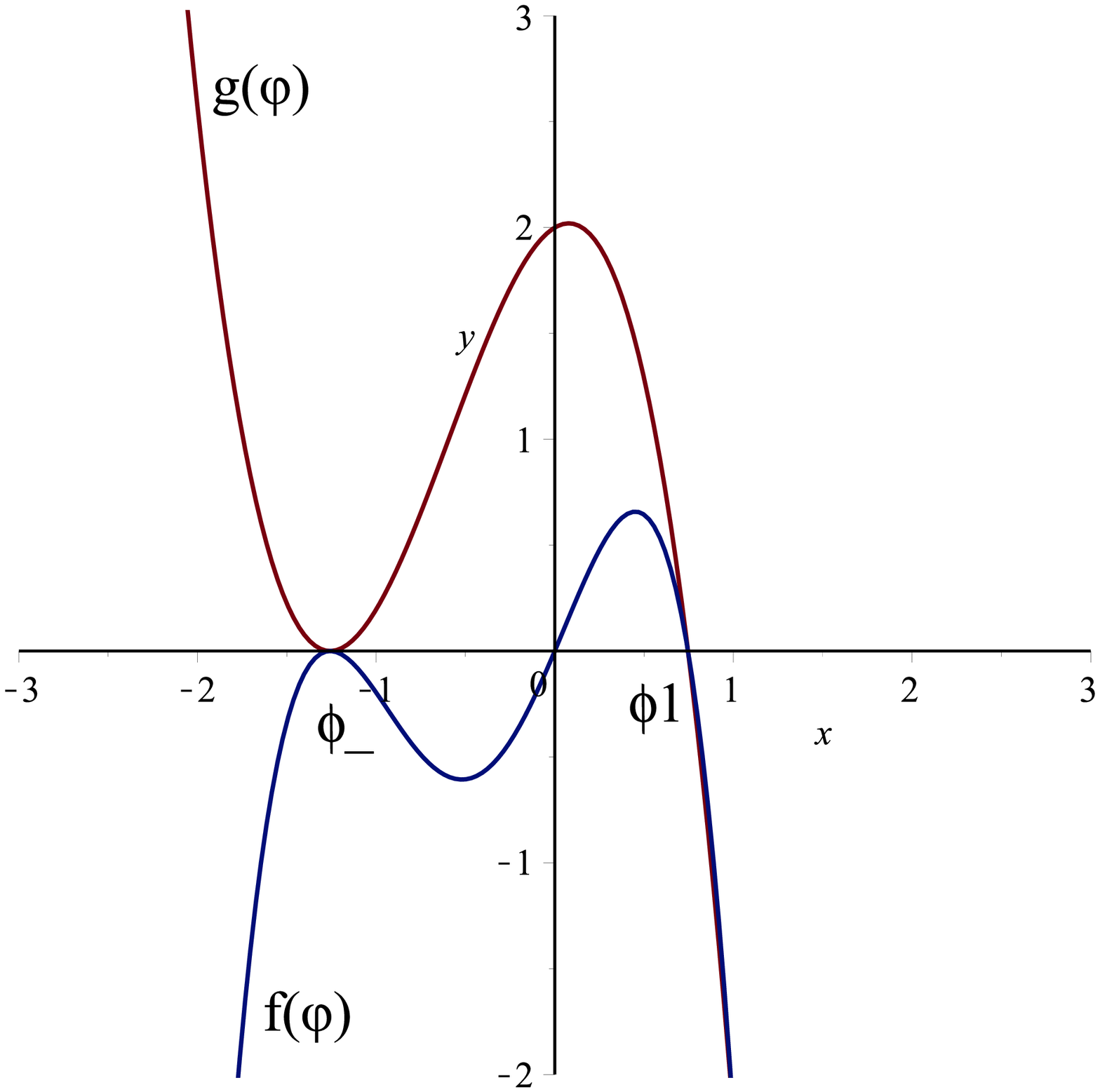}&
		\epsfxsize=5cm \epsfysize=5cm \epsffile{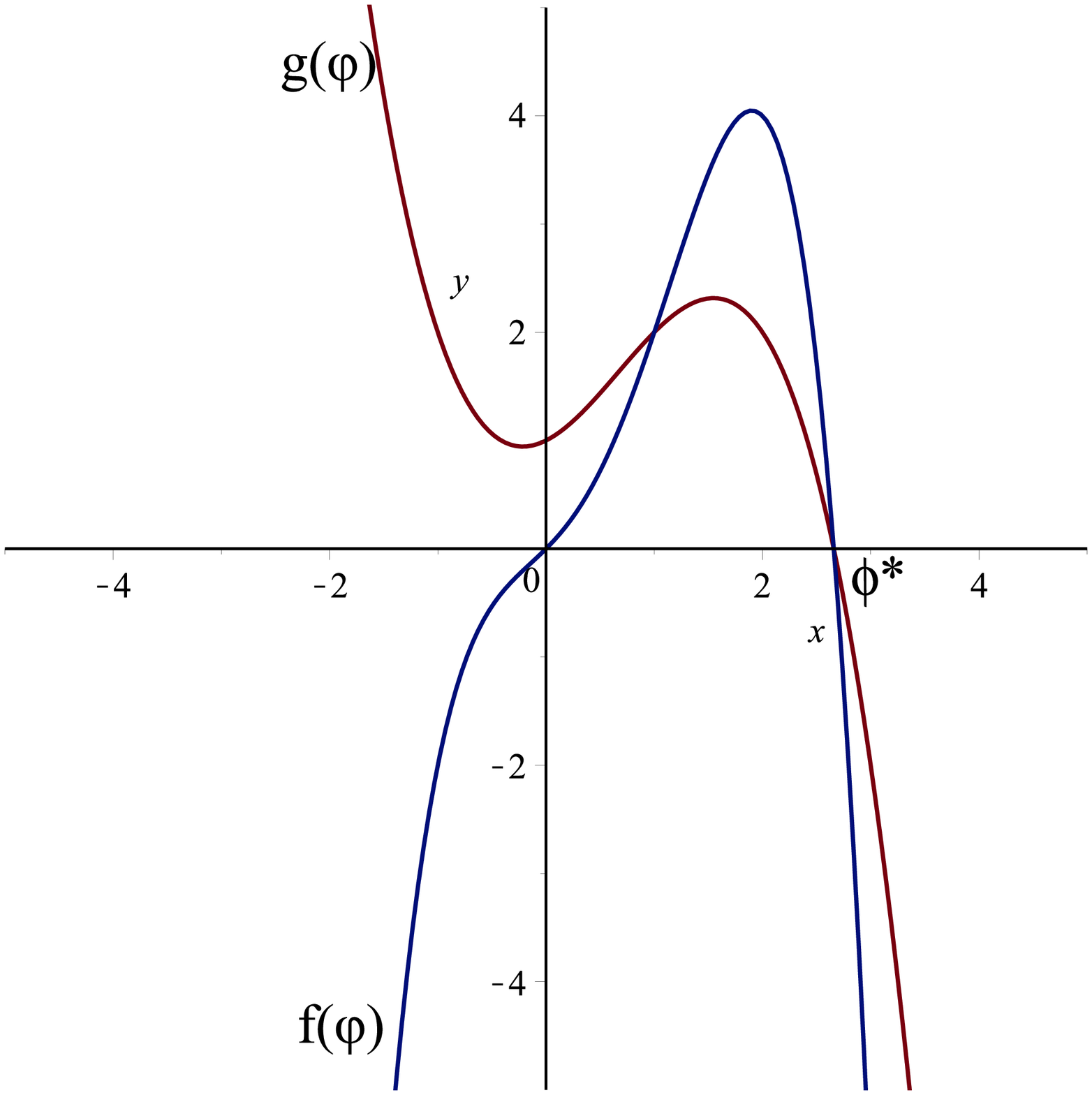}&
		\epsfxsize=5cm \epsfysize=5cm \epsffile{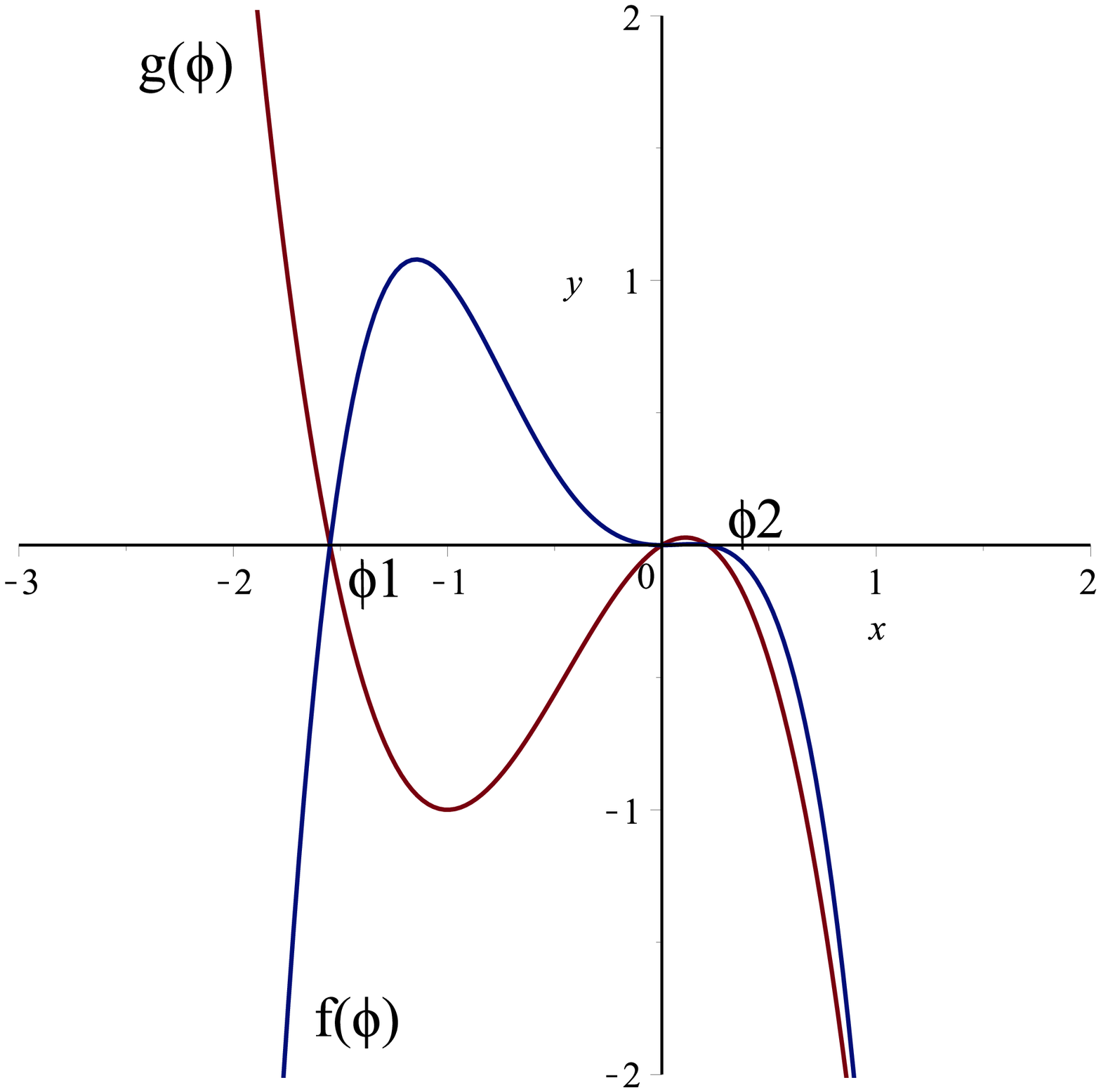}\\
		\footnotesize{(c) $K\neq0, \Delta_{g'}>0, g(\tilde\phi_{-})=0$.} & \footnotesize{ (d) $K\neq0, \Delta_{g'}>0, g(\tilde\phi_{-})>0$. }& \footnotesize{ (d) $K=0, C^2_2>2C_3$. }
	\end{tabular}
\end{center}

\begin{center}
	{\small Fig.1.  Cases 1(i)-(v) and 3(i)  }
\end{center}

\subsubsection{Classification of singular points and phase portraits of system (\ref{planar system23})}

Let $M_1(\phi,y)$ be the matrix of the linearized system of (\ref{planar system23}),

\begin{equation}\label{matrix}
	M_1(\phi,y)=
	\left(
	\begin{array}{cc}
		\frac{1}{4}y & \frac{1}{4}\phi-C_1  \\
		f^{'}(\phi) & -\frac{1}{2}y\\
	\end{array}
	\right).
\end{equation}

with
\begin{equation}
	J_1(\phi,y)=\det{M}=-\frac{1}{8}y^2-(\frac{1}{4}\phi-C_1)f^{'}(\phi) \end{equation}

Particularly,
\begin{equation}
	J_1(4C_1,y_{\pm})=-\frac{1}{8}(y_{\pm})^2
\end{equation}
and
\begin{equation}
	J_1(\phi_i,0)=-(\frac{1}{4}\phi_i-C_1)f^{'}(\phi_i)
\end{equation}

Based on the theory of planar dynamical system (see\cite{Li-Book}), for an equilibrium point of a planar integrable system, the equilibrium point is a saddle point if $J<0$; the equilibrium point is a center point (a node point) if $J>0$ and $(trM)^{2}-4J<(>)0$; the equilibrium point is a cusp if $J=0$ and the Poincar\'{e} index of the equilibrium point is $0$. For example, when they exist, the singular points $y_\pm$ are always saddle points.

Let $h_i = H(\phi_i, 0)$, $h_0 = H(0,0)$ and $h_s = H( 4C_1,y_\pm)$, where
H is given by (\ref{Hamiltinian2}).
For the aforementioned cases, we let the singular line $\phi=4C_1$ move from right to left in the $(\phi, y)-$phase plane and obtain
the following topological phase portraits of system (\ref{planar system23}) (see Fig.2-Fig.6).

\begin{center}
	\begin{tabular}{ccc}
		\epsfxsize=5cm \epsfysize=5cm \epsffile{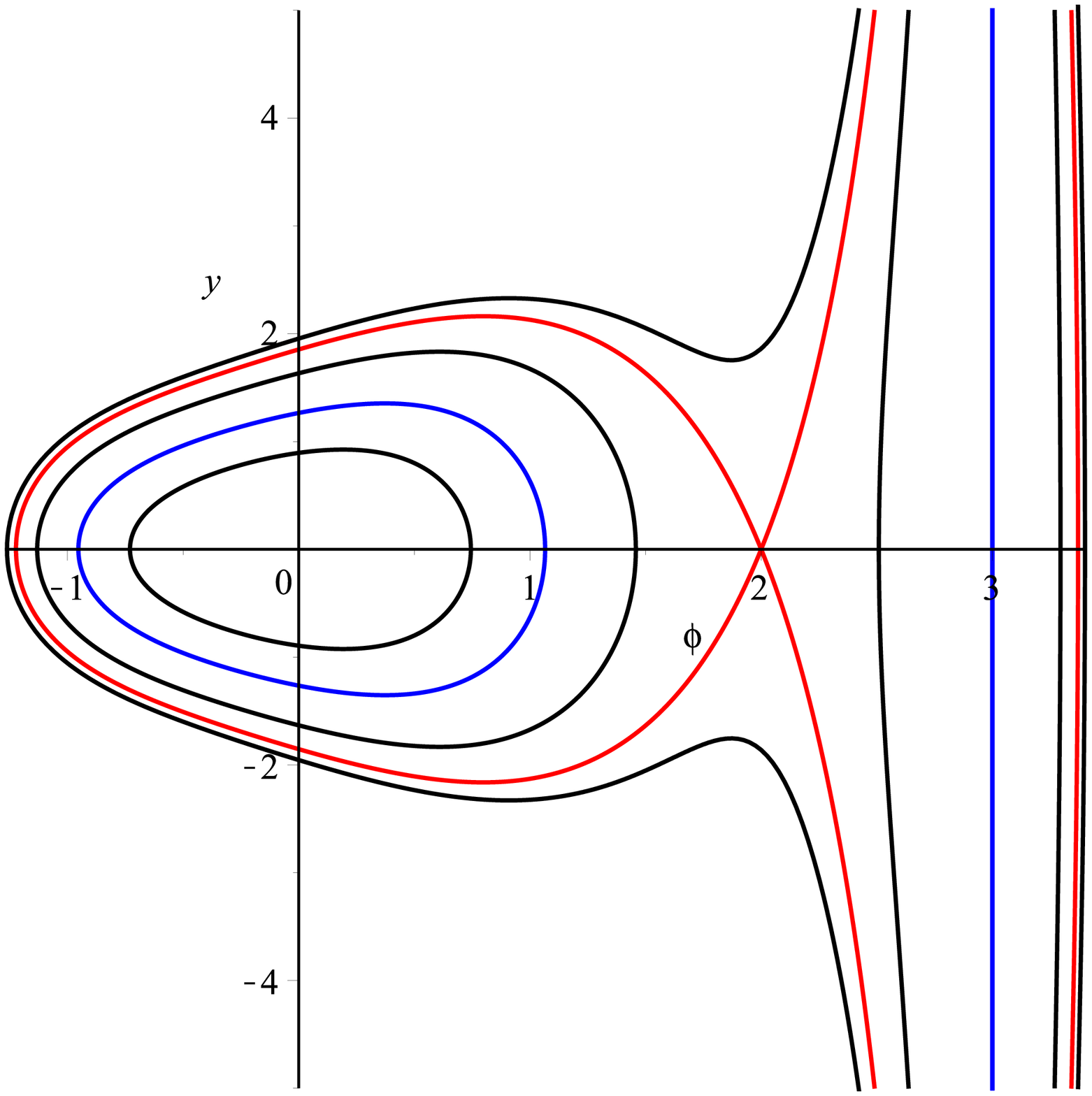}&
		\epsfxsize=5cm \epsfysize=5cm \epsffile{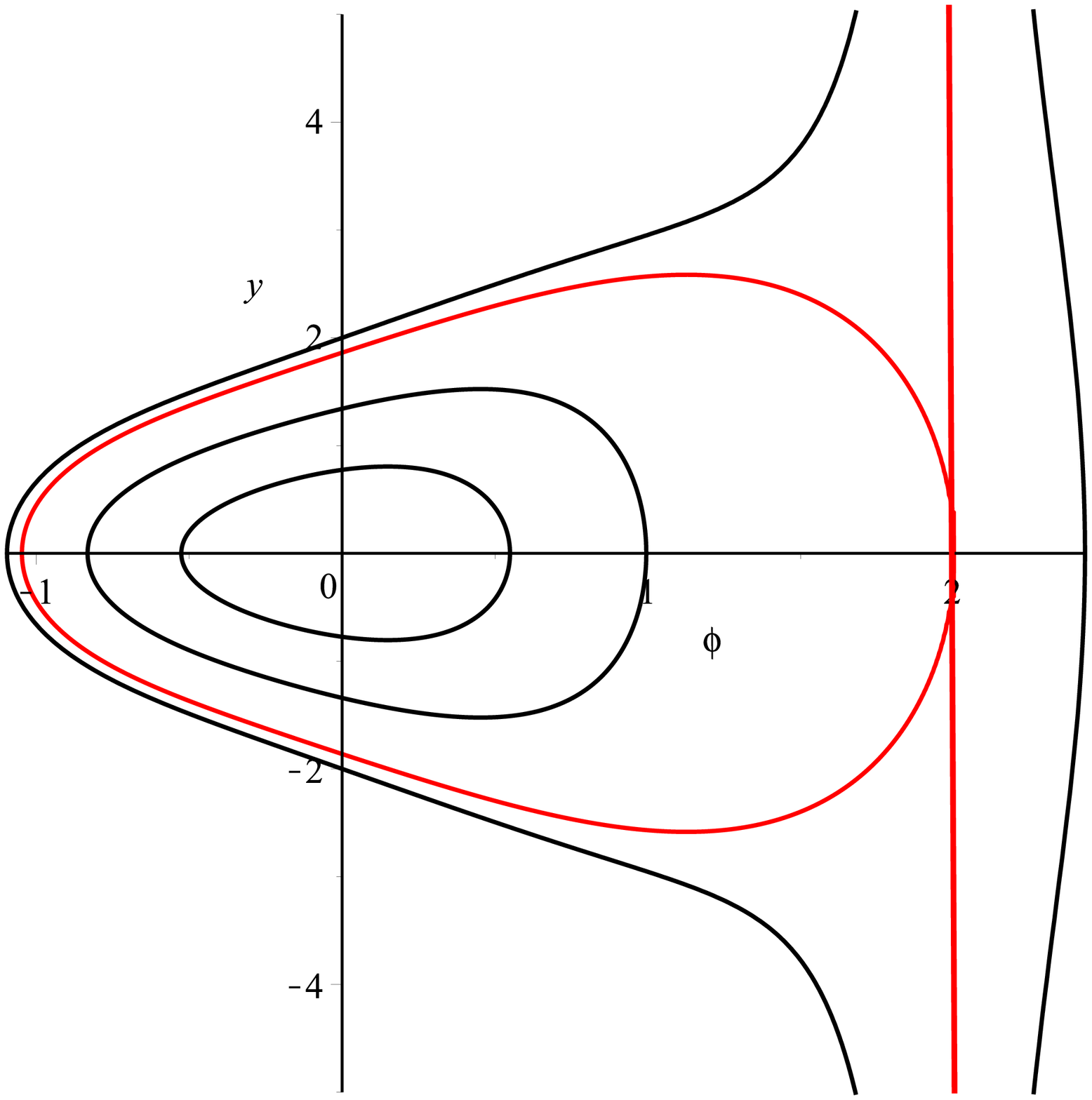} &
		\epsfxsize=5cm \epsfysize=5cm \epsffile{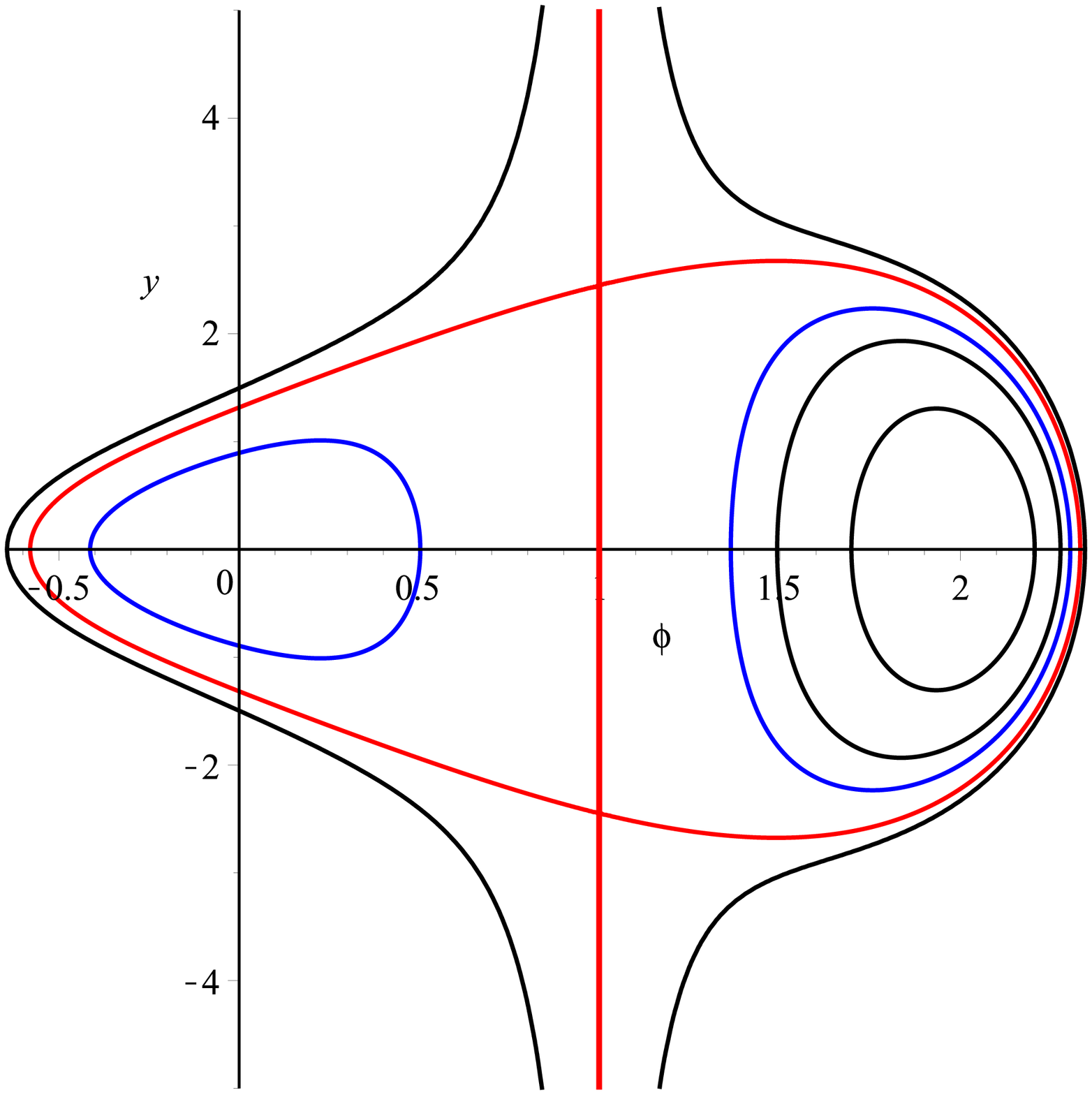}\\
		
		\footnotesize{ (a) $4C_1>\phi_1$ } & \footnotesize{(b) $4C_1=\phi_1$ }&
		\footnotesize{(c) $0<4C_1<\phi_1$ }
	\end{tabular}
\end{center}
\begin{center}
	\begin{tabular}{cc}
		\epsfxsize=5cm
		\epsfysize=5cm \epsffile{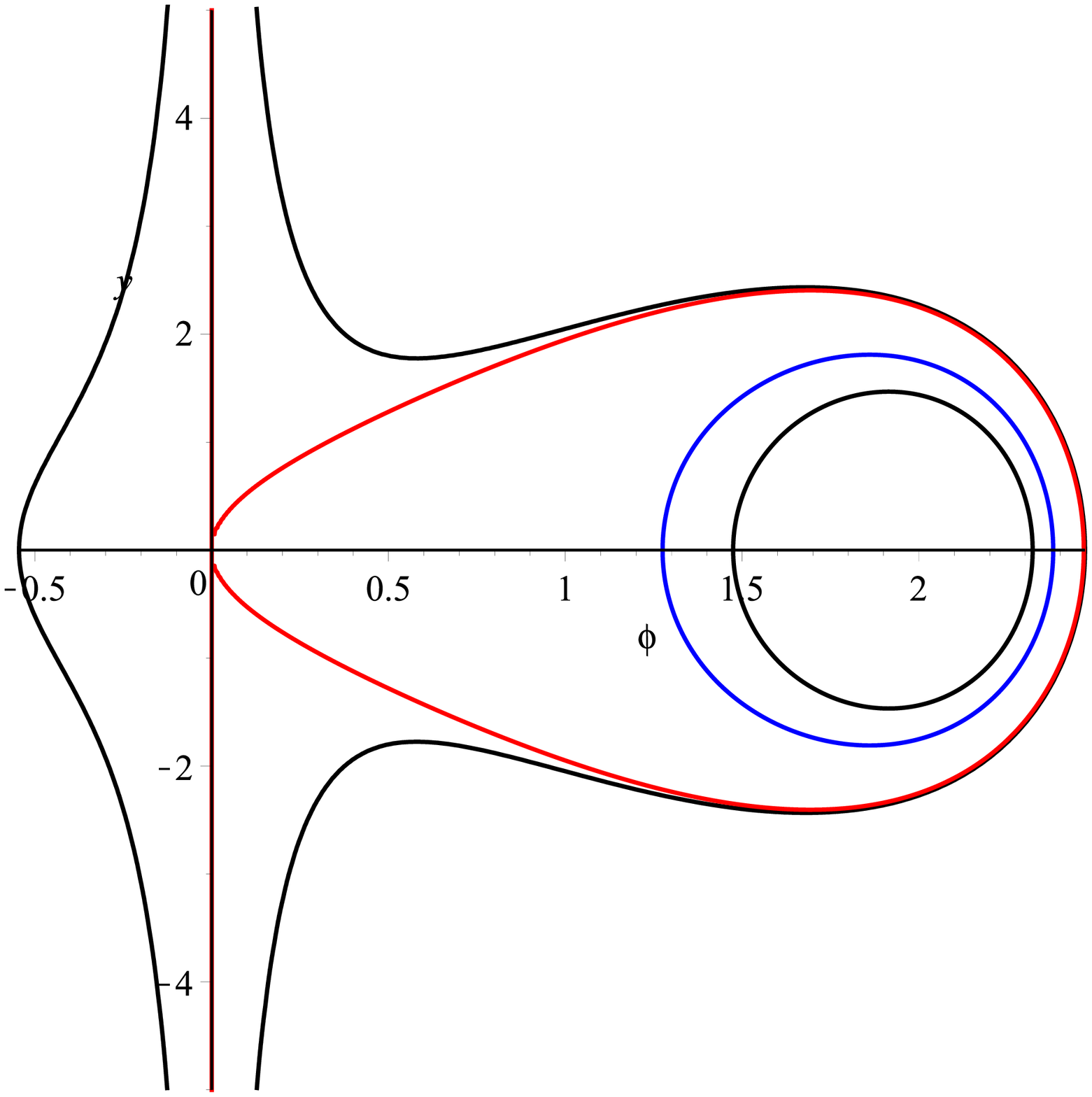}&
		\epsfxsize=5cm \epsfysize=5cm \epsffile{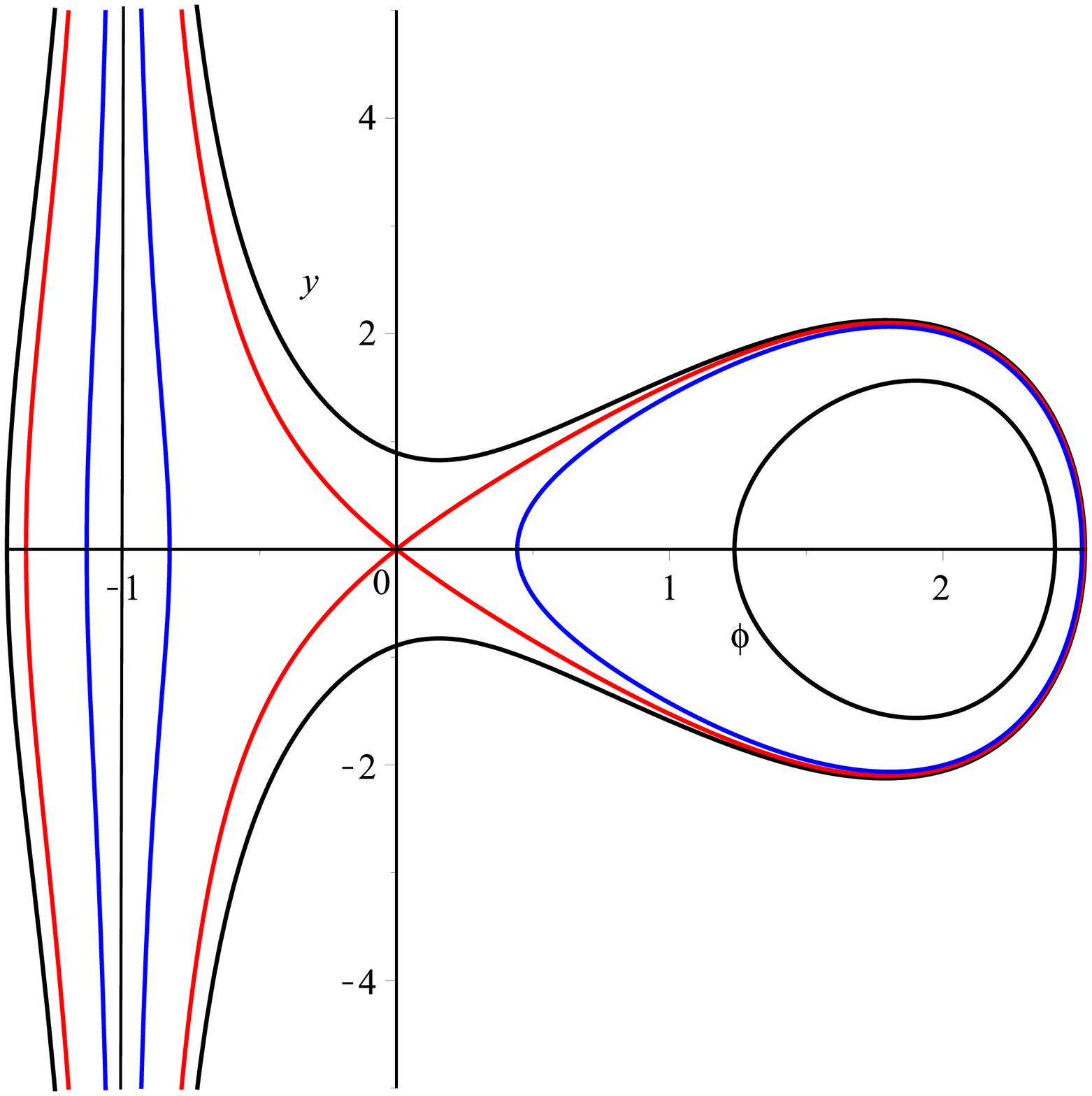}\\
		\footnotesize{ (d) $4C_1=0$ } & \footnotesize{(e) $4C_1<0$ }
	\end{tabular}
\end{center}

\begin{center}
	{\small Fig.2  The function $g(\phi)$ admits only one zero, i.e $(\Delta_g)>0$ and $g(\phi_-)>0$.}
\end{center}

\begin{center}
	\begin{tabular}{ccc}
		\epsfxsize=5cm
		\epsfysize=5cm \epsffile{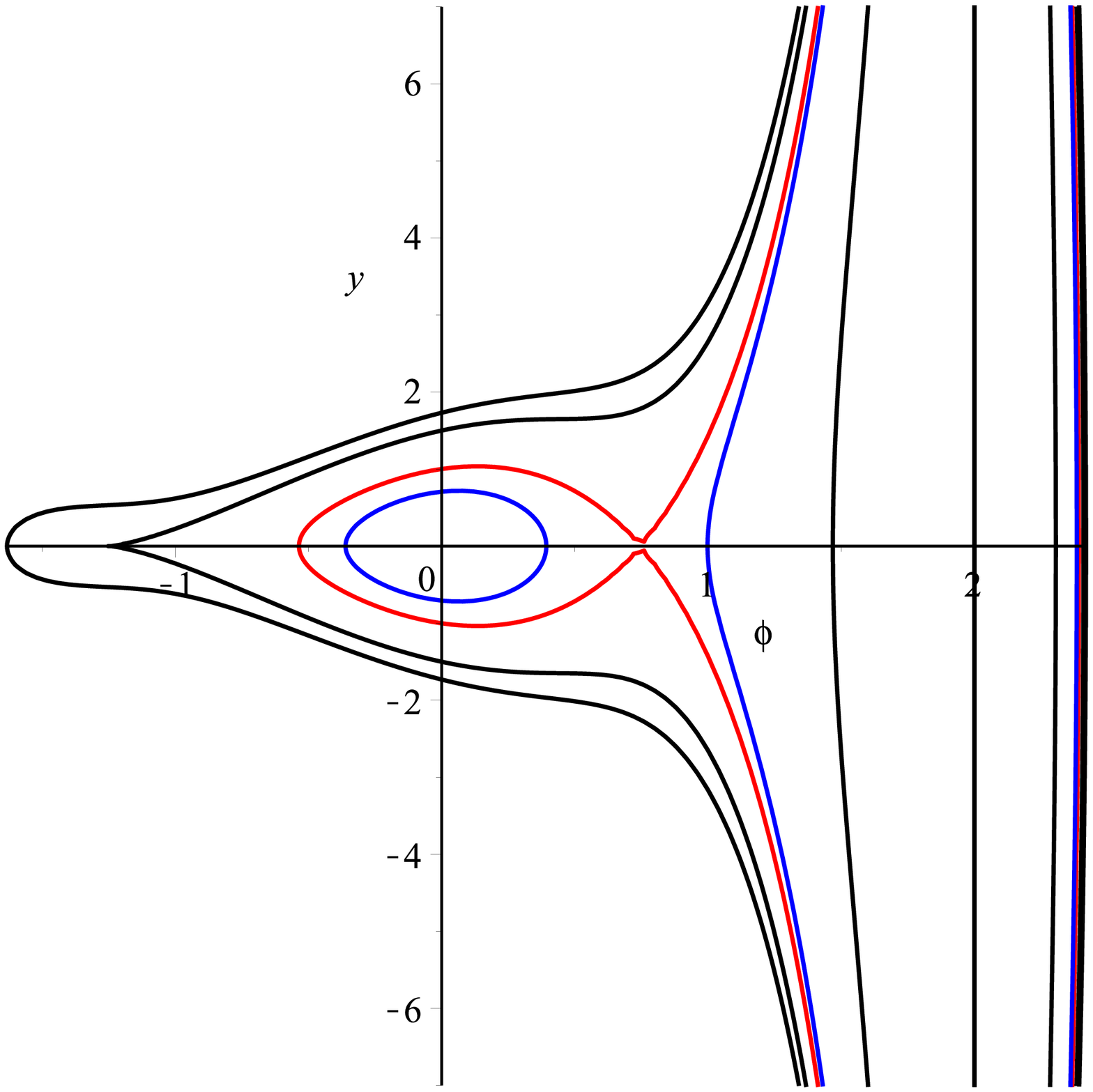}&
		\epsfxsize=5cm \epsfysize=5cm \epsffile{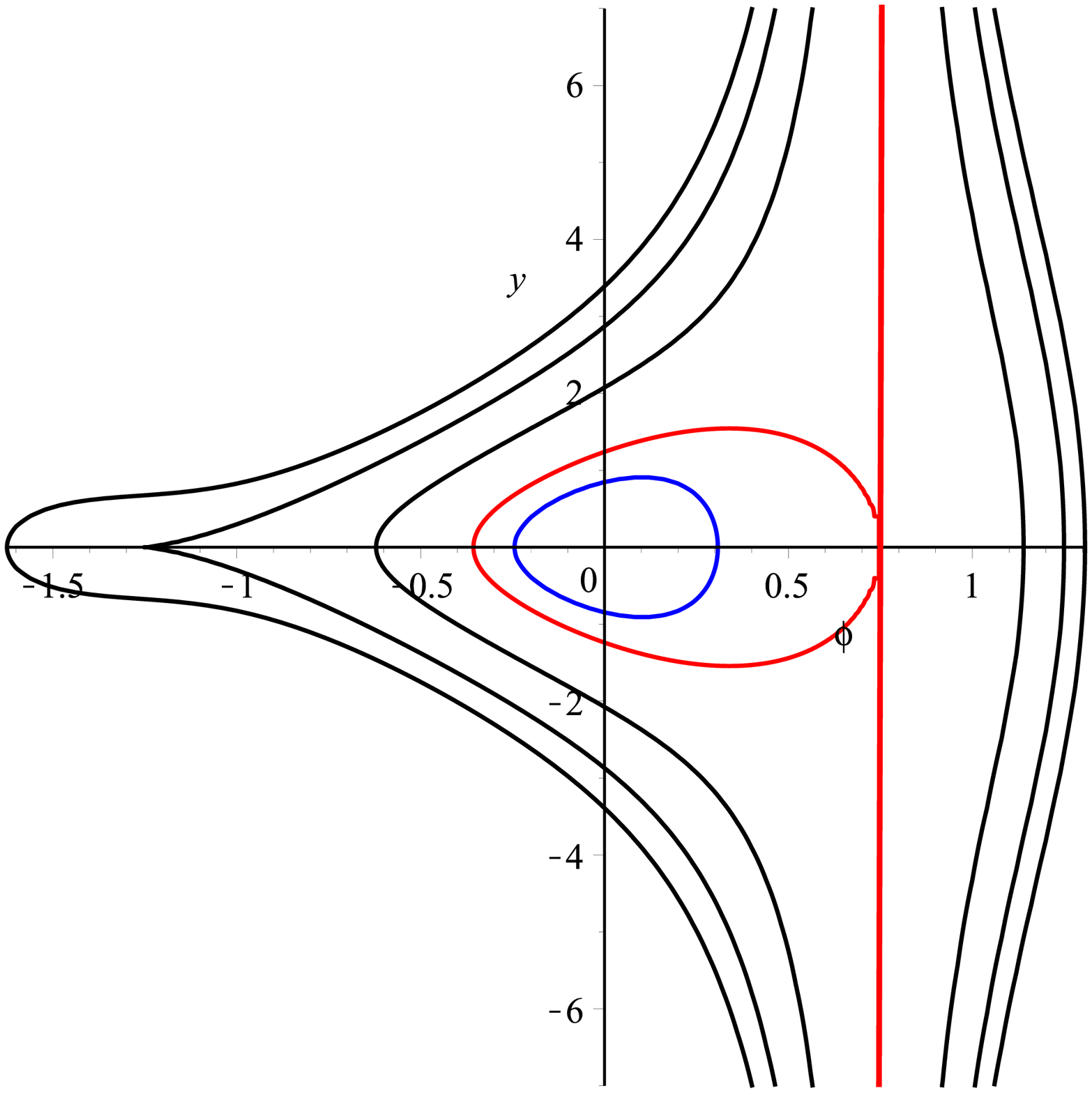}&
		\epsfxsize=5cm
		\epsfysize=5cm \epsffile{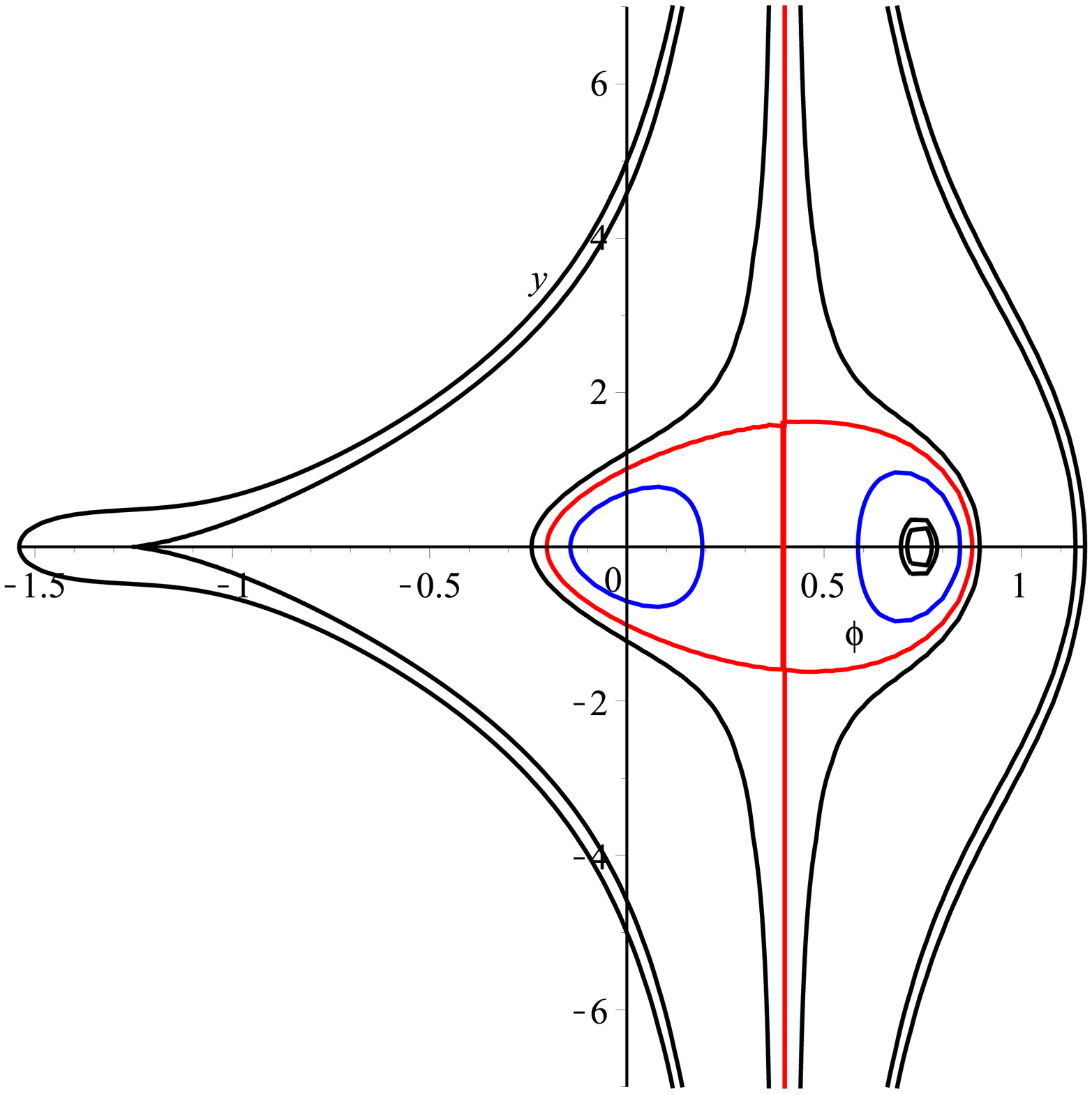}\\
		\footnotesize{ (a) $4C_1>\phi_1$ } & \footnotesize{(b) $4C_1=\phi_1$ }&
		\footnotesize{(c) $0<4C_1<\phi_1$ }
	\end{tabular}
\end{center}
\begin{center}
	\begin{tabular}{ccc}
		\epsfxsize=5cm
		\epsfysize=5cm \epsffile{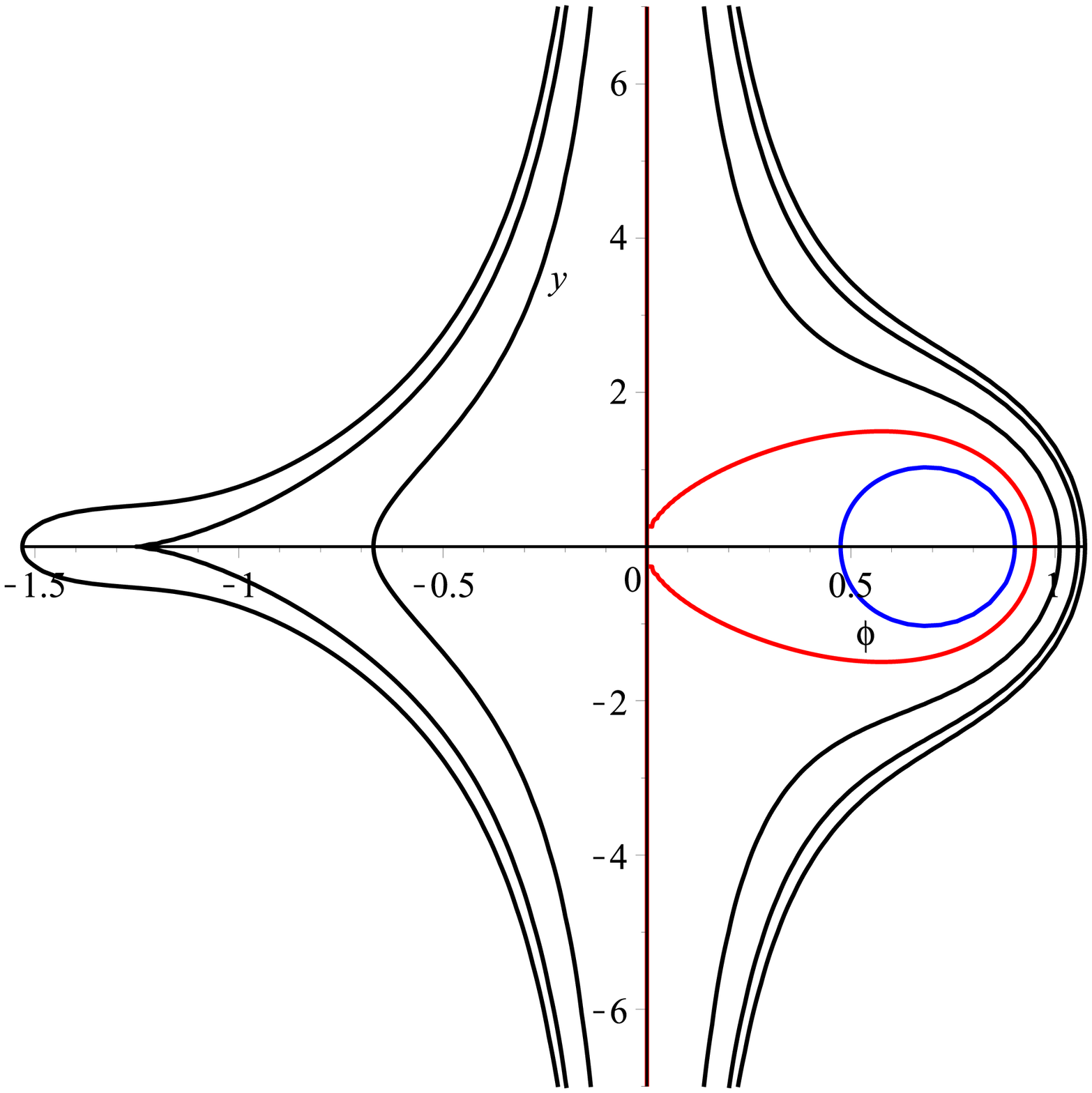}&
		\epsfxsize=5cm \epsfysize=5cm \epsffile{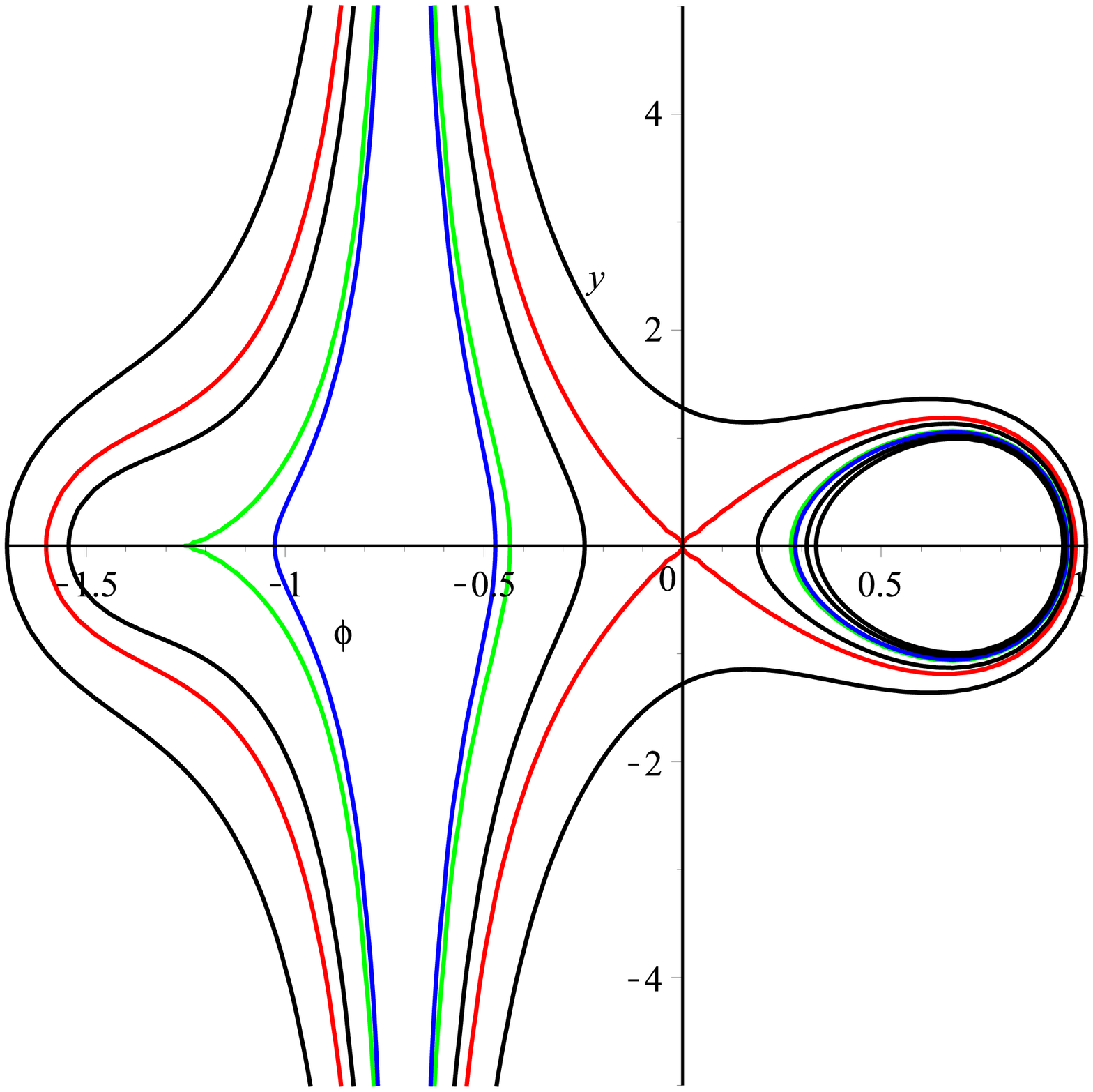}&
		\epsfxsize=5cm \epsfysize=5cm
		\epsffile{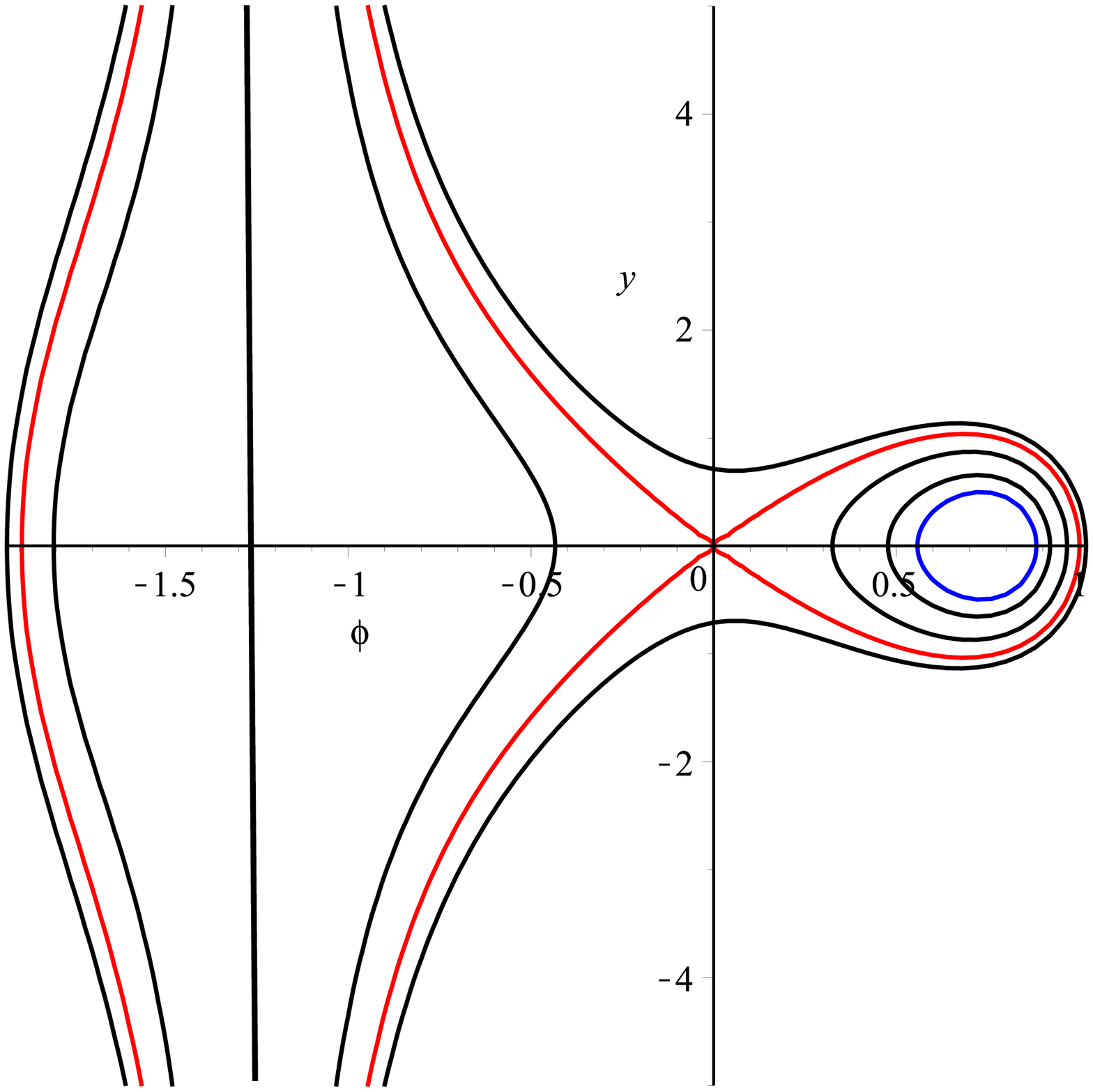}\\
		\footnotesize{ (d) $4C_1=0$ } & \footnotesize{(e) $\phi_2<4C_1<0$ }&
		\footnotesize{(f) $4C_1=\phi_2$ }
	\end{tabular}
\end{center}

\begin{center}
	\begin{tabular}{cc}
		\epsfxsize=5cm
		\epsfysize=5cm \epsffile{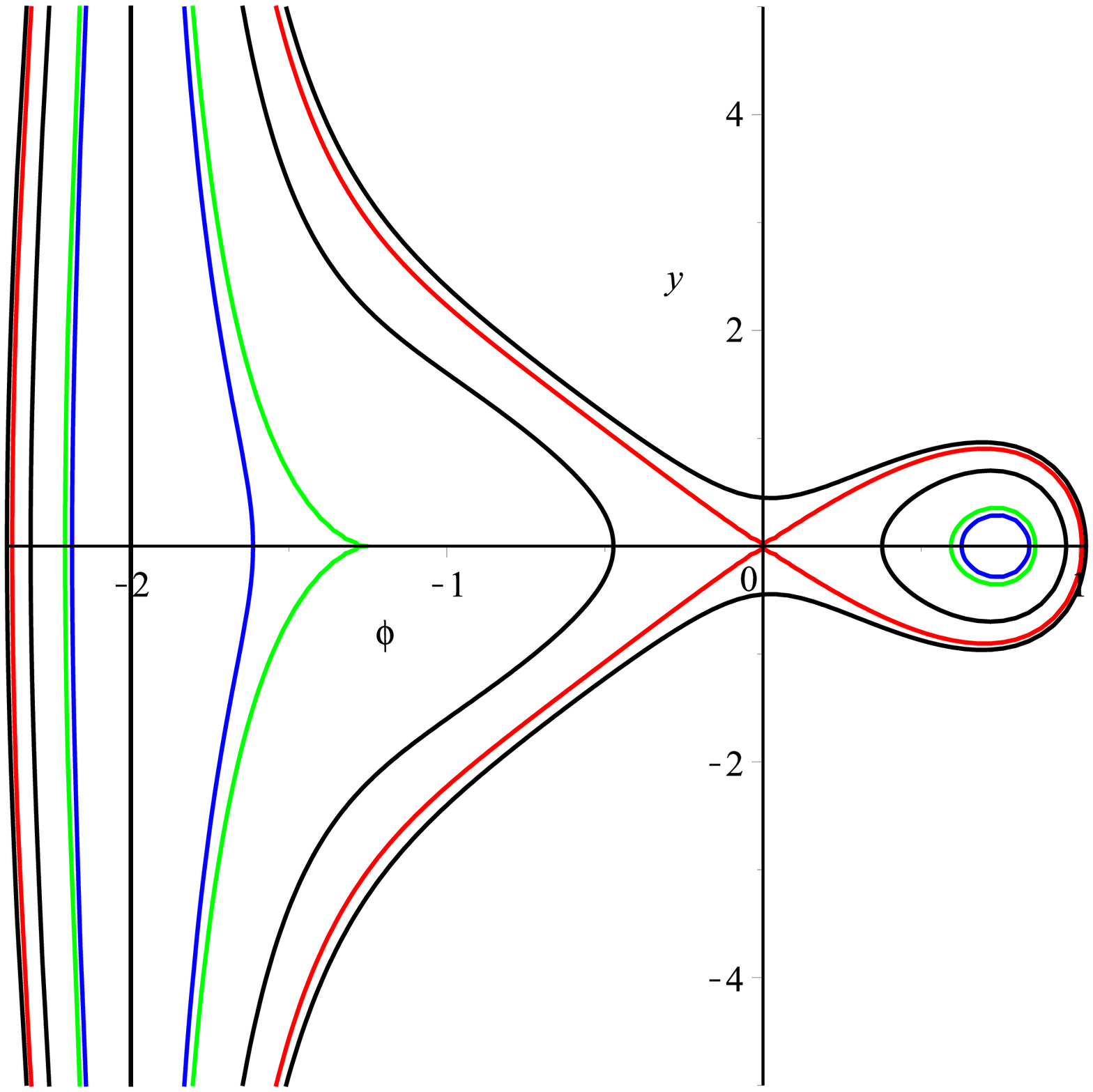}\\
		\footnotesize{ (g) $4C_1<\phi_2$ }
	\end{tabular}
\end{center}
\begin{center}
	{\small Fig.3 The function $g(\phi)$ admits a double zero , and $\tilde\phi_-<0<\phi_1$.       }
\end{center}

\begin{center}
	\begin{tabular}{ccc}
		\epsfxsize=5cm\epsfysize=5cm \epsffile{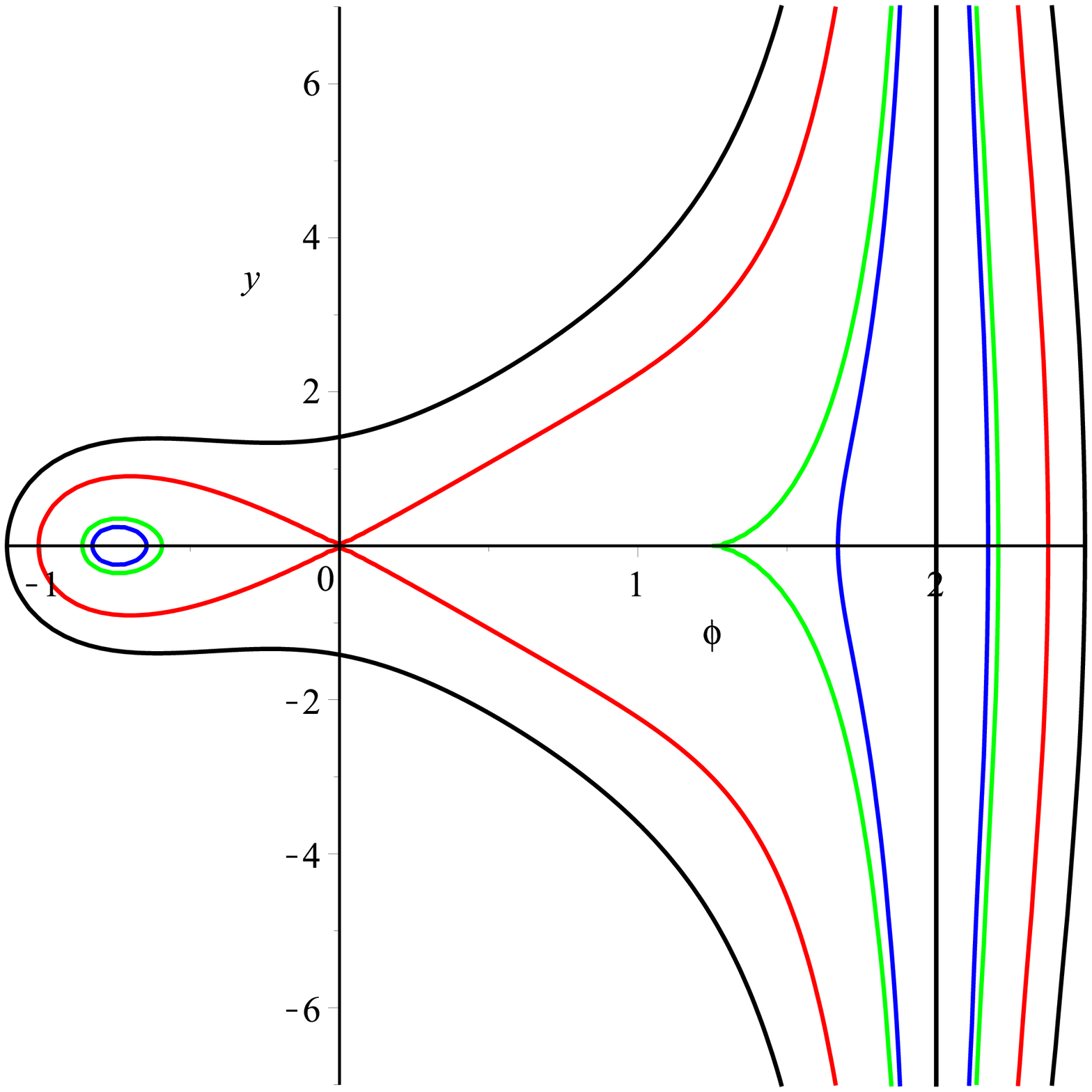}&
		\epsfxsize=5cm \epsfysize=5cm \epsffile{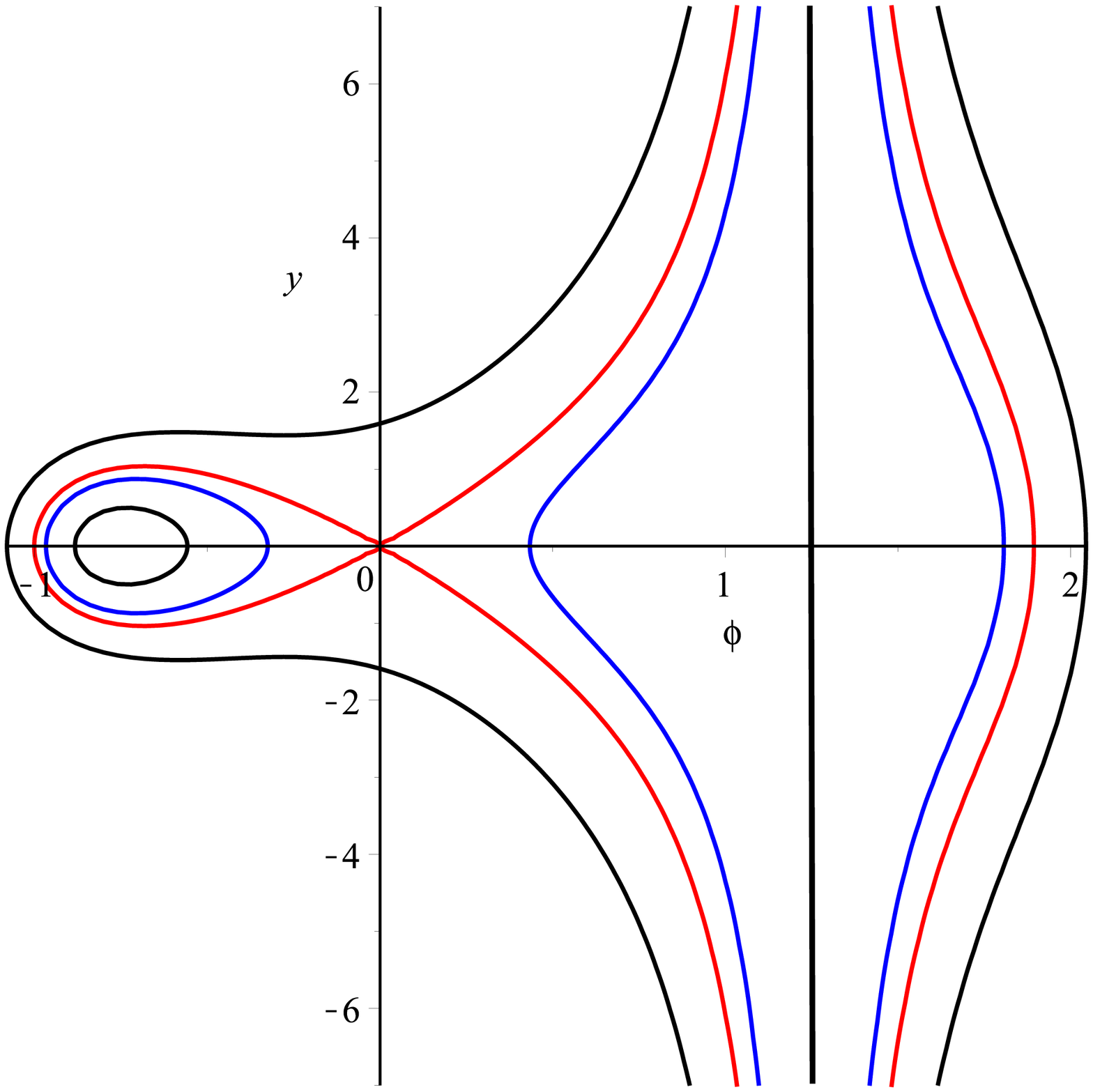}&
		\epsfxsize=5cm \epsfysize=5cm \epsffile{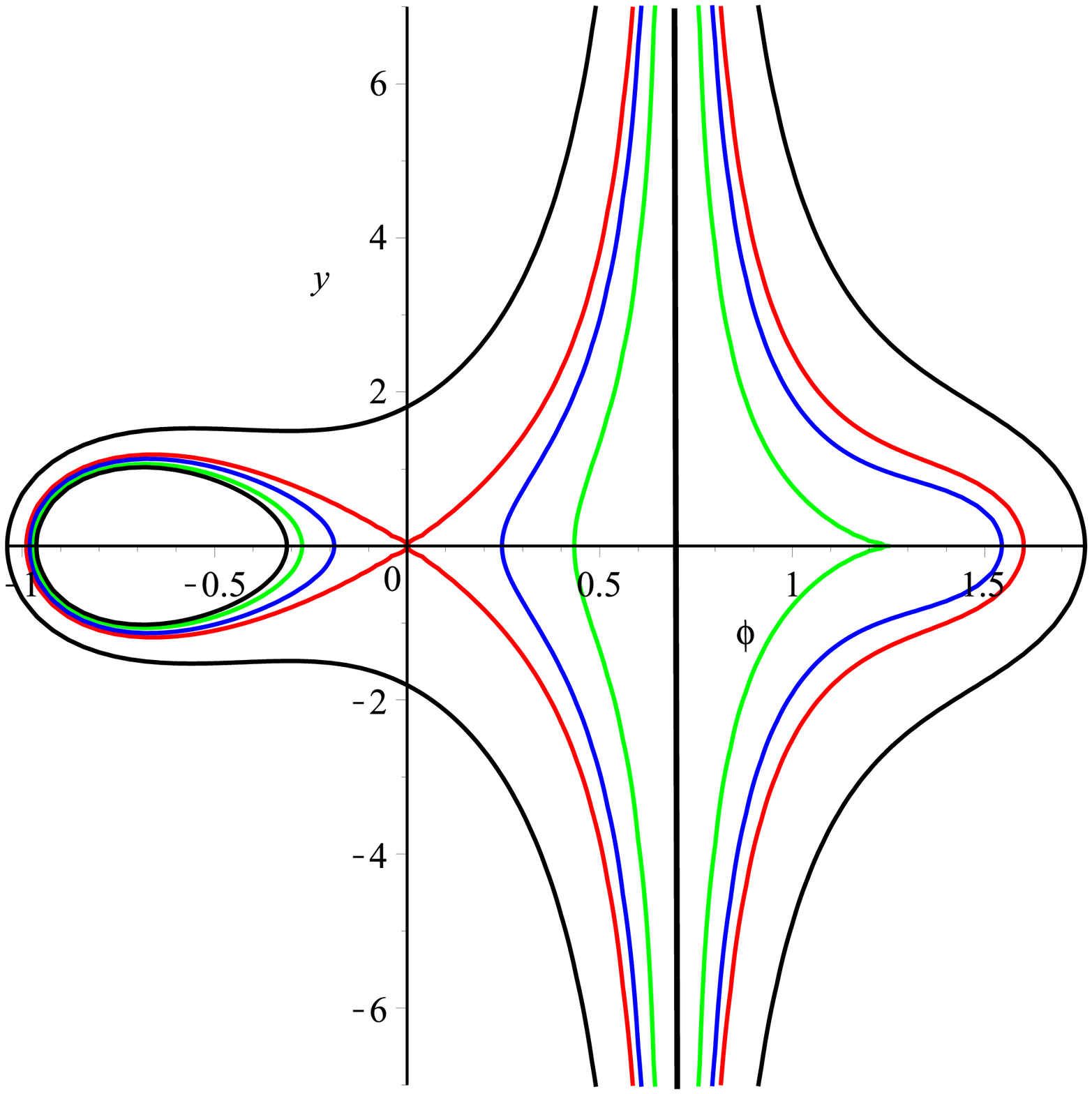}\\
		\footnotesize{ (a) $4C_1>\phi_1$ } & \footnotesize{(b) $4C_1=\phi_1$ }&
		\footnotesize{(c) $0<4C_1<\phi_1$ }
	\end{tabular}
\end{center}
\begin{center}
	\begin{tabular}{ccc}
		\epsfxsize=5cm
		\epsfysize=5cm \epsffile{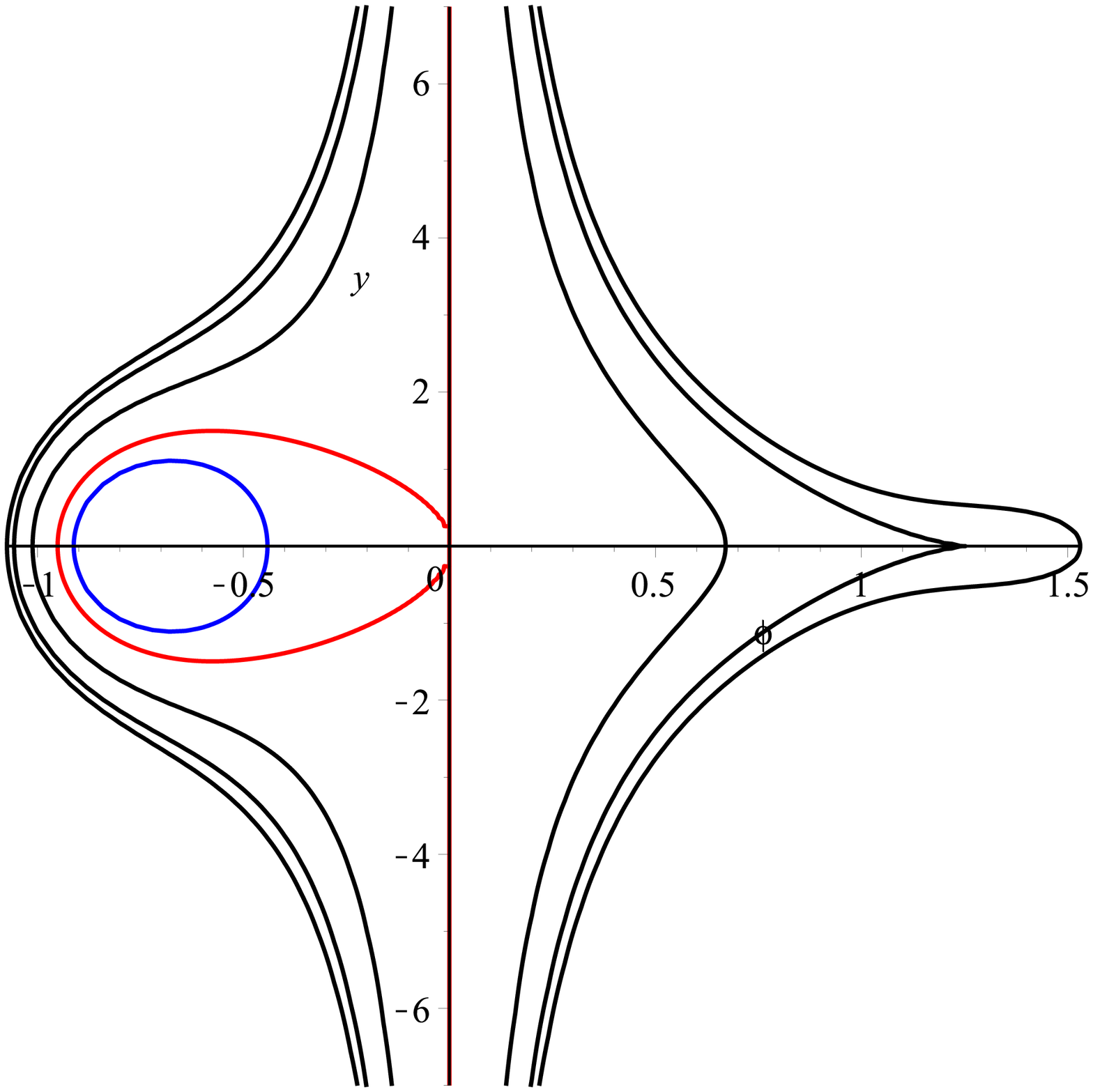}&
		\epsfxsize=5cm \epsfysize=5cm \epsffile{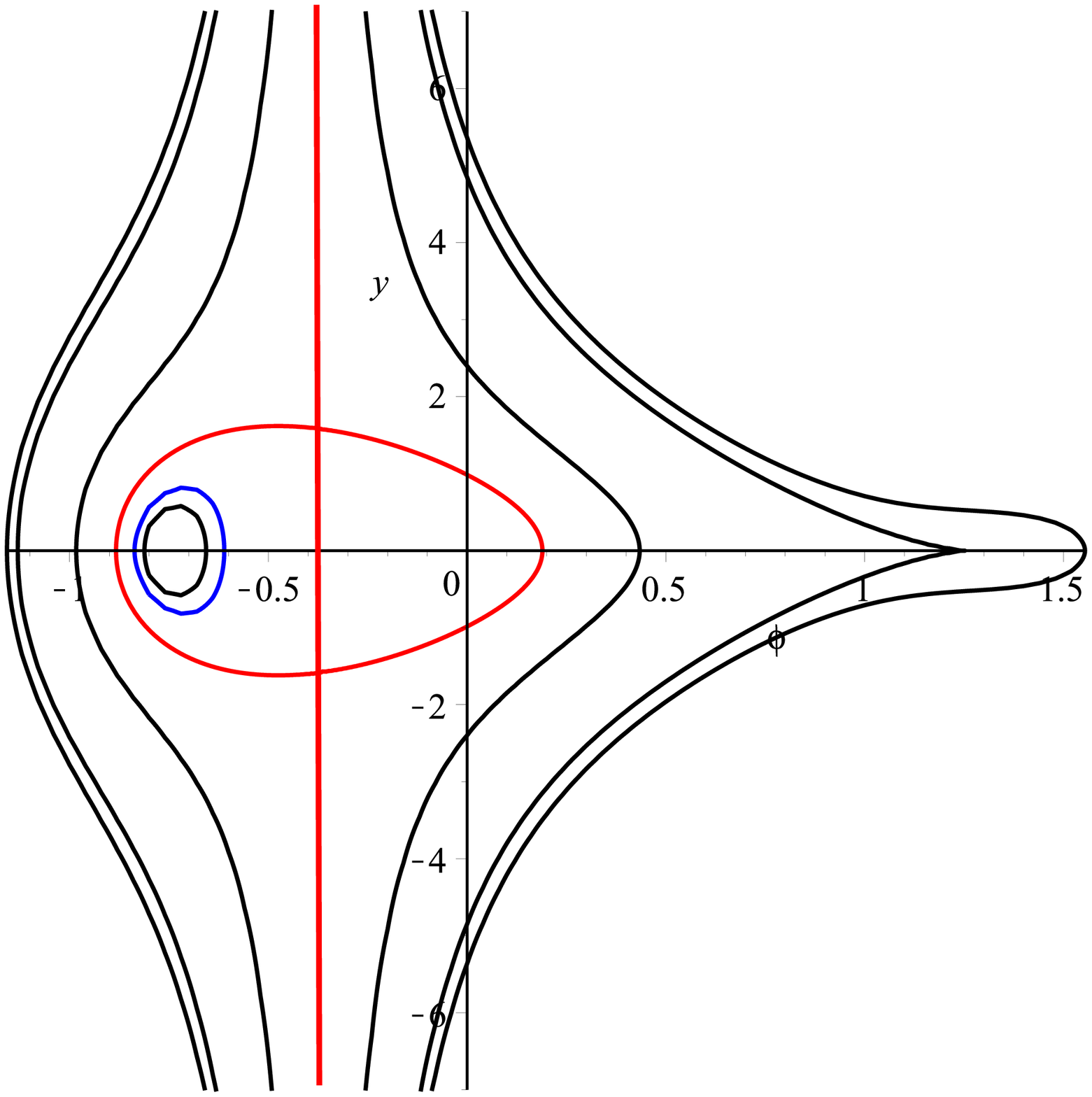}&
		\epsfxsize=5cm \epsfysize=5cm \epsffile{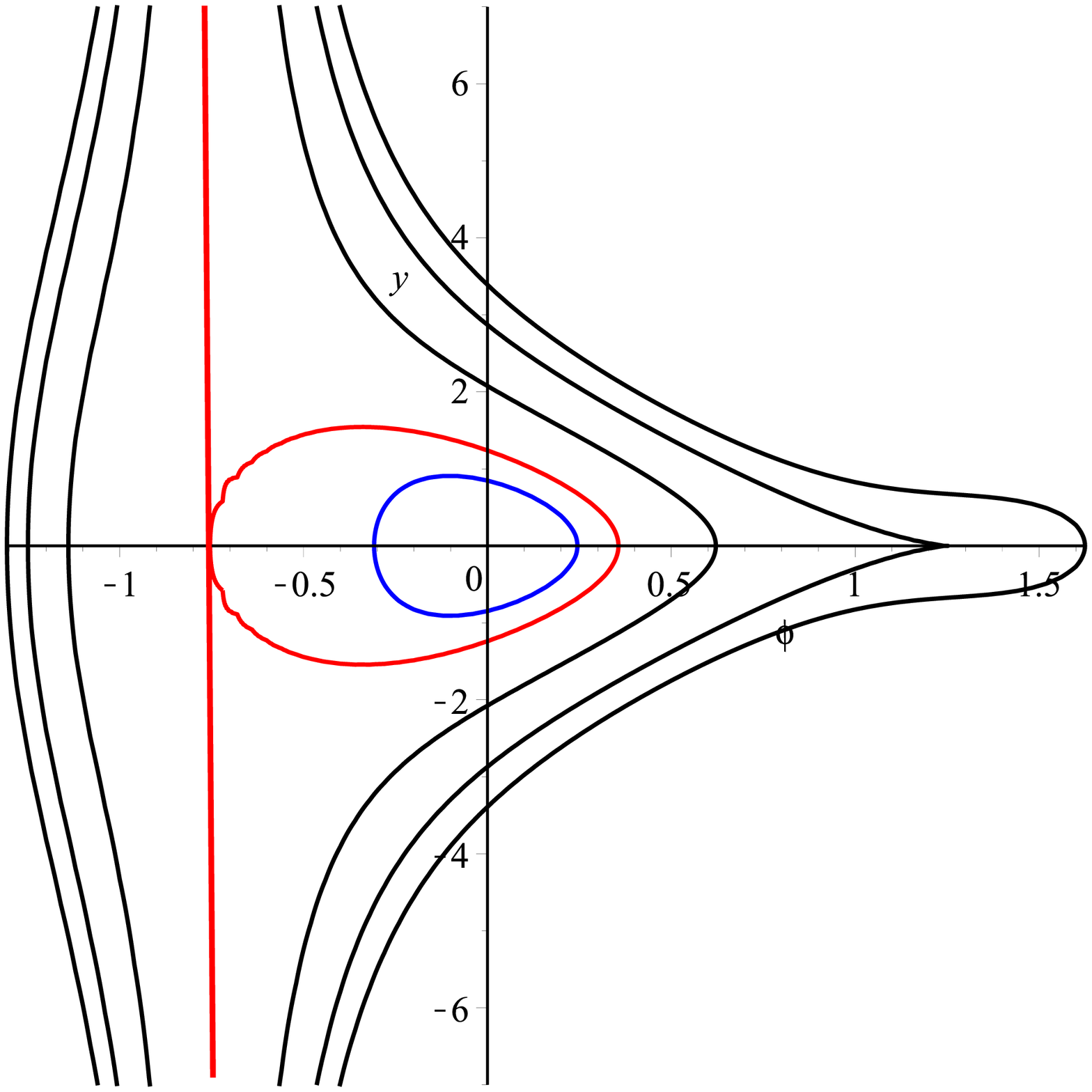}\\
		\footnotesize{ (d) $4C_1=0$ } & \footnotesize{(e) $\phi_2<4C_1<0$ }&
		\footnotesize{(f) $4C_1=\phi_2$ }\\
	\end{tabular}
\end{center}
\begin{center}
	\begin{tabular}{cc}
		\epsfxsize=5cm
		\epsfysize=5cm \epsffile{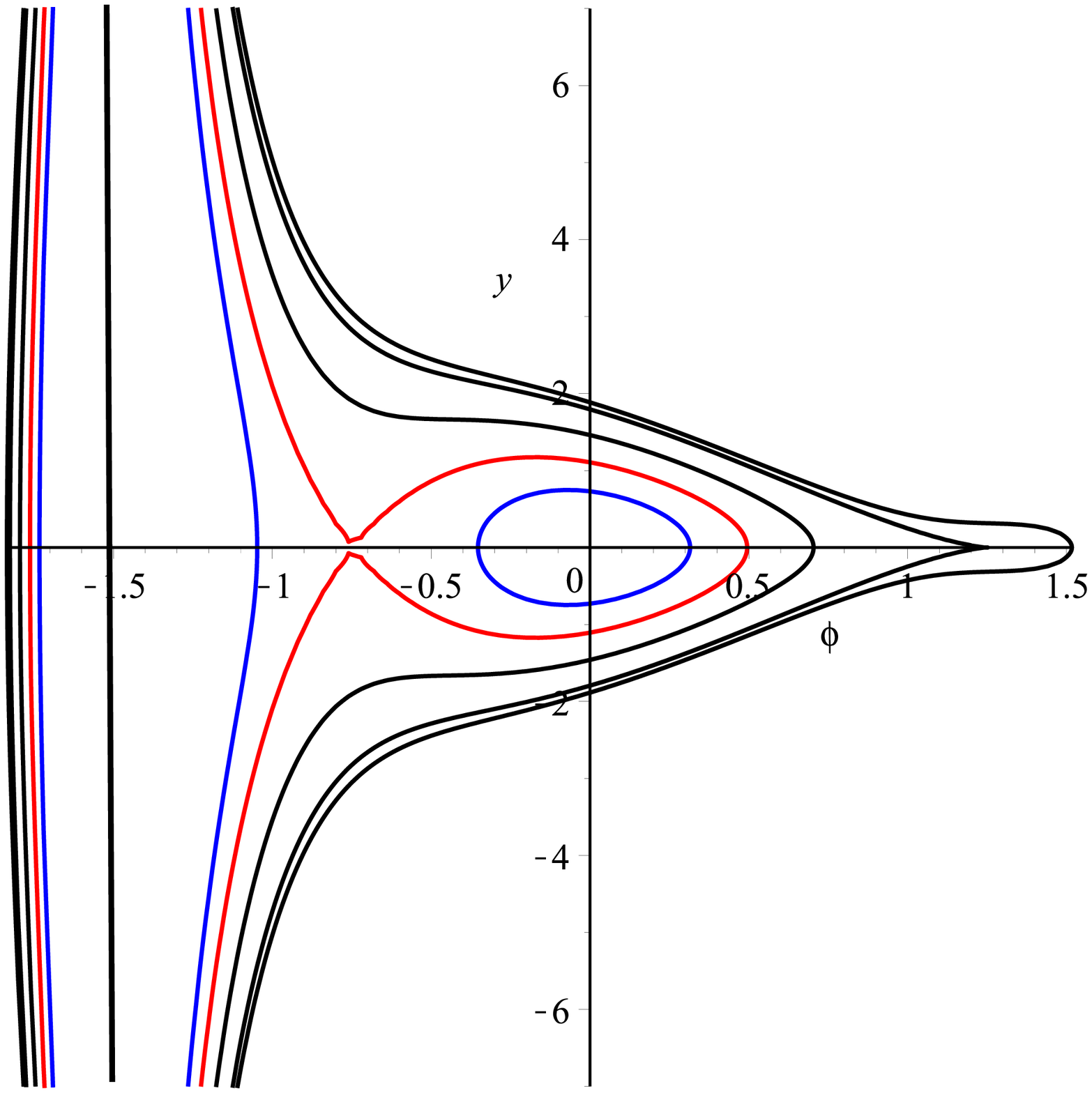}\\
		\footnotesize{ (g) $4C_1<\phi_2$ }
	\end{tabular}
\end{center}
\begin{center}
	{\small Fig.4 The function $g(\phi)$ admits a double zero , and $\phi_1<0<\tilde\phi_+$.)       }
\end{center}

\begin{center}
	\begin{tabular}{ccc}
		\epsfxsize=5cm
		\epsfysize=5cm \epsffile{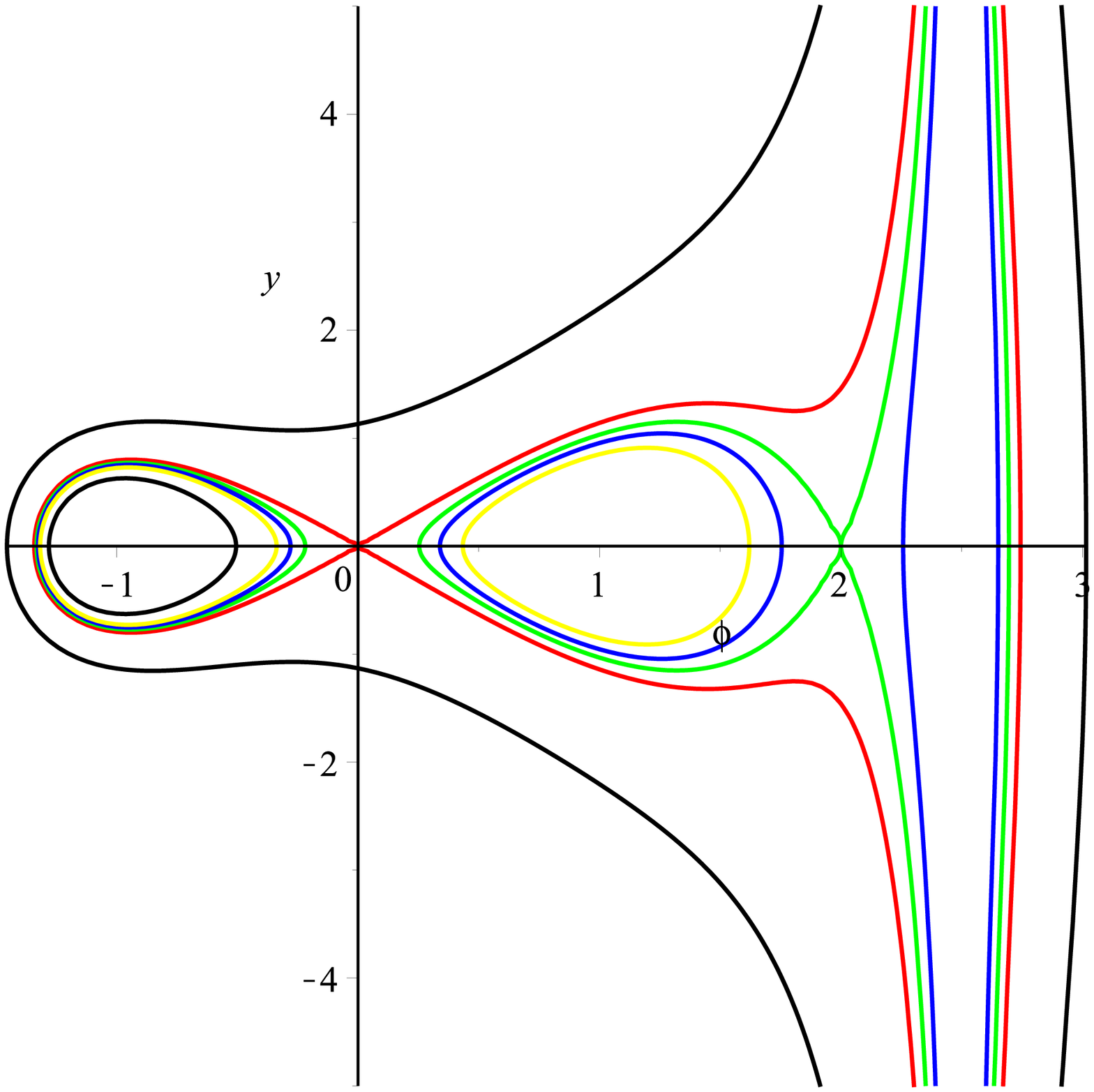}&
		\epsfxsize=5cm \epsfysize=5cm \epsffile{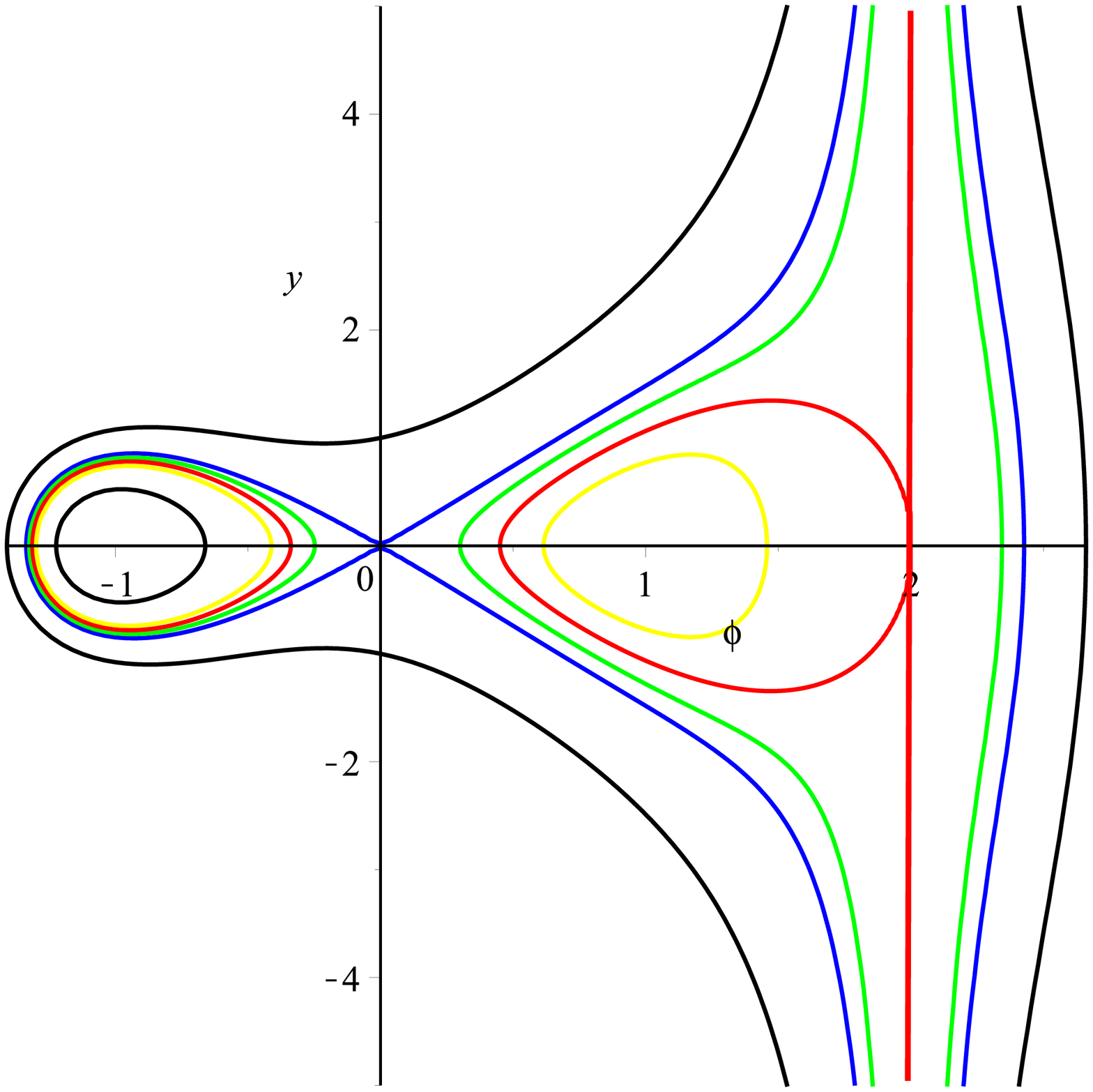}&
		\epsfxsize=5cm \epsfysize=5cm \epsffile{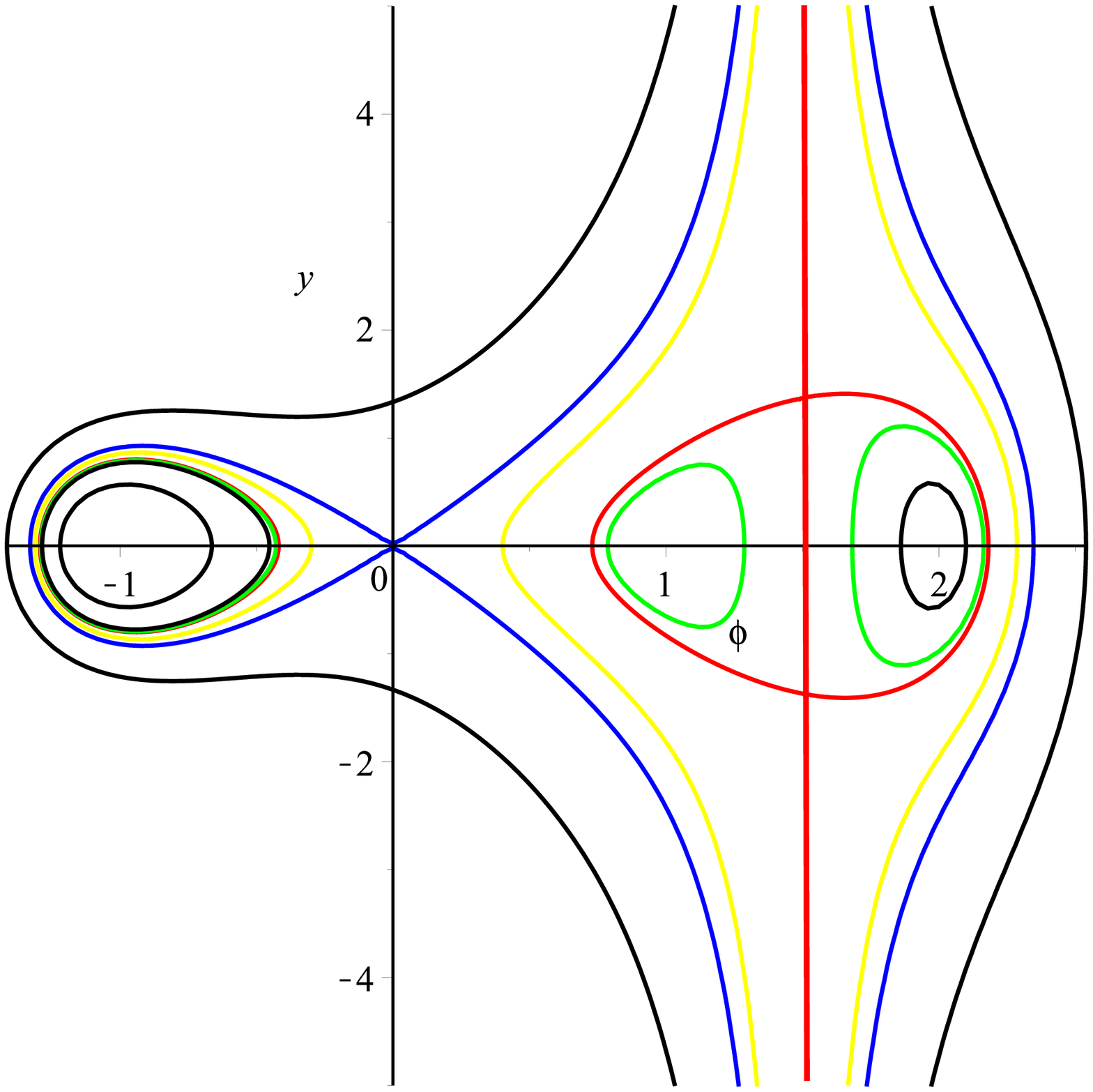}\\
		\footnotesize{ (a) $4C_1>\phi_1$ } & \footnotesize{(b) $4C_1=\phi_1$ }&
		\footnotesize{(c) $0<4C_1<\phi_1$ }
	\end{tabular}
\end{center}

\begin{center}
	\begin{tabular}{ccc}
		\epsfxsize=5cm
		\epsfysize=5cm \epsffile{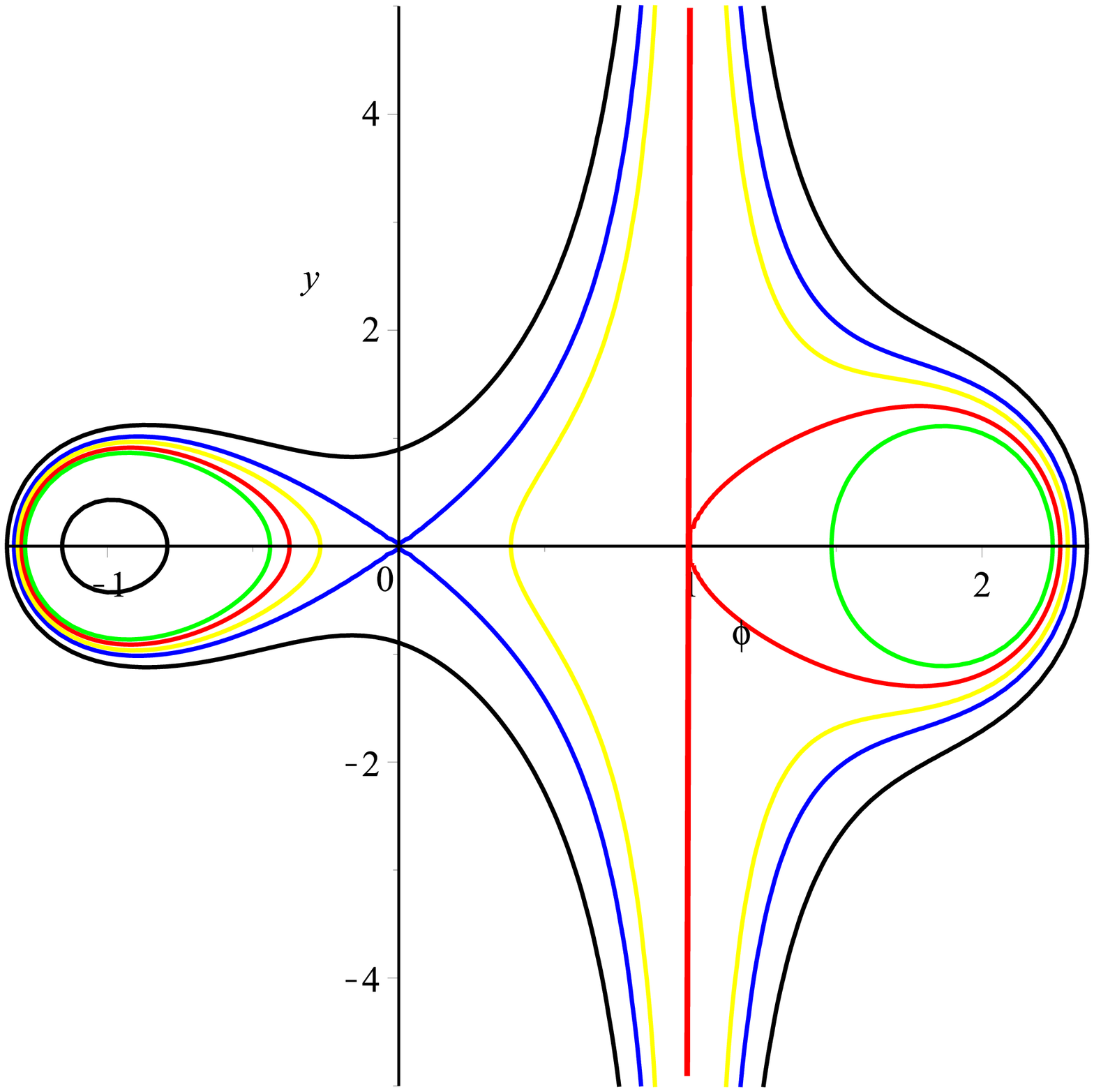}&
		\epsfxsize=5cm \epsfysize=5cm \epsffile{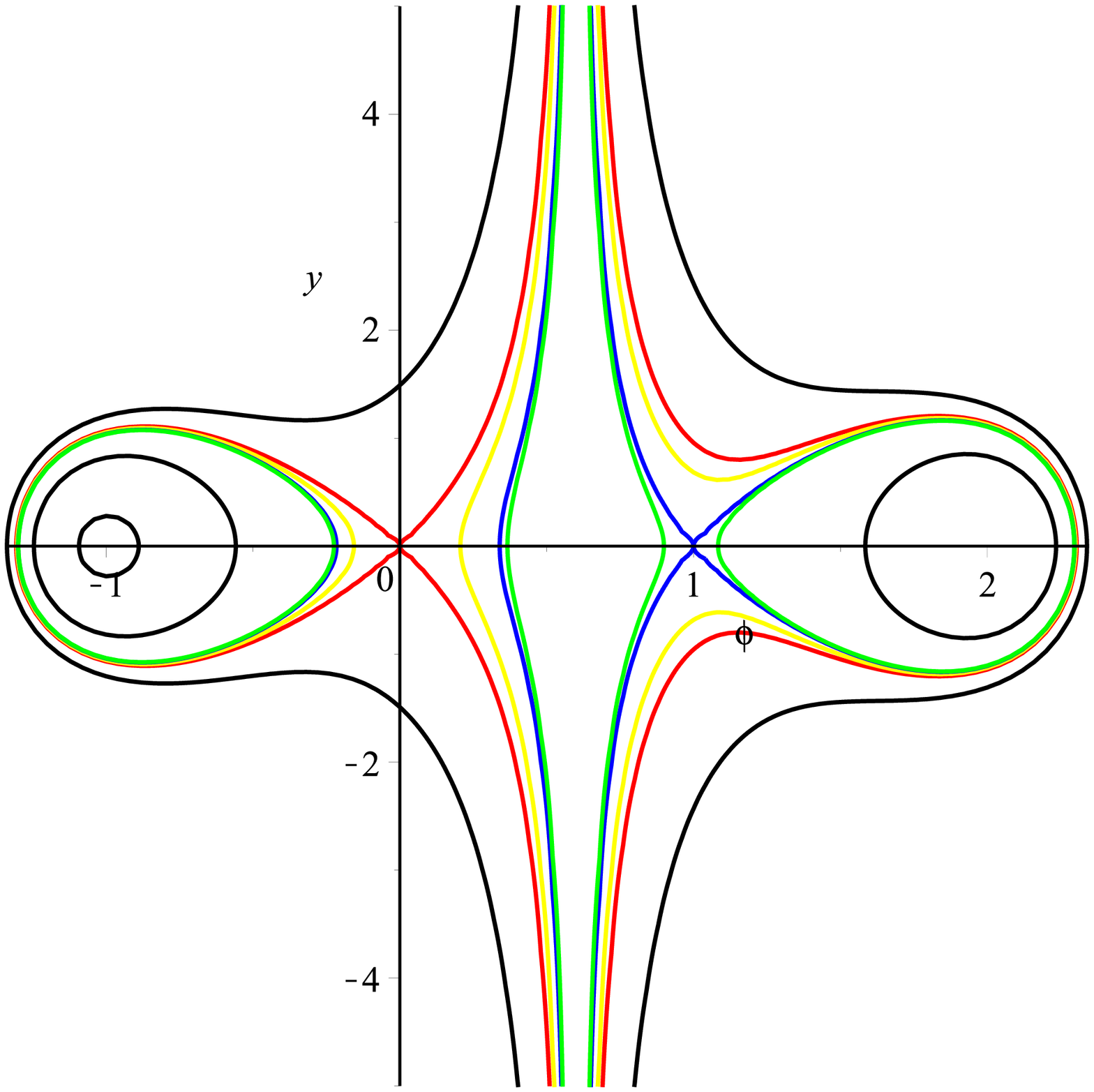}&
		\epsfxsize=5cm \epsfysize=5cm \epsffile{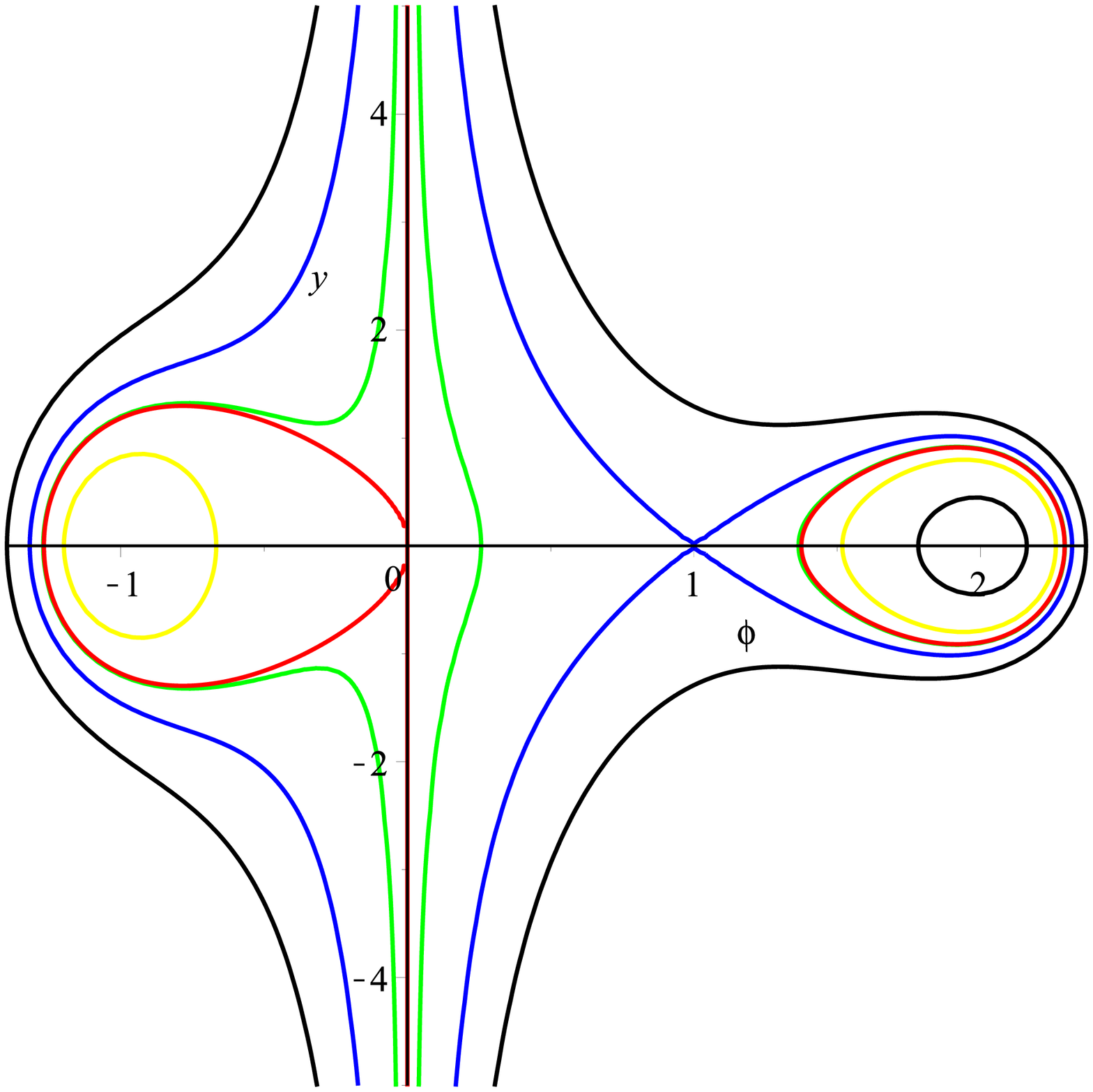}\\
		\footnotesize{ (d) $4C_1=0$ } & \footnotesize{(e) $\phi_2<4C_1<0$ }&
		\footnotesize{(f) $4C_1=\phi_2$ }
	\end{tabular}
\end{center}

\begin{center}
	\begin{tabular}{ccc}
		\epsfxsize=5cm
		\epsfysize=5cm \epsffile{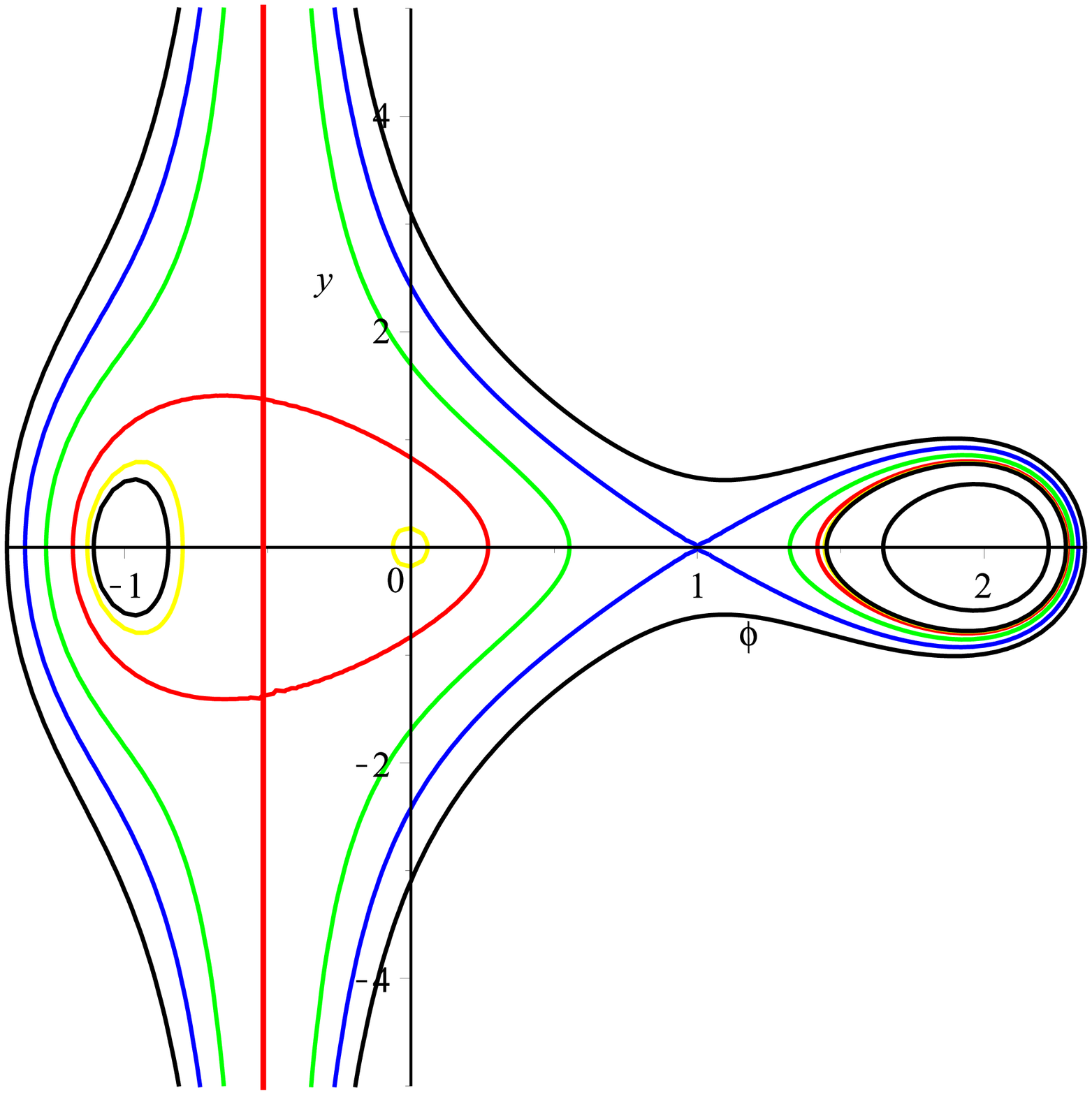}&
		\epsfxsize=5cm \epsfysize=5cm \epsffile{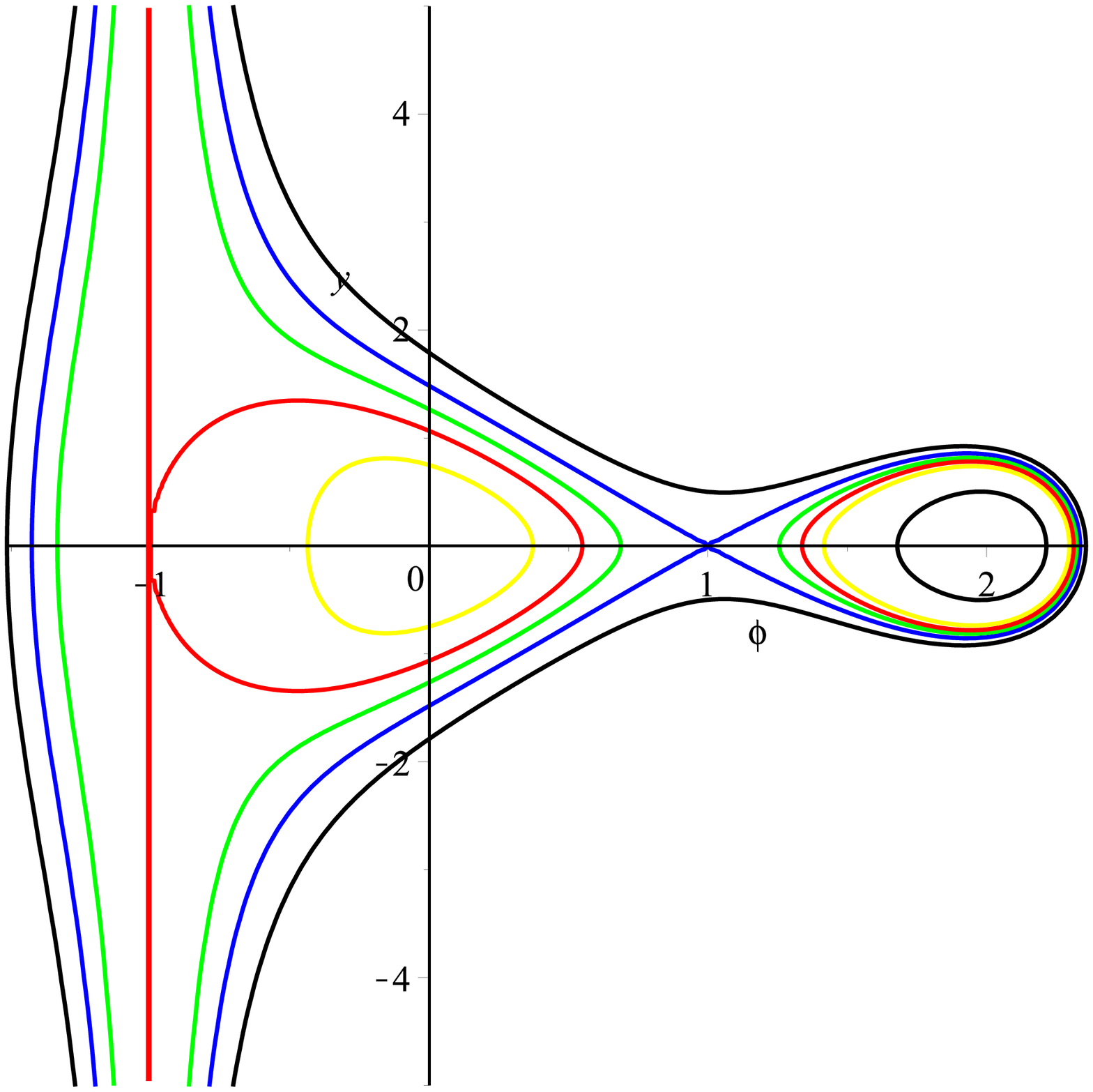}&
		\epsfxsize=5cm \epsfysize=5cm \epsffile{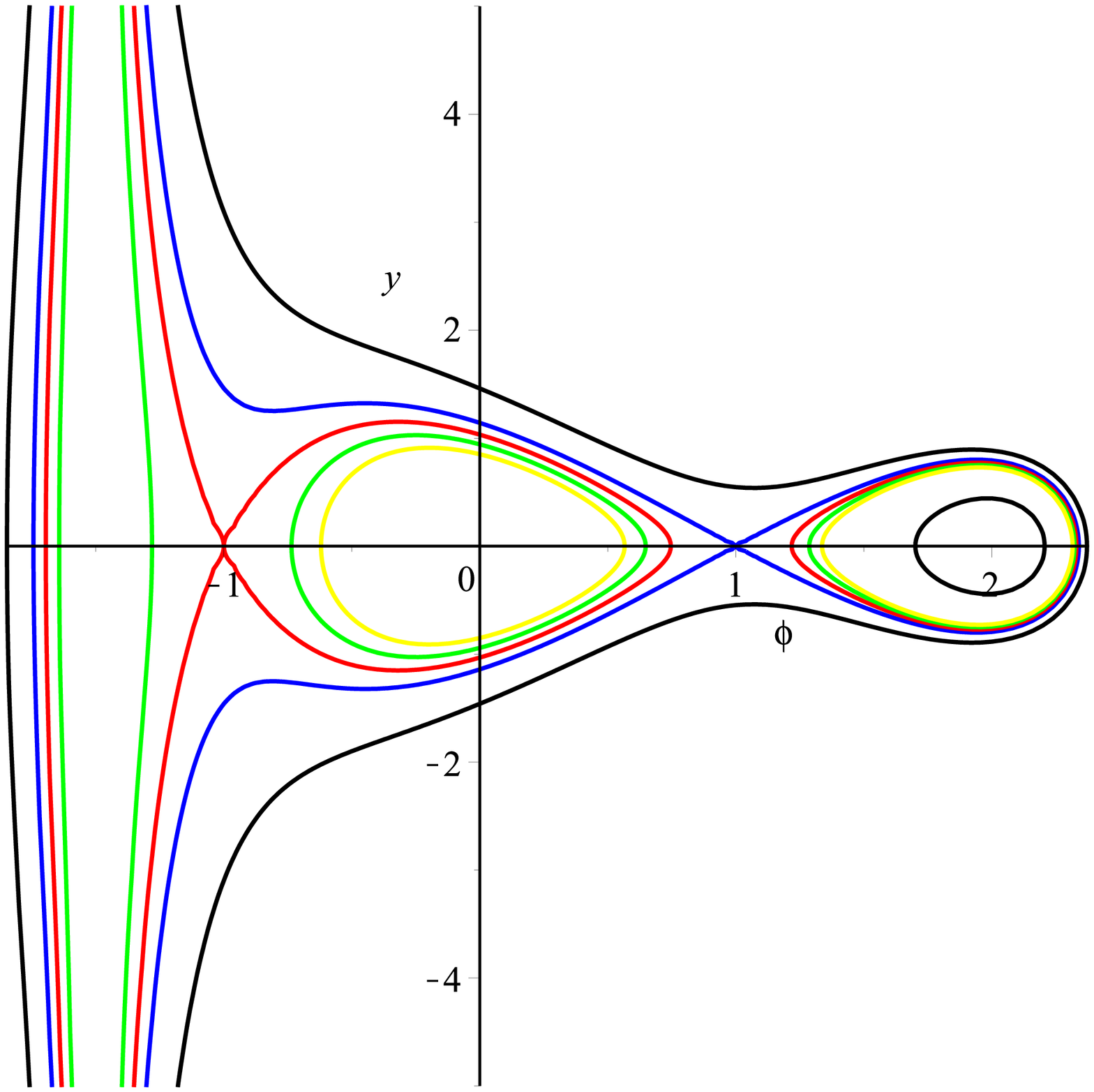}\\
		\footnotesize{ (g) $\phi_3<4C_1<\phi_2$ } & \footnotesize{(h) $4C_1=\phi_3$ }&
		\footnotesize{(i) $4C_1<\phi_3$ }
	\end{tabular}
\end{center}

\begin{center}
	{\small Fig.5 The function $g(\phi)$ admits three real zeros and $\phi_1<\phi_2<0<\phi_3$.       }
\end{center}

\begin{center}
	\begin{tabular}{ccc}
		\epsfxsize=5cm
		\epsfysize=5cm \epsffile{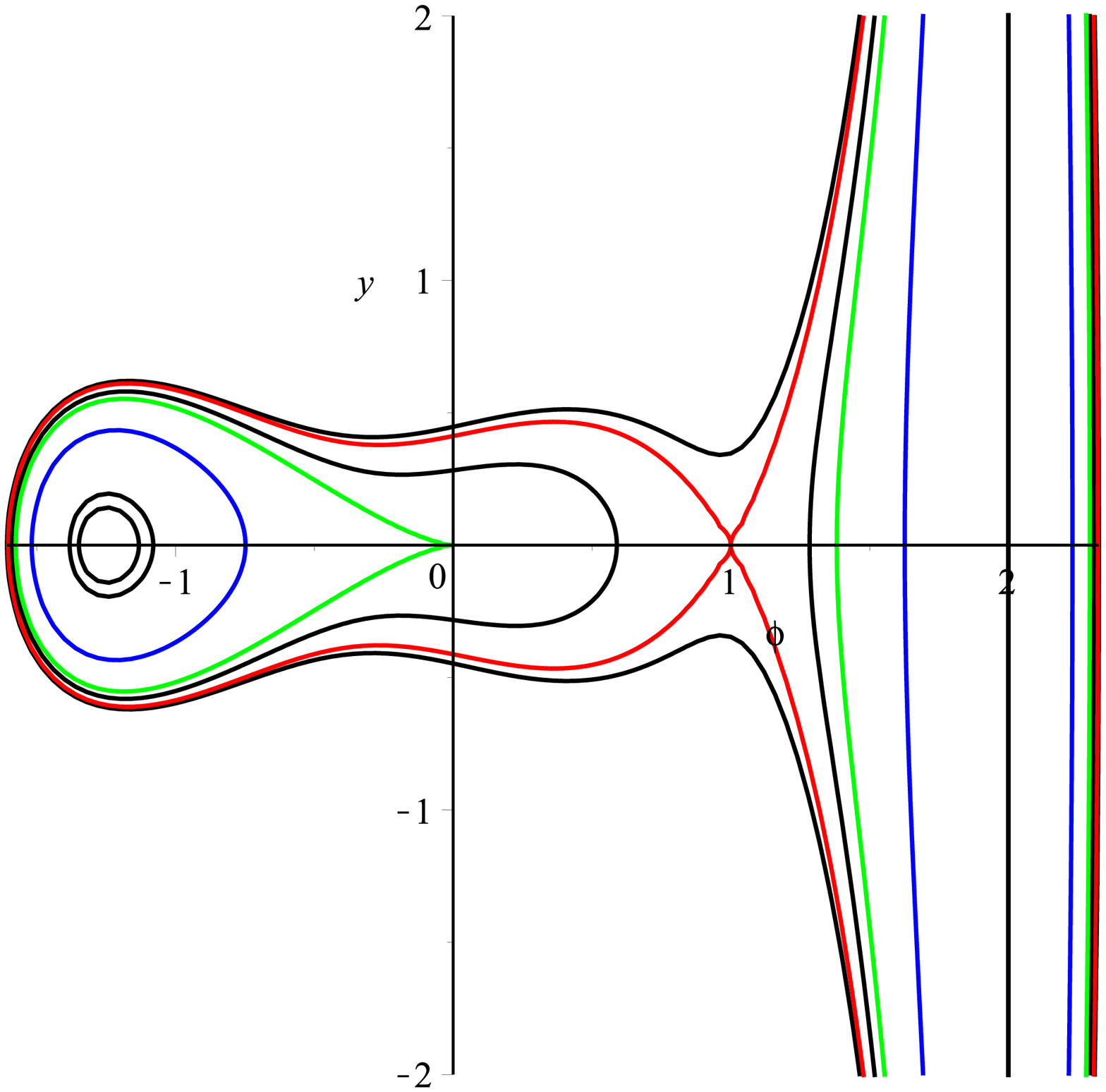}&
		\epsfxsize=5cm \epsfysize=5cm \epsffile{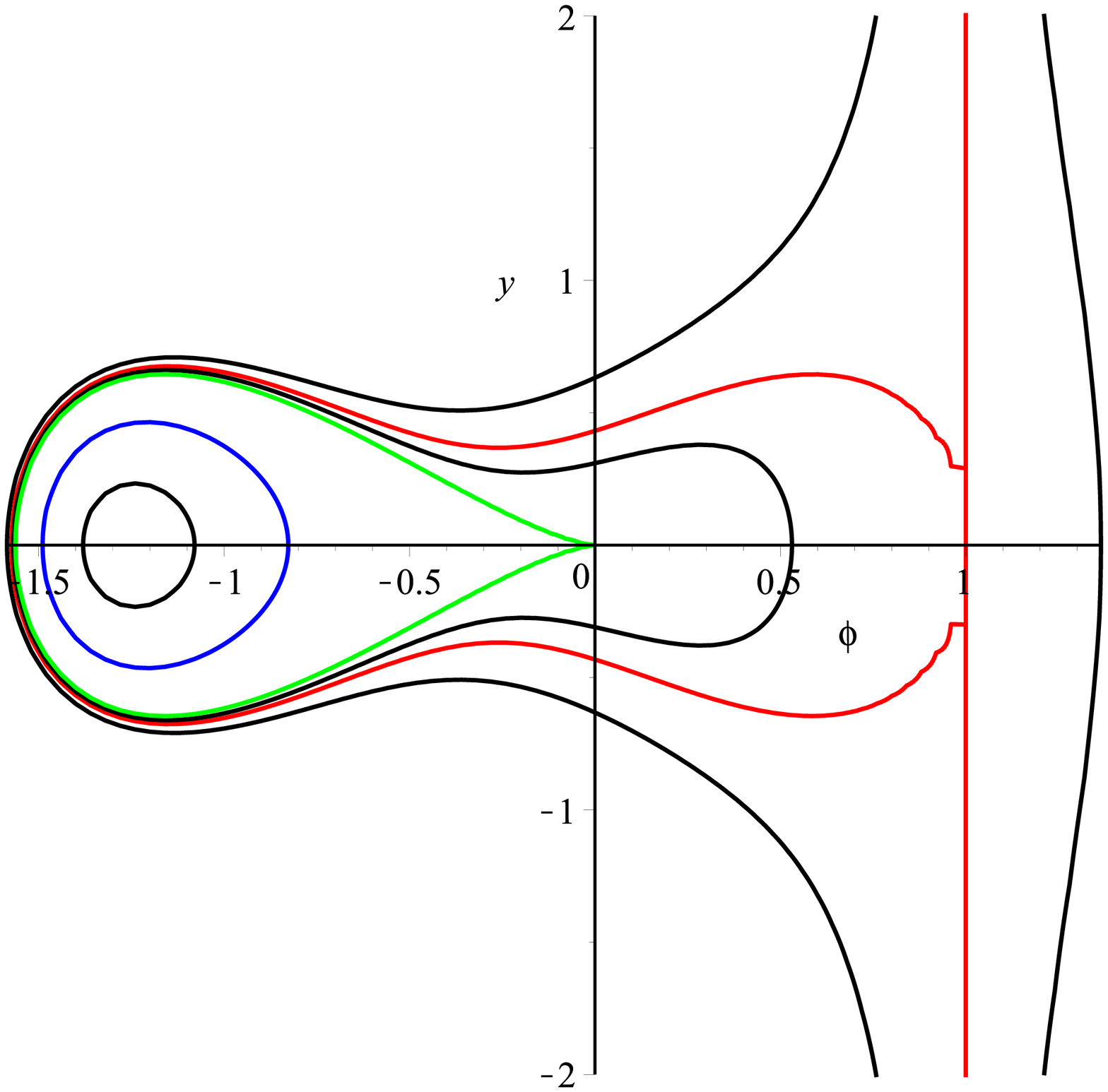}&
		\epsfxsize=5cm \epsfysize=5cm \epsffile{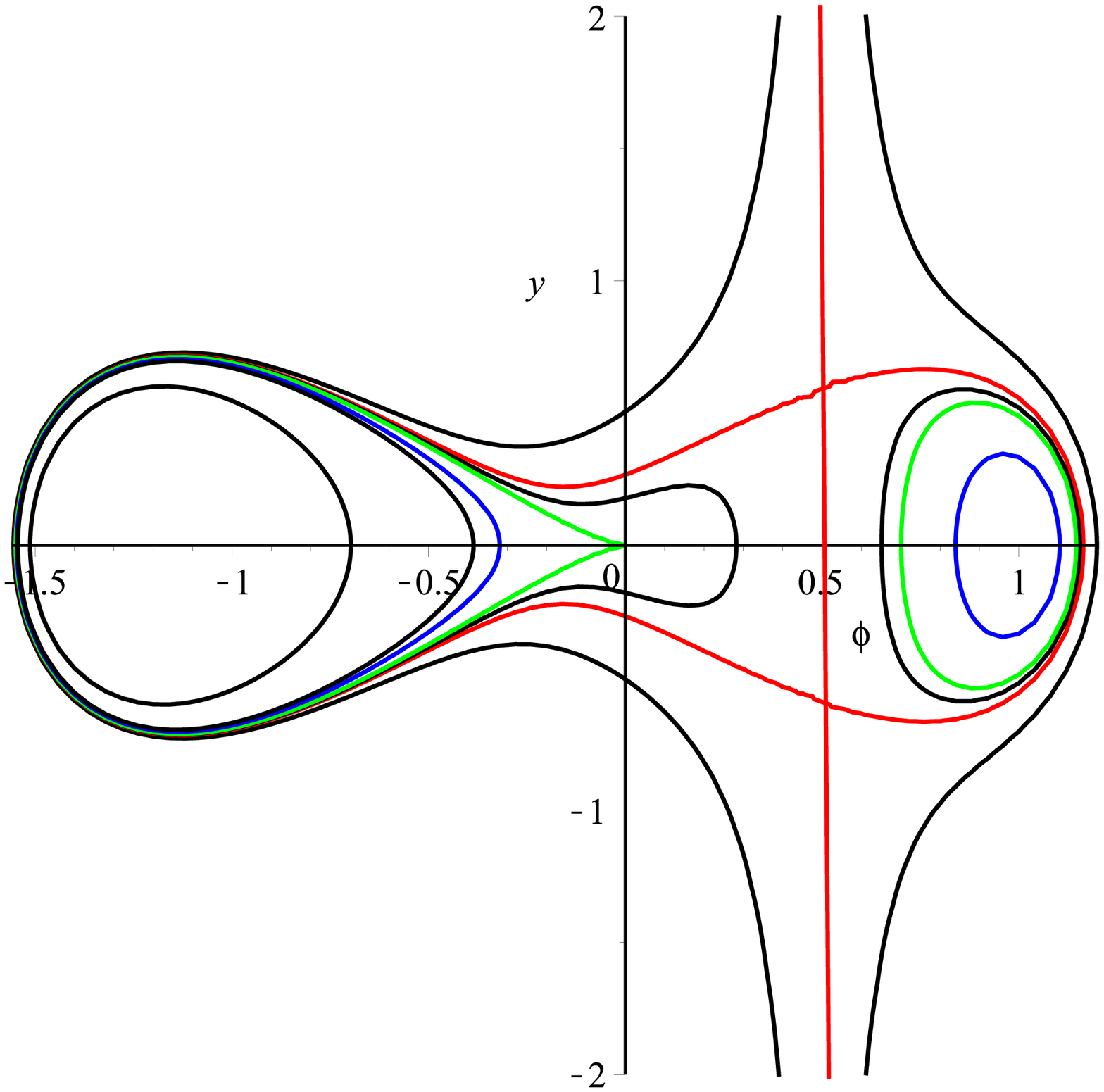}\\
		\footnotesize{ (a) $4C_1>\phi_1$ } & \footnotesize{(b) $4C_1=\phi_1$ }&
		\footnotesize{(c) $0<4C_1<\phi_1$ }
	\end{tabular}
\end{center}

\begin{center}
	\begin{tabular}{ccc}
		\epsfxsize=5cm
		\epsfysize=5cm \epsffile{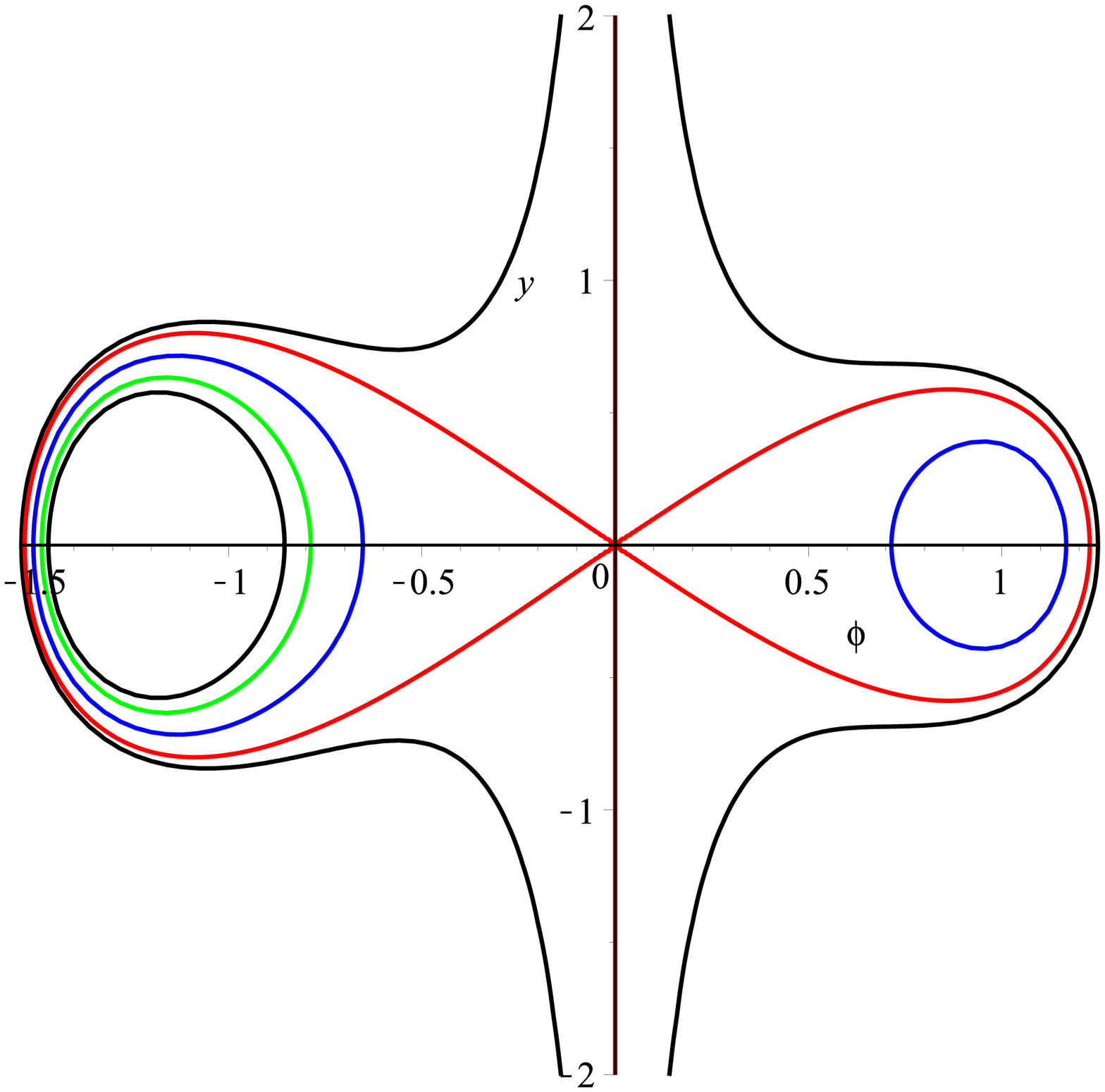}&
		\epsfxsize=5cm \epsfysize=5cm \epsffile{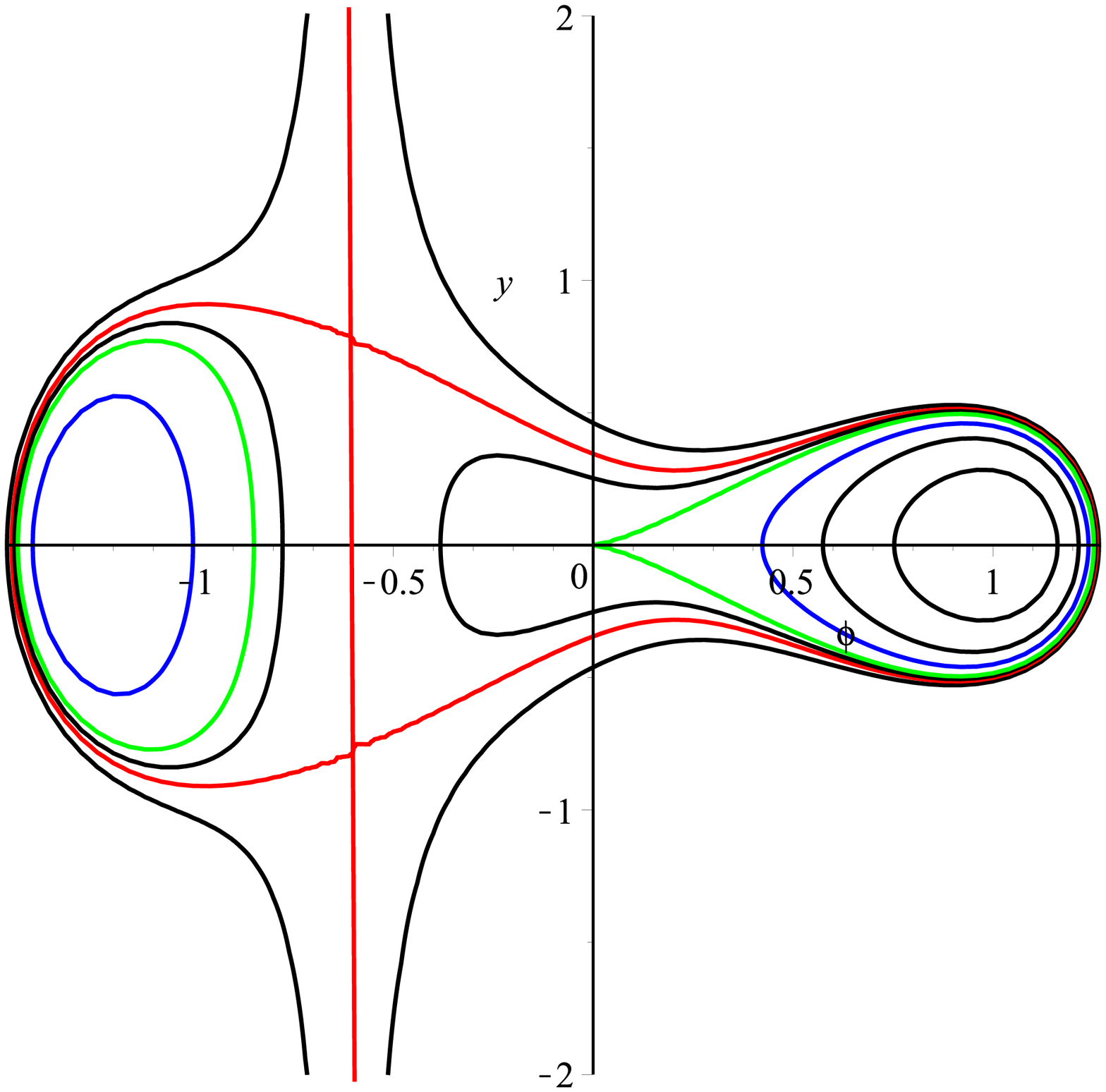}&
		\epsfxsize=5cm \epsfysize=5cm \epsffile{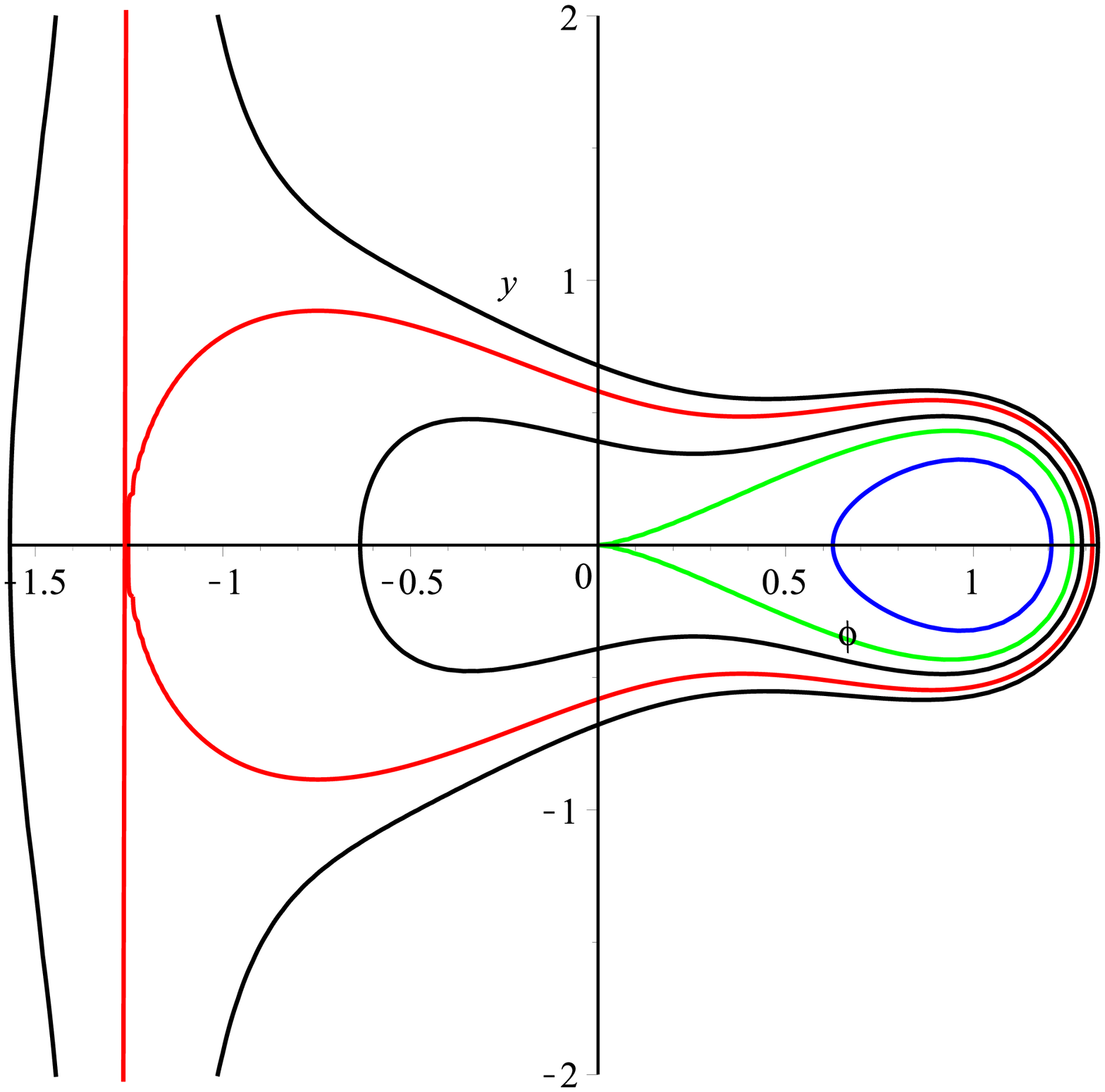}\\
		\footnotesize{ (d) $4C_1=0$ } & \footnotesize{(e) $\phi_2<4C_1<0$ }&
		\footnotesize{(f)$4C_1=\phi_2$}
	\end{tabular}
\end{center}
\begin{center}
	\begin{tabular}{c}
		\epsfxsize=5cm
		\epsfysize=5cm \epsffile{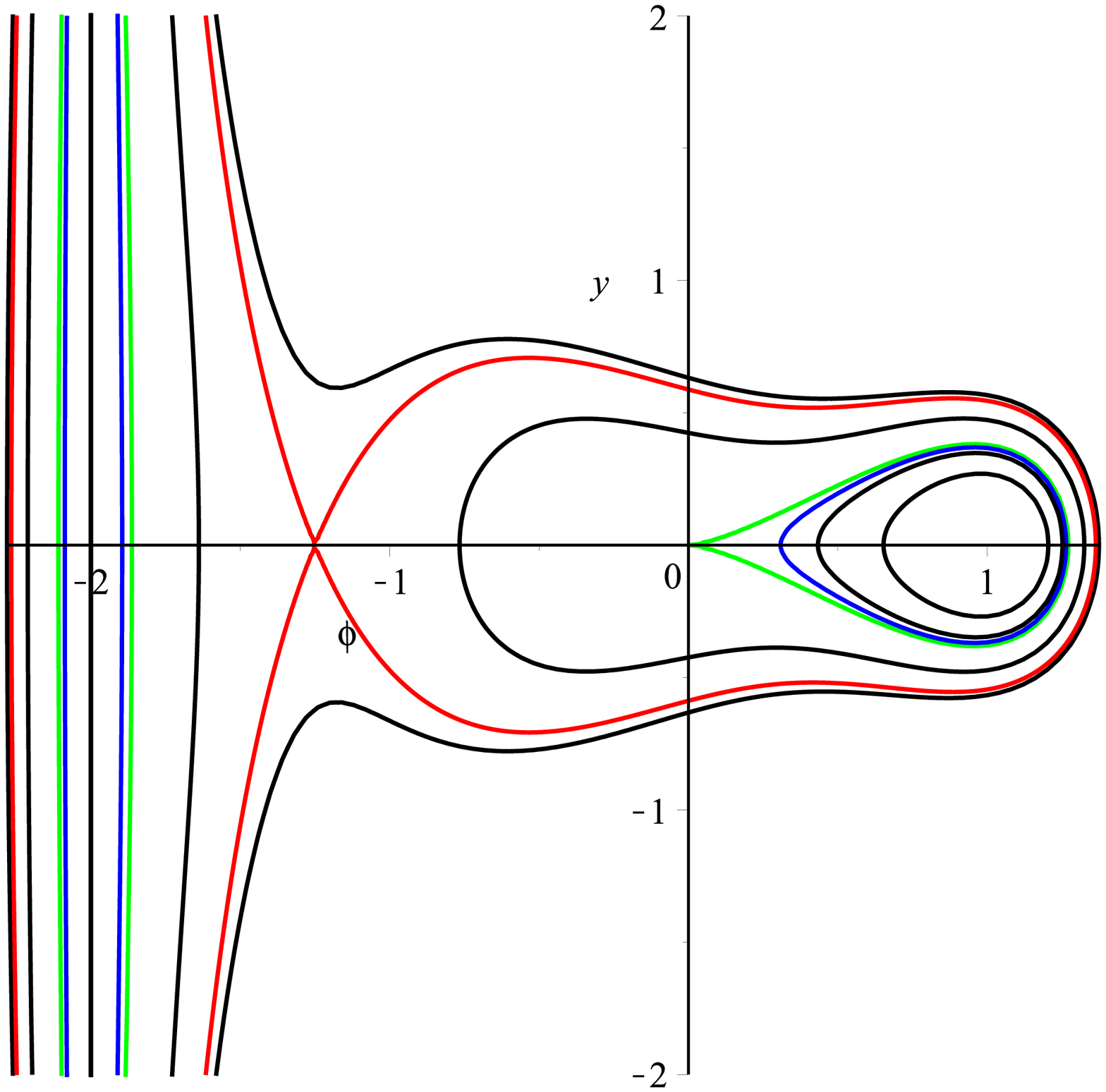}\\
		\footnotesize{(f) $4C_1<\phi_2$}
	\end{tabular}
\end{center}

\begin{center}
	{\small Fig.6 The special case of k=0, $\phi_1<0<\phi_2$.       }
\end{center}

\subsection{Case of \textbf{$\theta=\frac{1}{2}$}}
If $\theta=\frac{1}{2}$, then (\ref{planar system}) takes the form

\begin{equation}\label{planar system24}
	\begin{cases}
		\frac{d\phi}{d\xi}=y,\\
		\frac{dy}{d\xi}=\frac{C_3\phi^4+C_2\phi^3+\frac{1}{2}\phi^2+K\phi}{\frac{1}{2}\phi-C_1},\\
	\end{cases}
\end{equation}
with the associated regular system
\begin{equation}\label{planar system25}
	\begin{cases}
		\frac{d\phi}{d\tau}=y(\frac{1}{2}\phi-C_1),\\
		\frac{dy}{d\tau}=C_3\phi^4+C_2\phi^3+\frac{1}{2}\phi^2+K\phi,\\
	\end{cases}
\end{equation}

and the Hamiltonian
\begin{equation}\label{Hamiltonian3}
	\begin{aligned}
		H(\phi,y)=-\frac{1}{2}(\phi-4C_1)^2y^2+\frac{1}{4}\phi^4+\frac{\alpha}{3}\phi^3+\frac{\beta}{2}\phi^2+\gamma\phi+\delta\ln(|\phi-2C_1|)=h,\\
	\end{aligned}
\end{equation}

where $\alpha= C_2+2C_1C_3$, $\beta= \frac{1}{2}+2C_1C_2+4C_1^2C_3$, $\gamma=K+C_1+4C_1^2C_2+8C_1^3C_3$, and $\delta=2C_1K+2C_1^2+8C_1^3C_2+16C_1^4C_3$.

\subsubsection{Qualitative analysis of singular points of system (\ref{planar system25})}
Similarly to system (\ref{planar system23}), the point O(0,0) is a singular point of system (\ref{planar system25}). There are no equilibrium points on the singular line $\phi=2C_1$. The other singular points $(\phi_i,0)$ of system(\ref{planar system25}) are obtained by solving the equation $f(\phi)=0$. Thus, the conclusions in section (2.2.1) hold.

\subsubsection{Classification of singular points and phase portraits of system (\ref{planar system25})}

Let $M_2(\phi,y)$ be the matrix of the linearized system of (\ref{planar system25}),

\begin{equation}\label{matrix}
	M_2(\phi,y)=
	\left(
	\begin{array}{cc}
		\frac{1}{2}y & \frac{1}{2}\phi-C_1  \\
		f^{'}(\phi) & 0\\
	\end{array}
	\right).
\end{equation}

with
\begin{equation}
	J_2(\phi,y)=\det{M_2}=-(\frac{1}{2}\phi-C_1)f^{'}(\phi) \end{equation}

We classify the equilibrium points based on the theory of dynamical system, and obtain the following phase portraits of system (\ref{planar system25}).

\begin{center}
	\begin{tabular}{ccc}
		\epsfxsize=5cm
		\epsfysize=5cm \epsffile{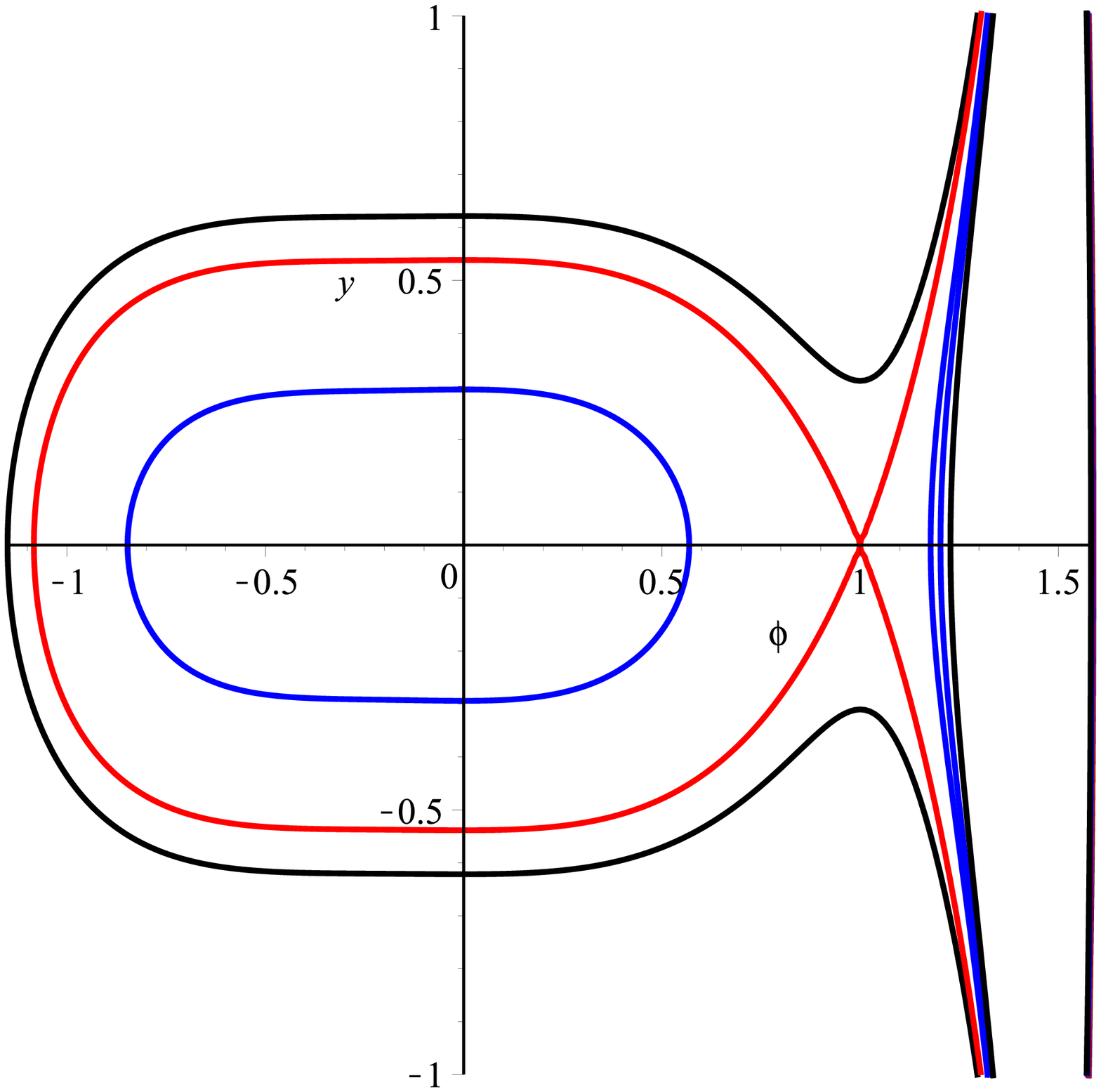}&
		\epsfxsize=5cm \epsfysize=5cm \epsffile{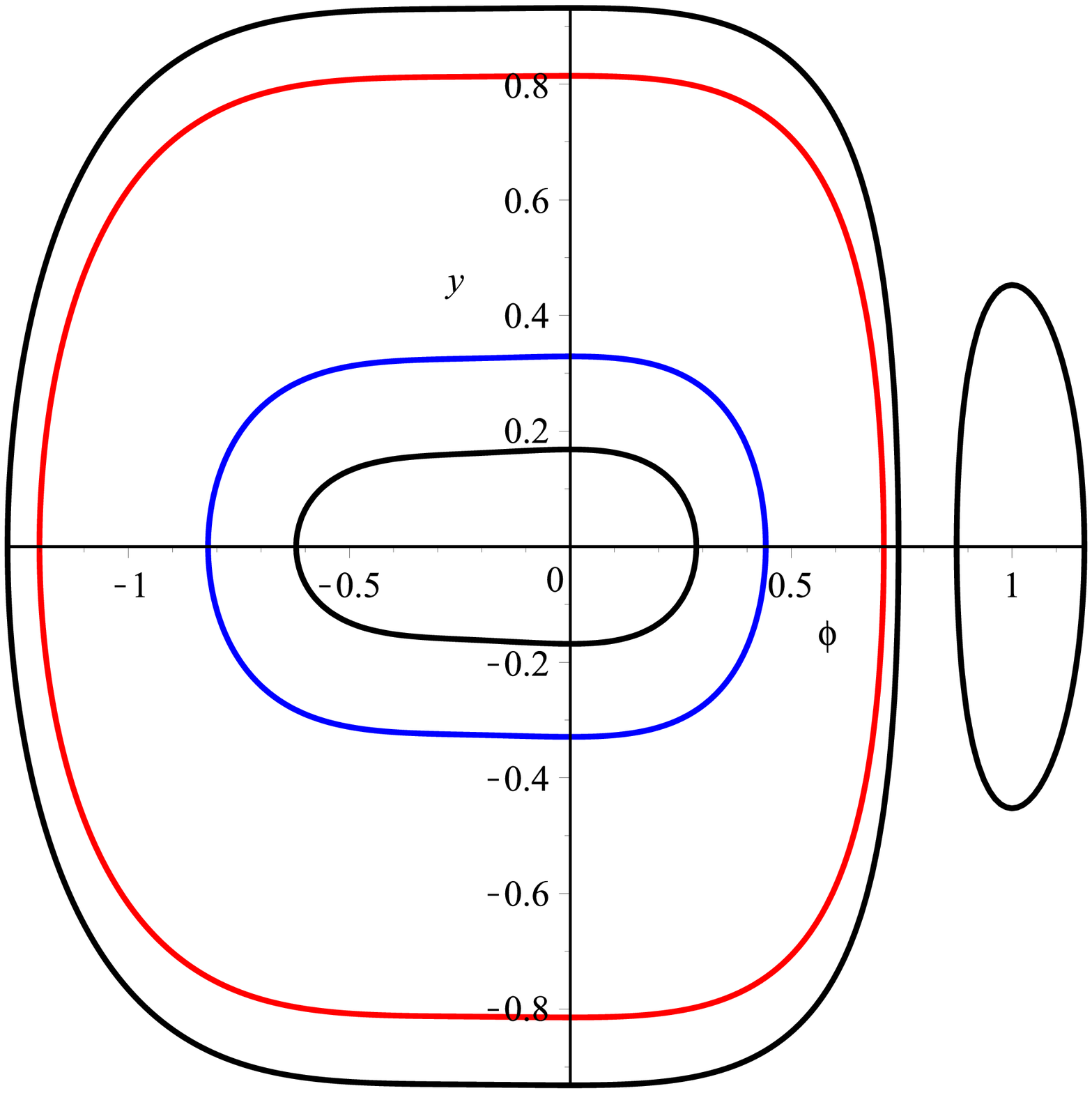}&
		\epsfxsize=5cm \epsfysize=5cm \epsffile{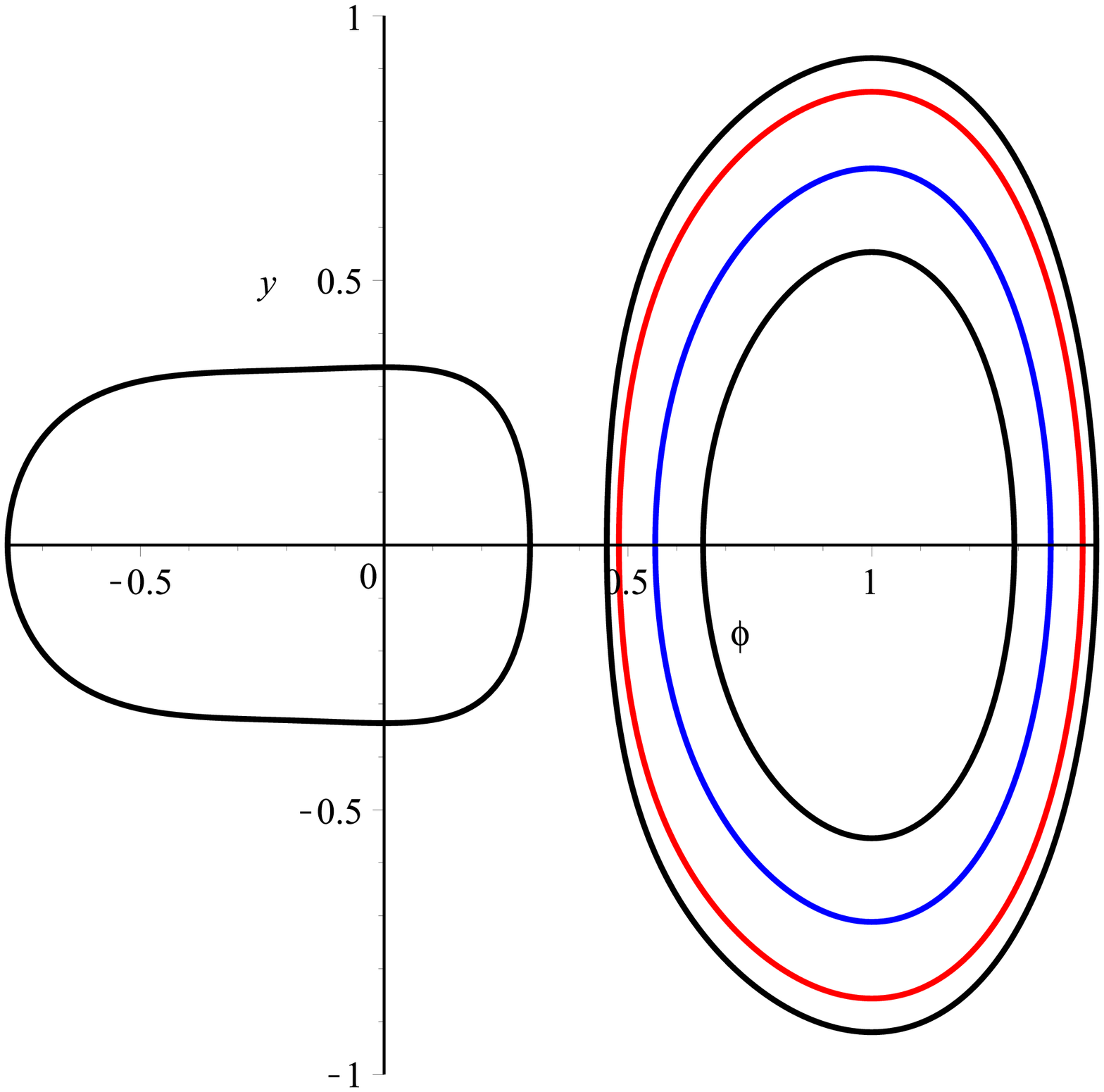}\\
		\footnotesize{ (a) $2C_1>\phi_1$ } & \footnotesize{(b) $0<2C_1<\phi_1$ }&
		\footnotesize{(c) $0<2C_1<\phi_1$ }
	\end{tabular}
\end{center}
\begin{center}
	\begin{tabular}{c}
		\epsfxsize=5cm
		\epsfysize=5cm \epsffile{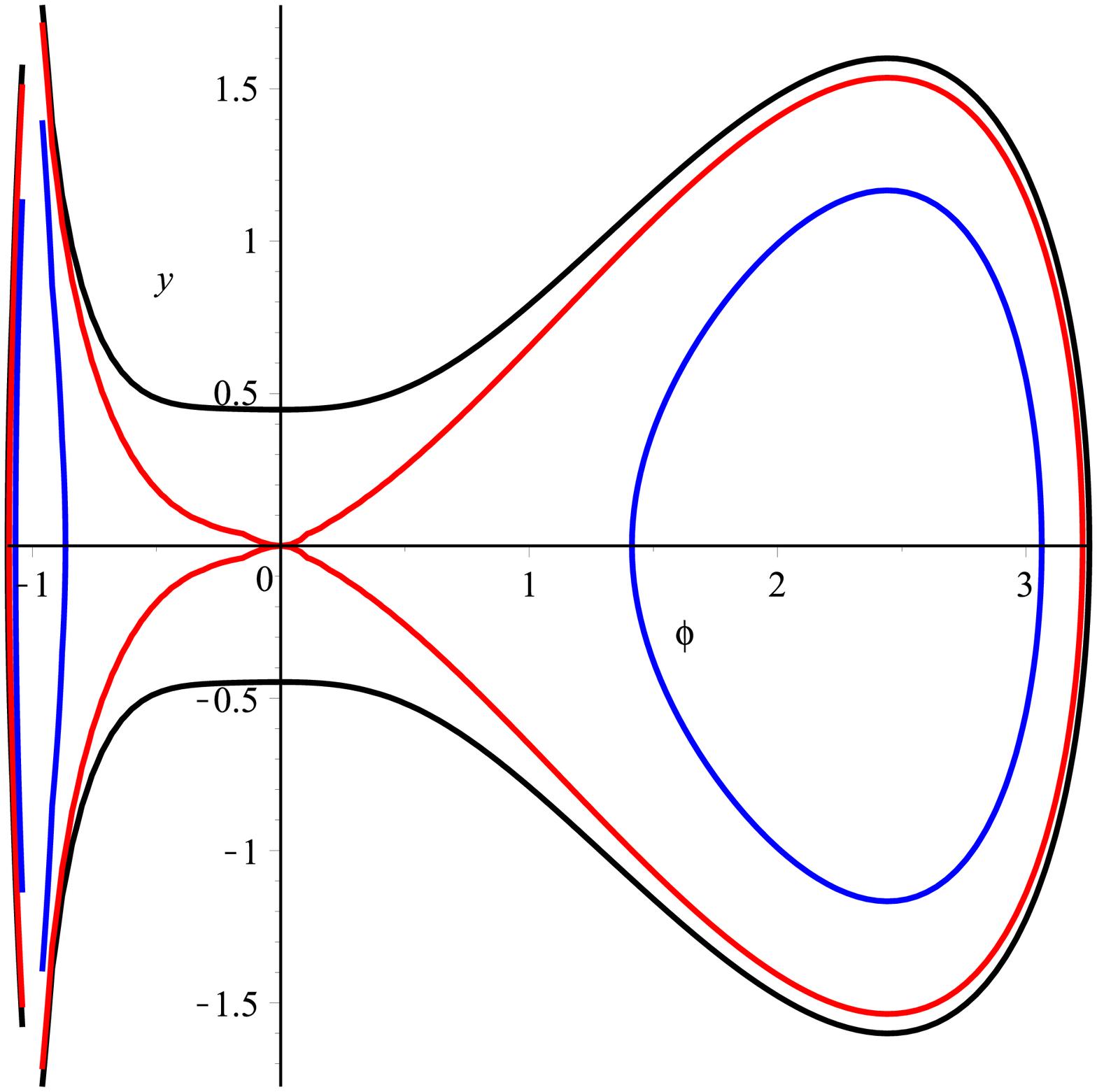}\\
		\footnotesize{(d) $2C_1<0$}
	\end{tabular}
\end{center}
\begin{center}
	{\small Fig.7  The function $g(\phi)$ admits only one zero, i.e $(\Delta_g)>0$ and $g(\phi_-)>0$.}
\end{center}

\begin{center}
	\begin{tabular}{ccc}
		\epsfxsize=5cm
		\epsfysize=5cm \epsffile{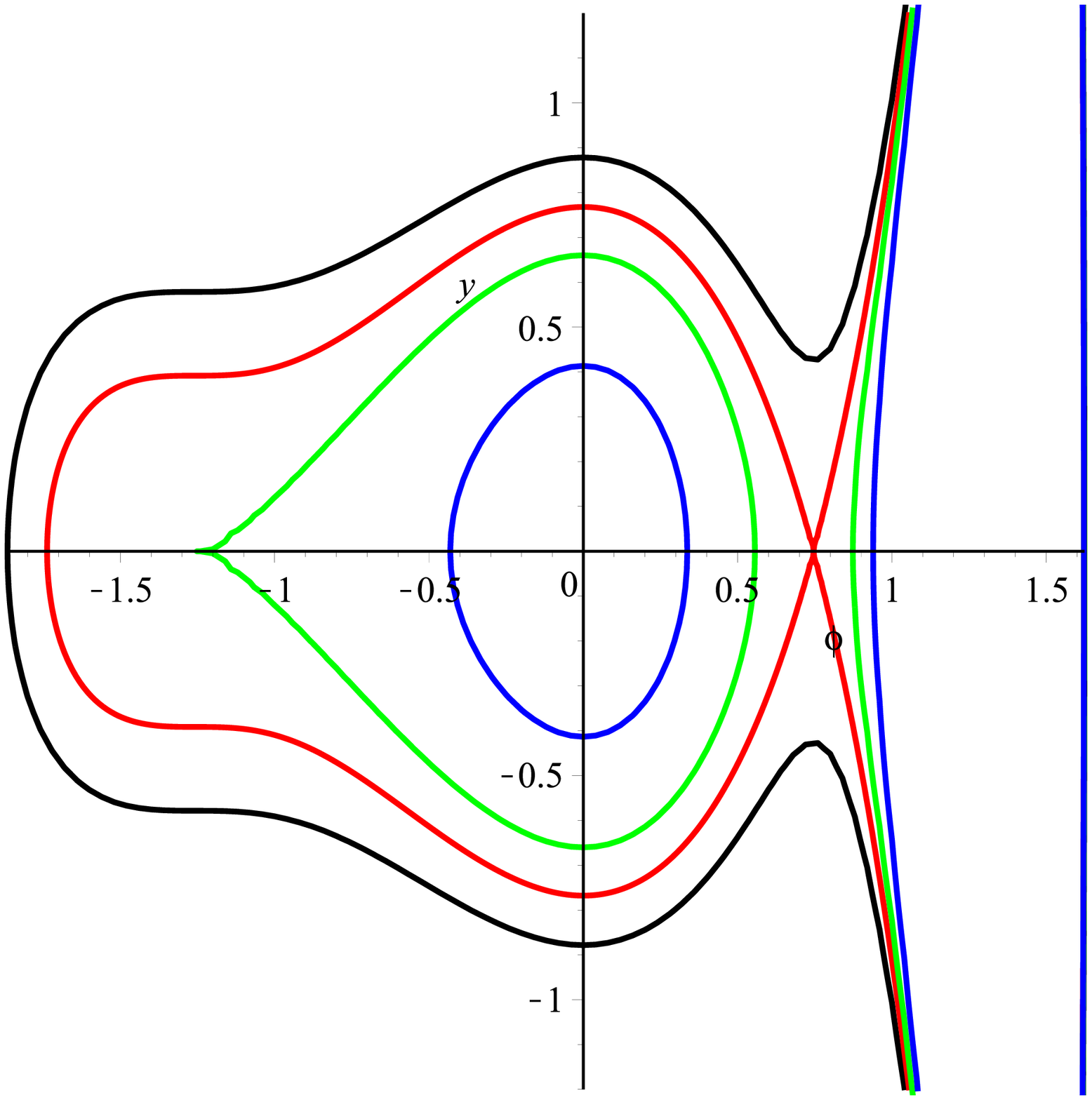}&
		\epsfxsize=5cm \epsfysize=5cm \epsffile{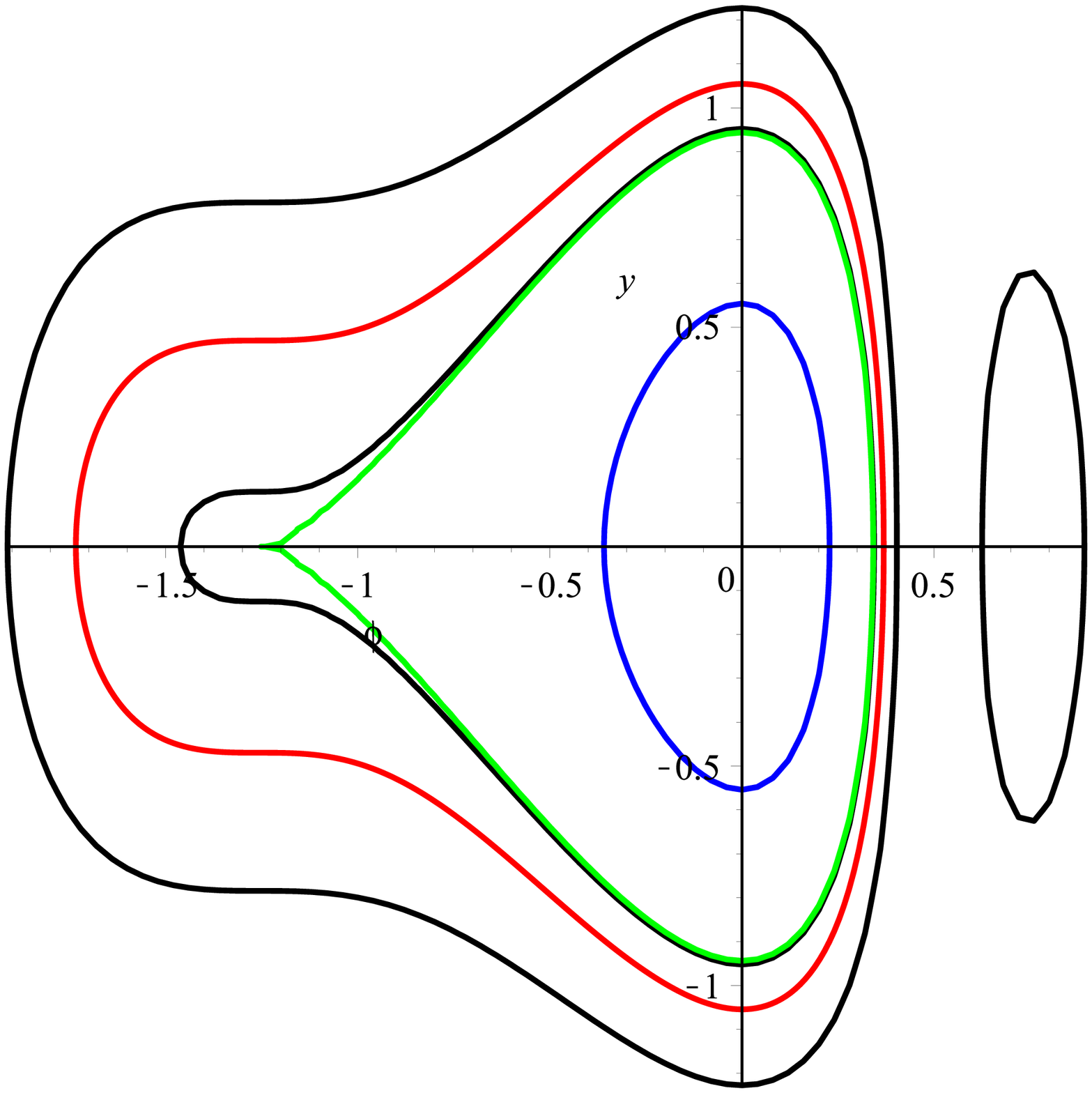}&
		\epsfxsize=5cm \epsfysize=5cm \epsffile{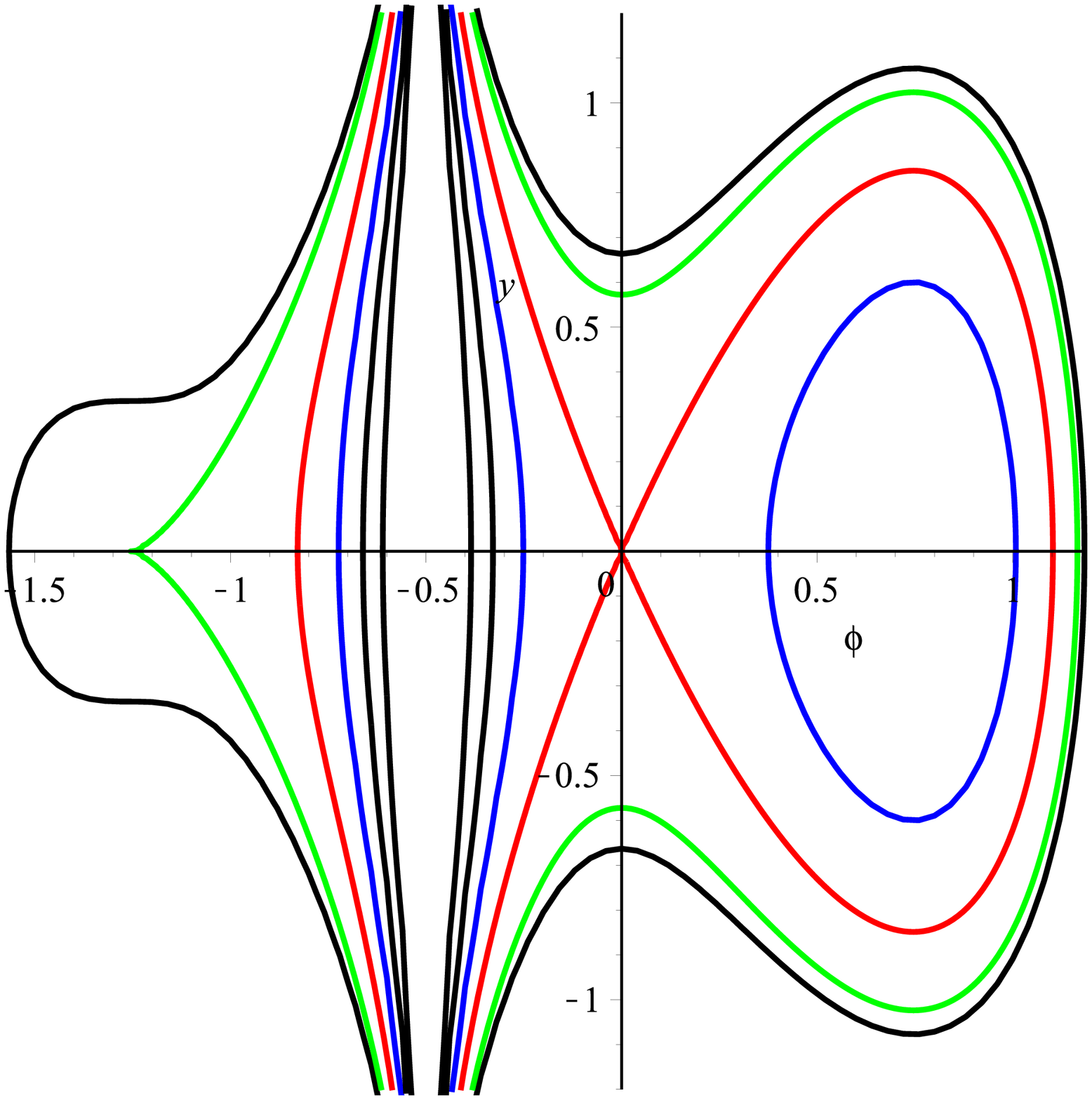}\\
		\footnotesize{ (a) $2C_1>\phi_1$ } & \footnotesize{(b) $0<2C_1<\phi_1$ }&
		\footnotesize{(c) $\phi_2<2C_1<0$ }
	\end{tabular}
\end{center}
\begin{center}
	\begin{tabular}{c}
		\epsfxsize=5cm
		\epsfysize=5cm \epsffile{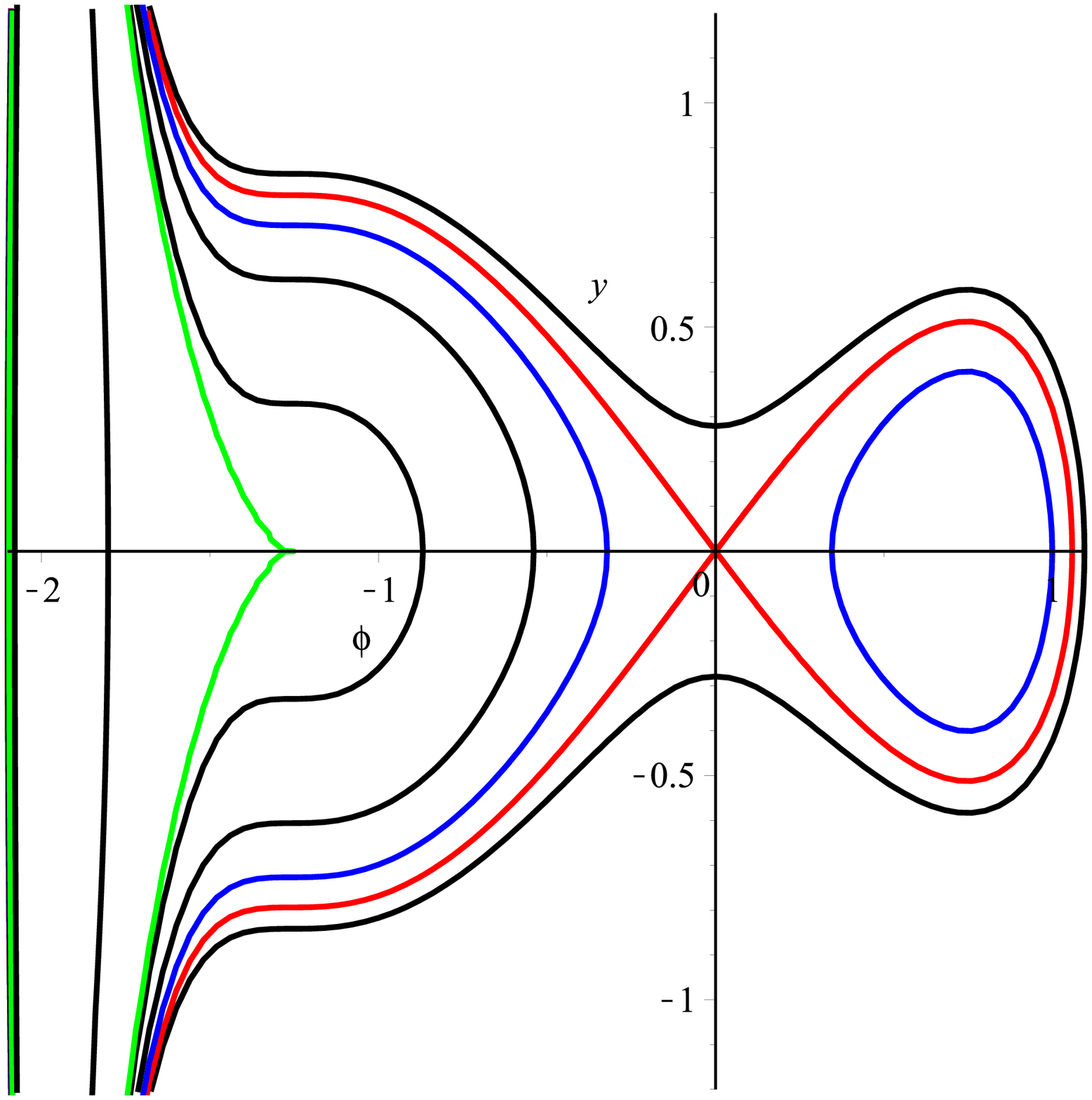}\\
		\footnotesize{(d) $2C_1<\phi_2$}
	\end{tabular}
\end{center}

\begin{center}
	{\small Fig.8 The function $g(\phi)$ admits a double zero , and $\tilde\phi_-<0<\phi_1$.       }
\end{center}

\begin{center}
	\begin{tabular}{ccc}
		\epsfxsize=5cm
		\epsfysize=5cm \epsffile{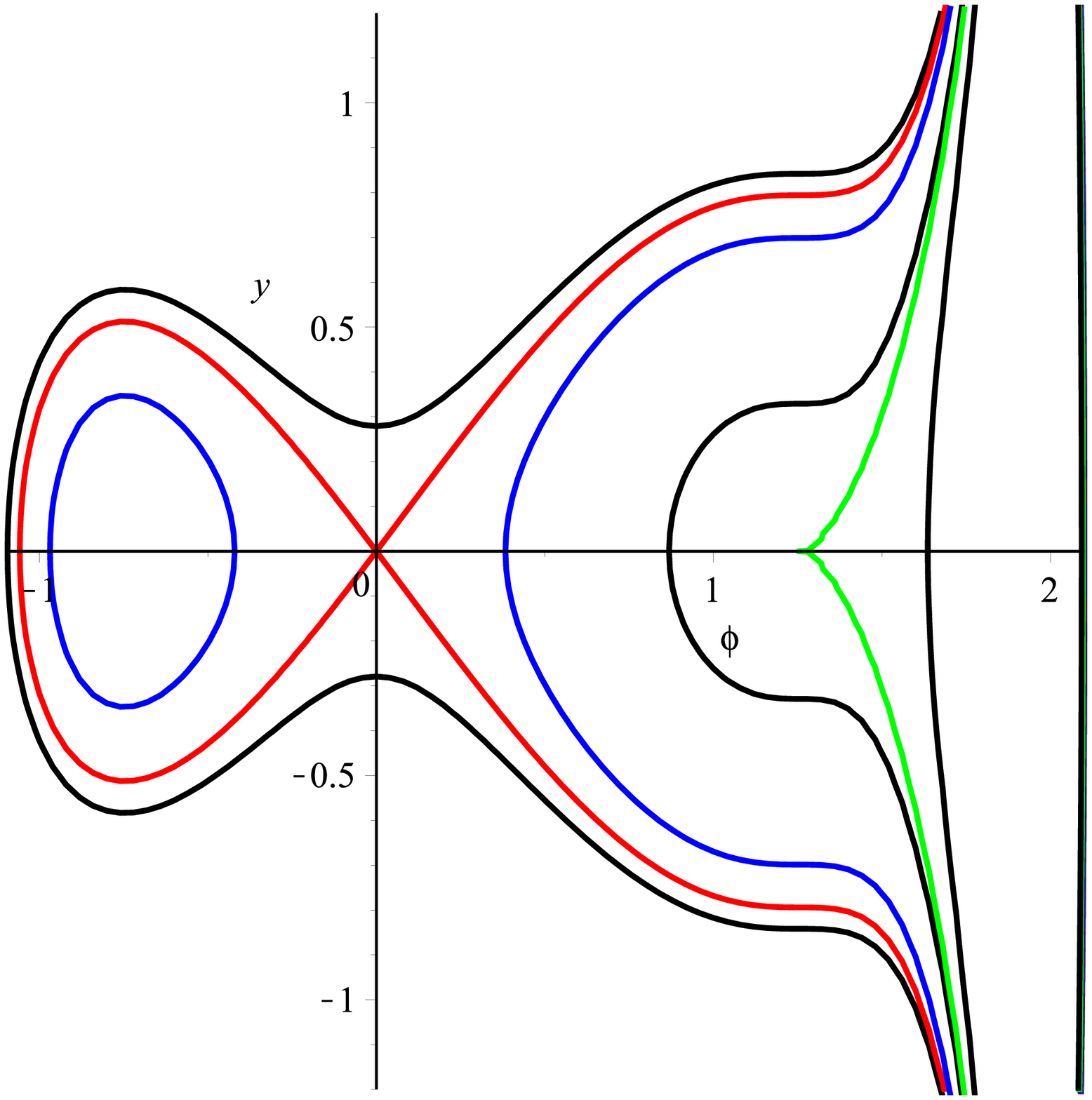}&
		\epsfxsize=5cm \epsfysize=5cm \epsffile{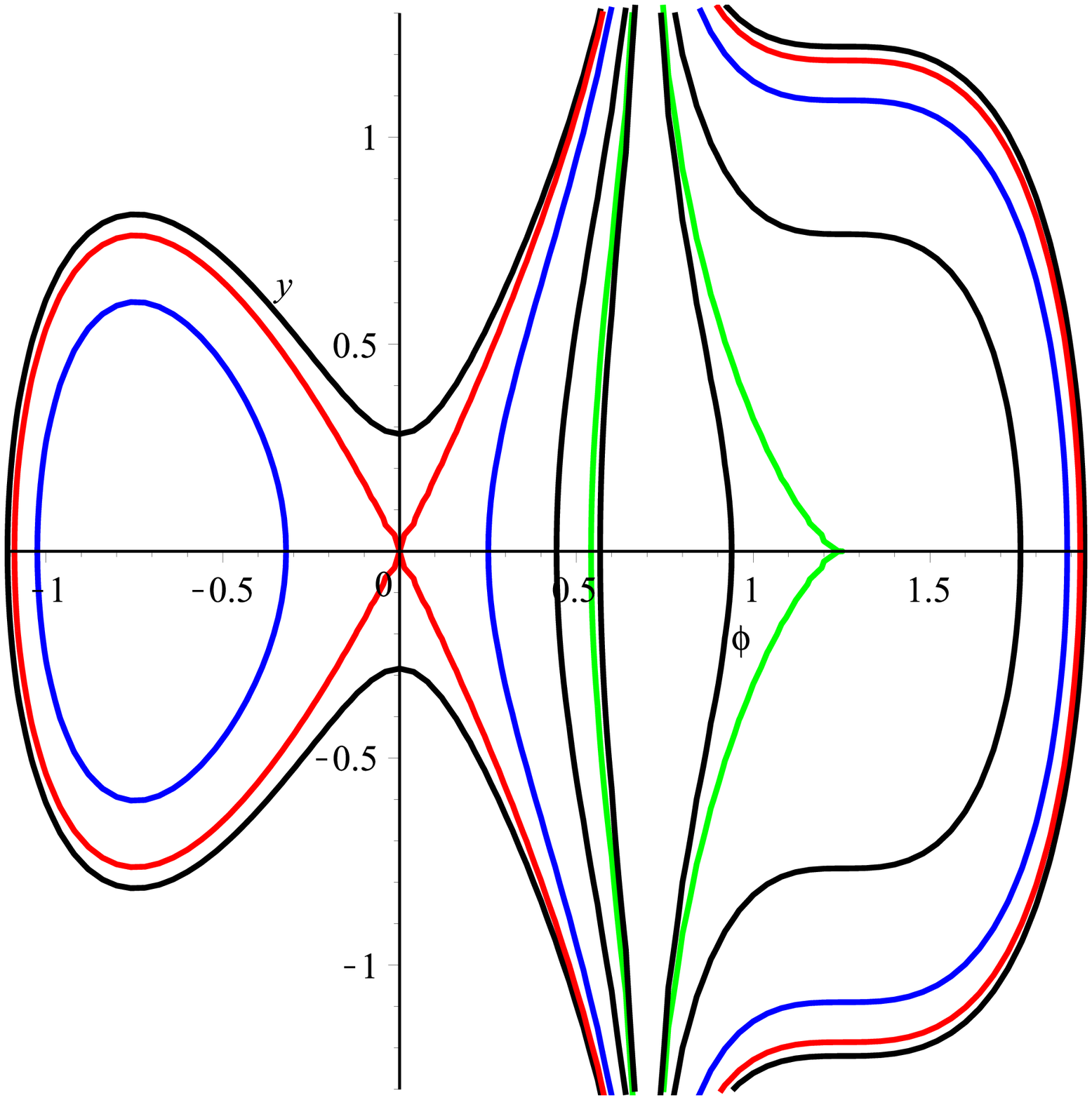}&
		\epsfxsize=5cm \epsfysize=5cm \epsffile{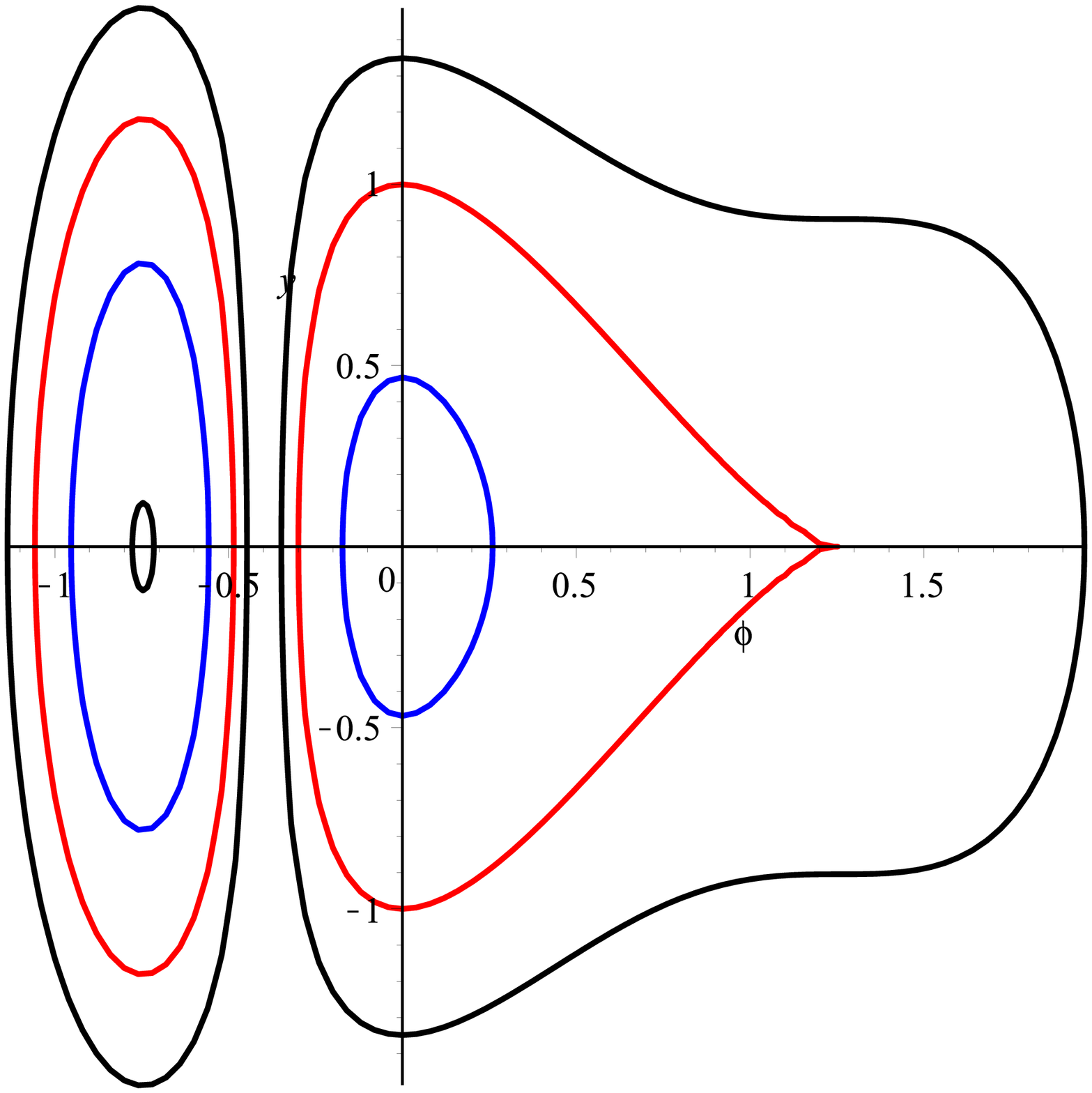}\\
		\footnotesize{ (a) $2C_1>\phi_1$ } & \footnotesize{(b) $0<2C_1<\phi_1$ }&
		\footnotesize{(c) $\phi_2<2C_1<0$ }
	\end{tabular}
\end{center}
\begin{center}
	\begin{tabular}{c}
		\epsfxsize=5cm
		\epsfysize=5cm \epsffile{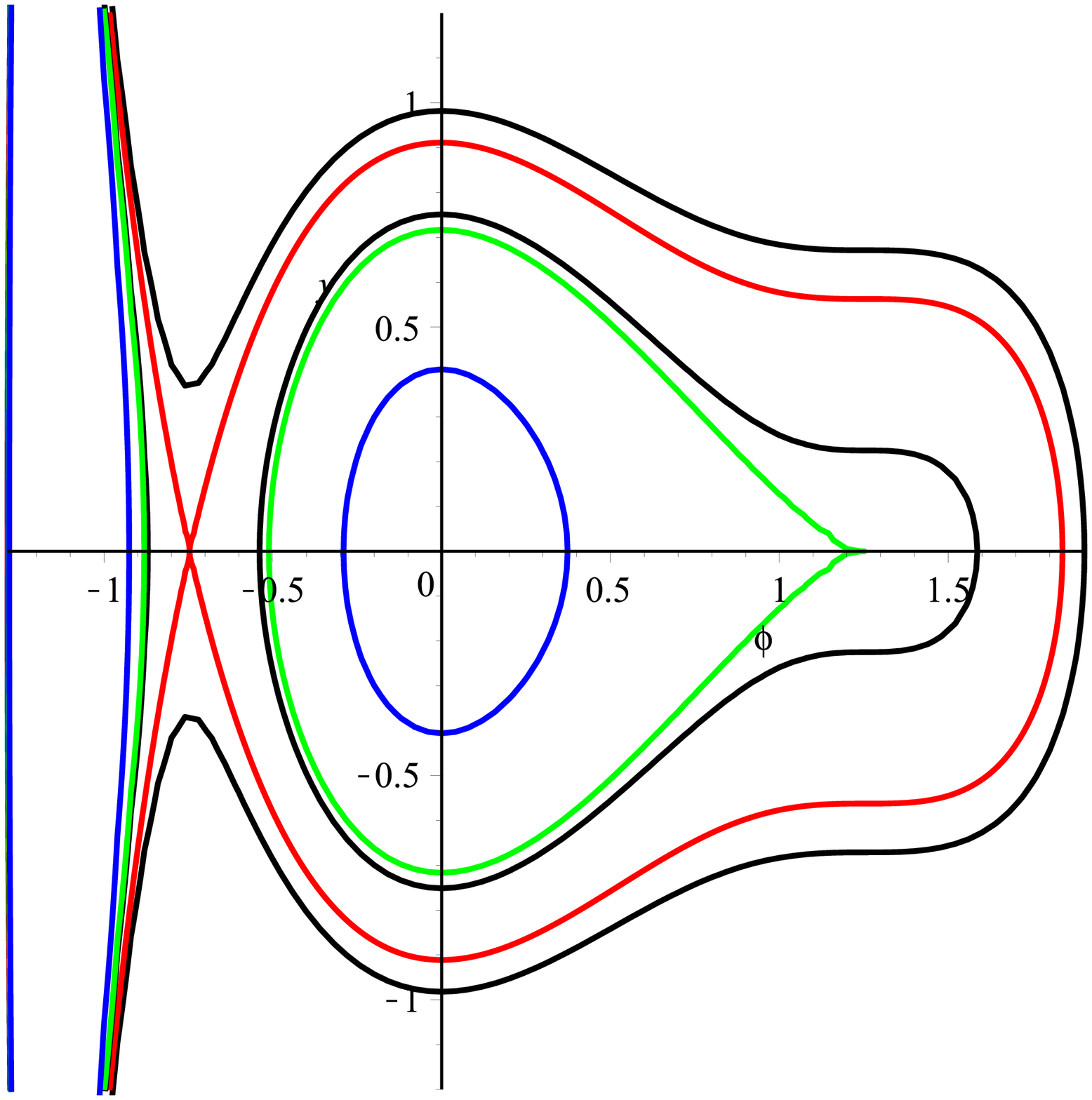}\\
		\footnotesize{(d) $2C_1<\phi_2$}
	\end{tabular}
\end{center}

\begin{center}
	{\small Fig.9 The function $g(\phi)$ admits a double zero , and $\phi_1<0<\tilde\phi_+$.)       }
\end{center}

\begin{center}
	\begin{tabular}{ccc}
		\epsfxsize=5cm
		\epsfysize=5cm \epsffile{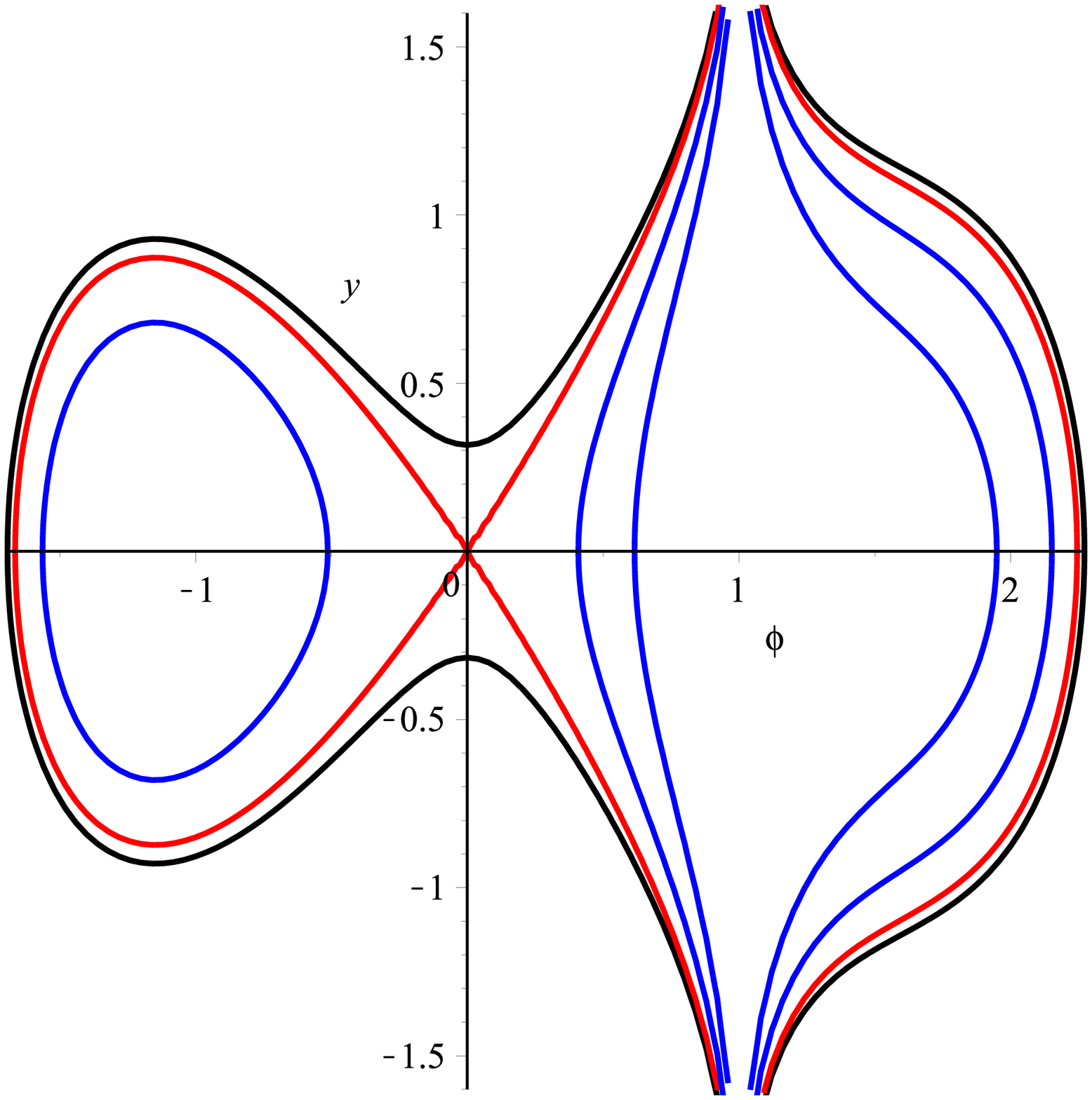}&
		\epsfxsize=5cm \epsfysize=5cm \epsffile{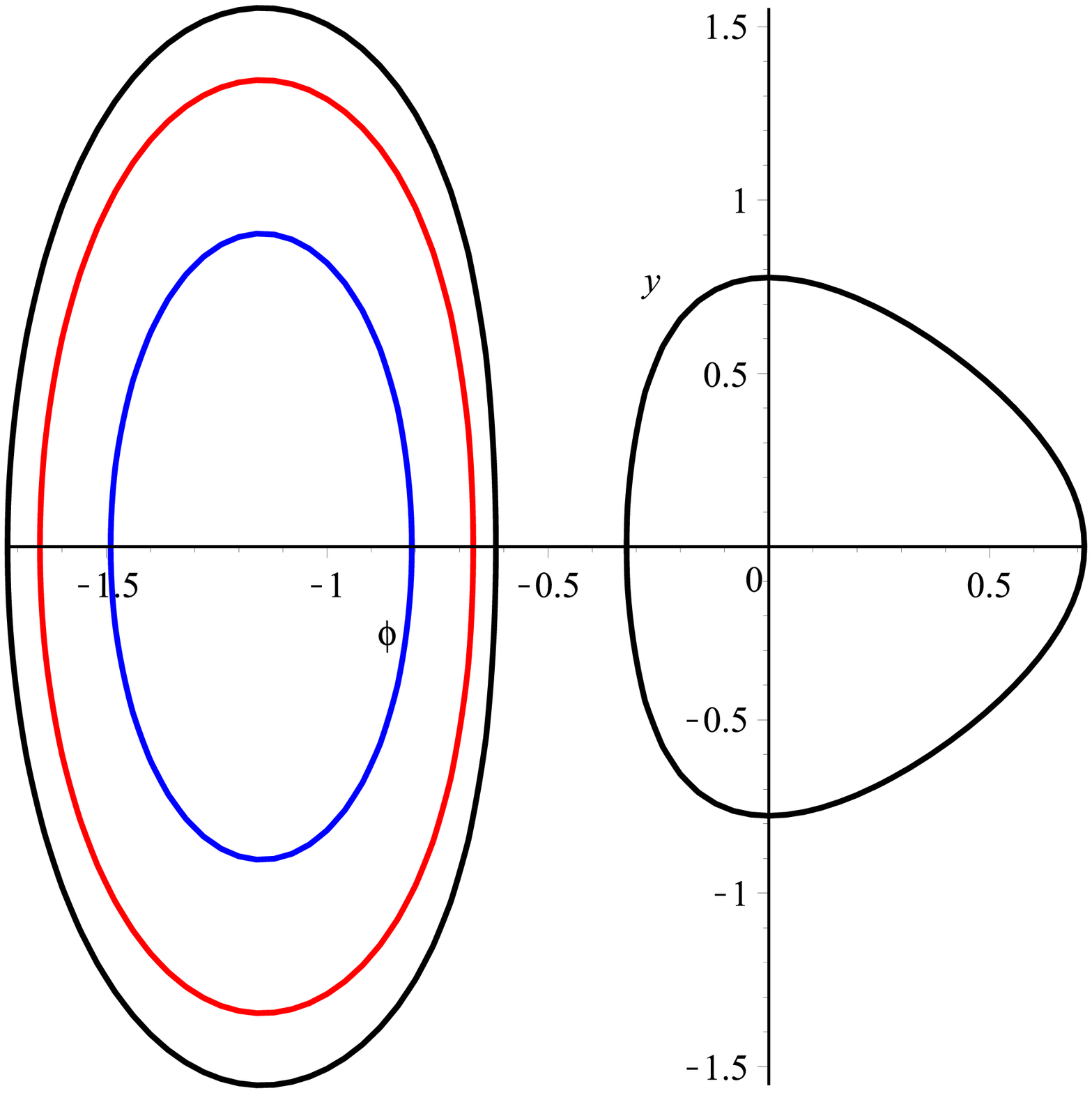}&
		\epsfxsize=5cm \epsfysize=5cm \epsffile{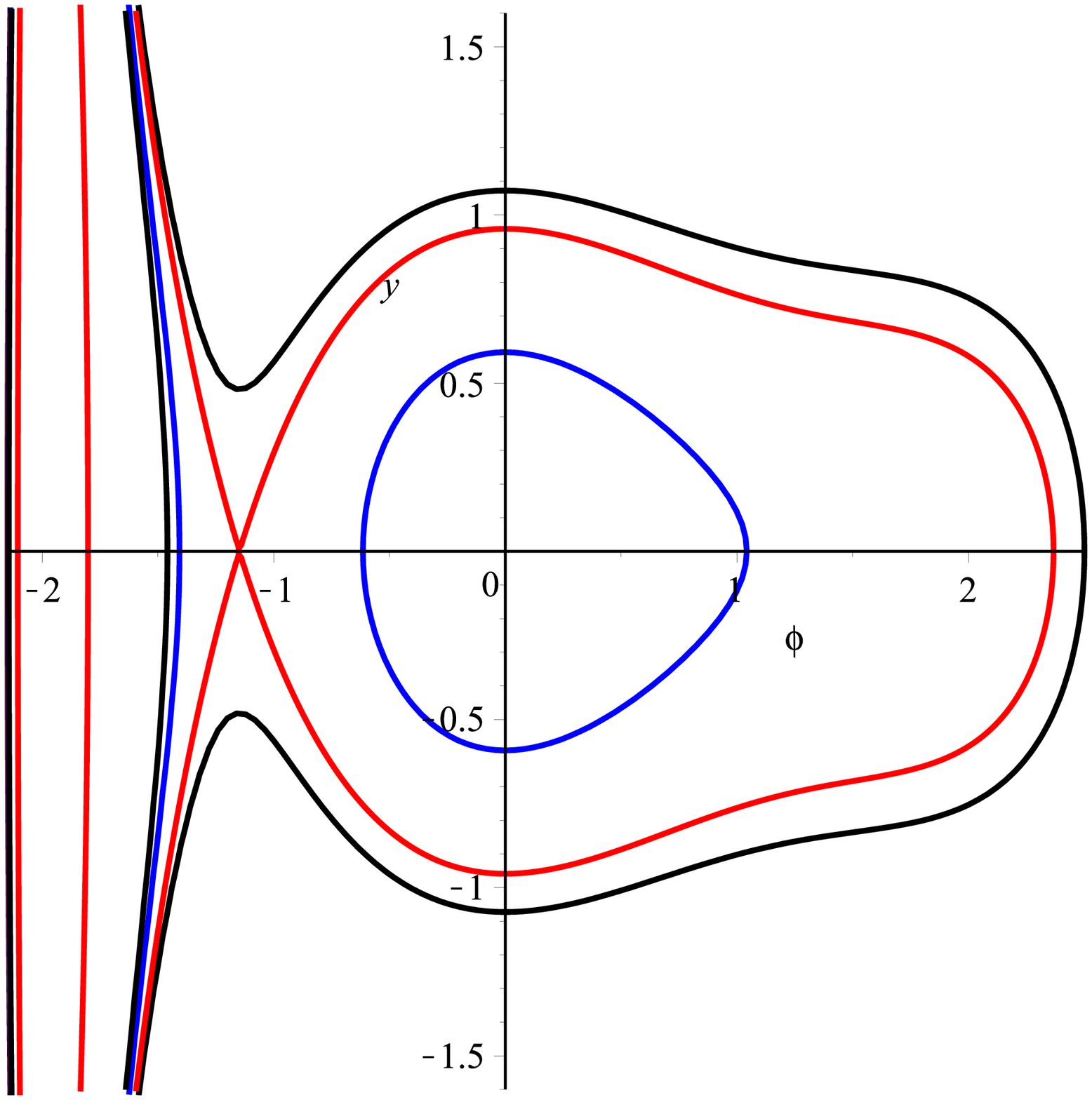}\\
		\footnotesize{ (a) $2C_1>0$ } & \footnotesize{(b) $\phi_1<2C_1<0$ }&
		\footnotesize{(c) $2C_1<\phi_1$ }
	\end{tabular}
\end{center}

\begin{center}
	{\small Fig.10  The function $g(\phi)$ admits only one zero, i.e $(\Delta_g)>0$ and $g(\phi_+)<0$.}
\end{center}

\begin{center}
	\begin{tabular}{ccc}
		\epsfxsize=5cm
		\epsfysize=5cm \epsffile{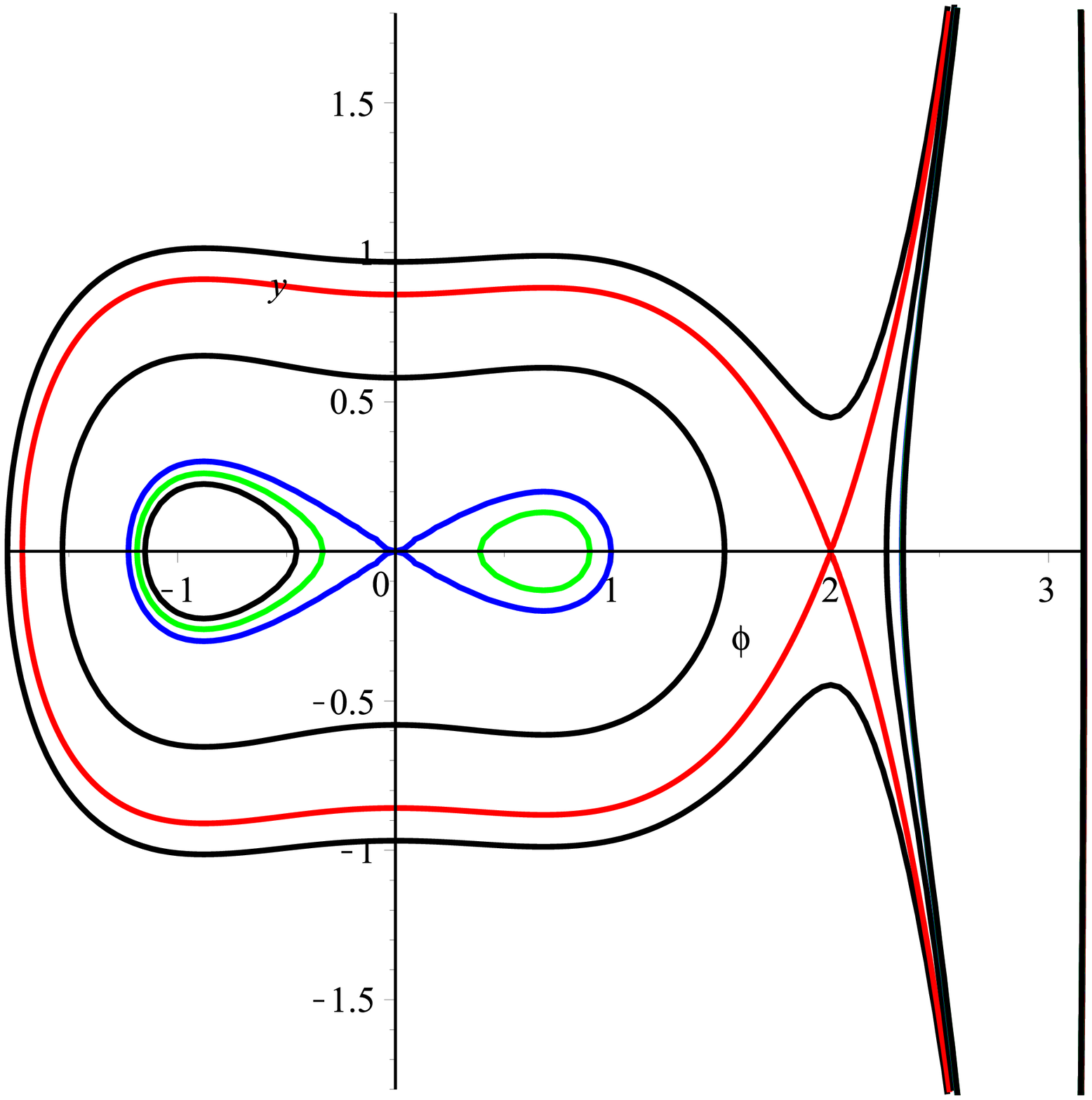}&
		\epsfxsize=5cm \epsfysize=5cm \epsffile{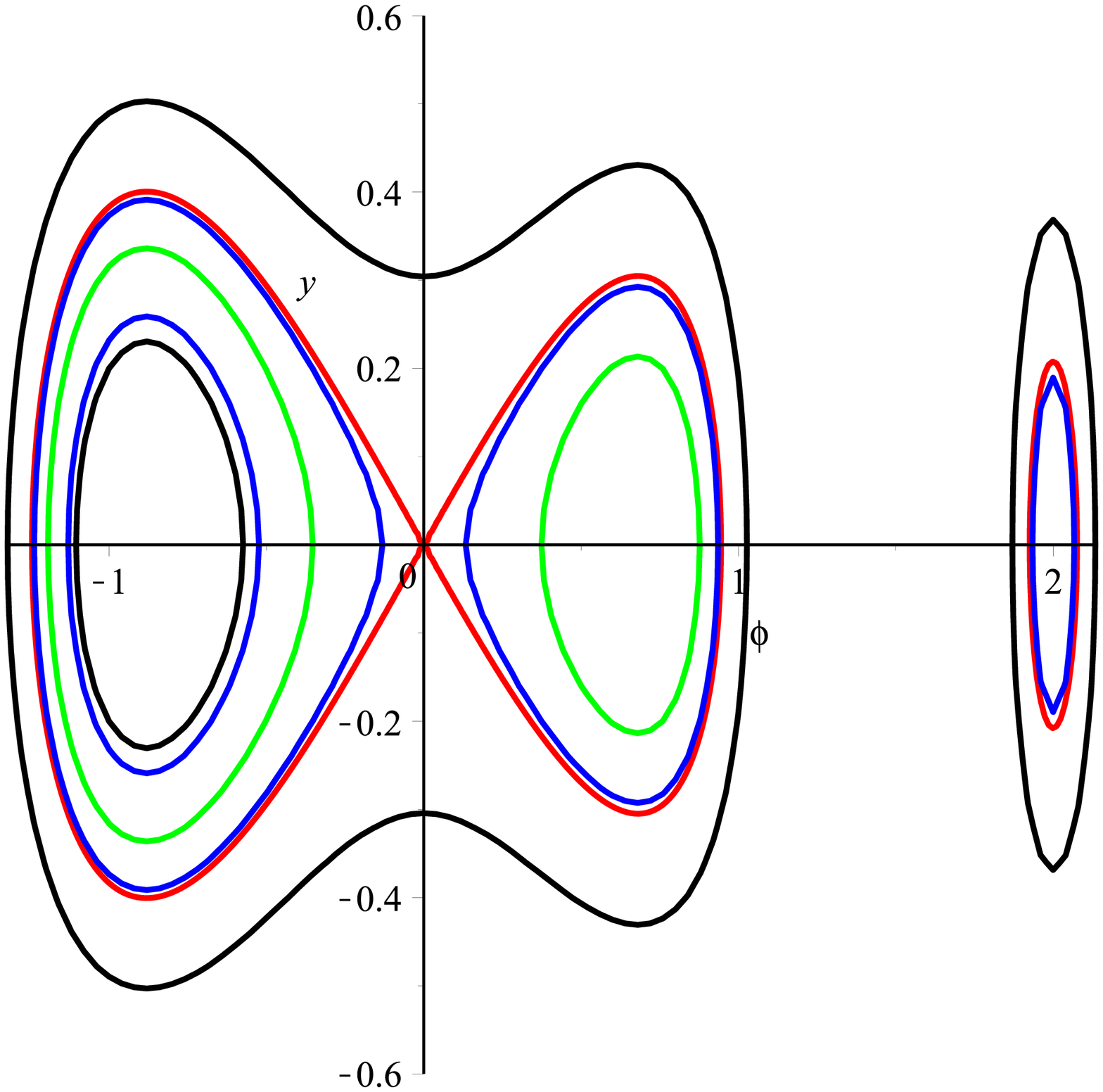}&
		\epsfxsize=5cm \epsfysize=5cm \epsffile{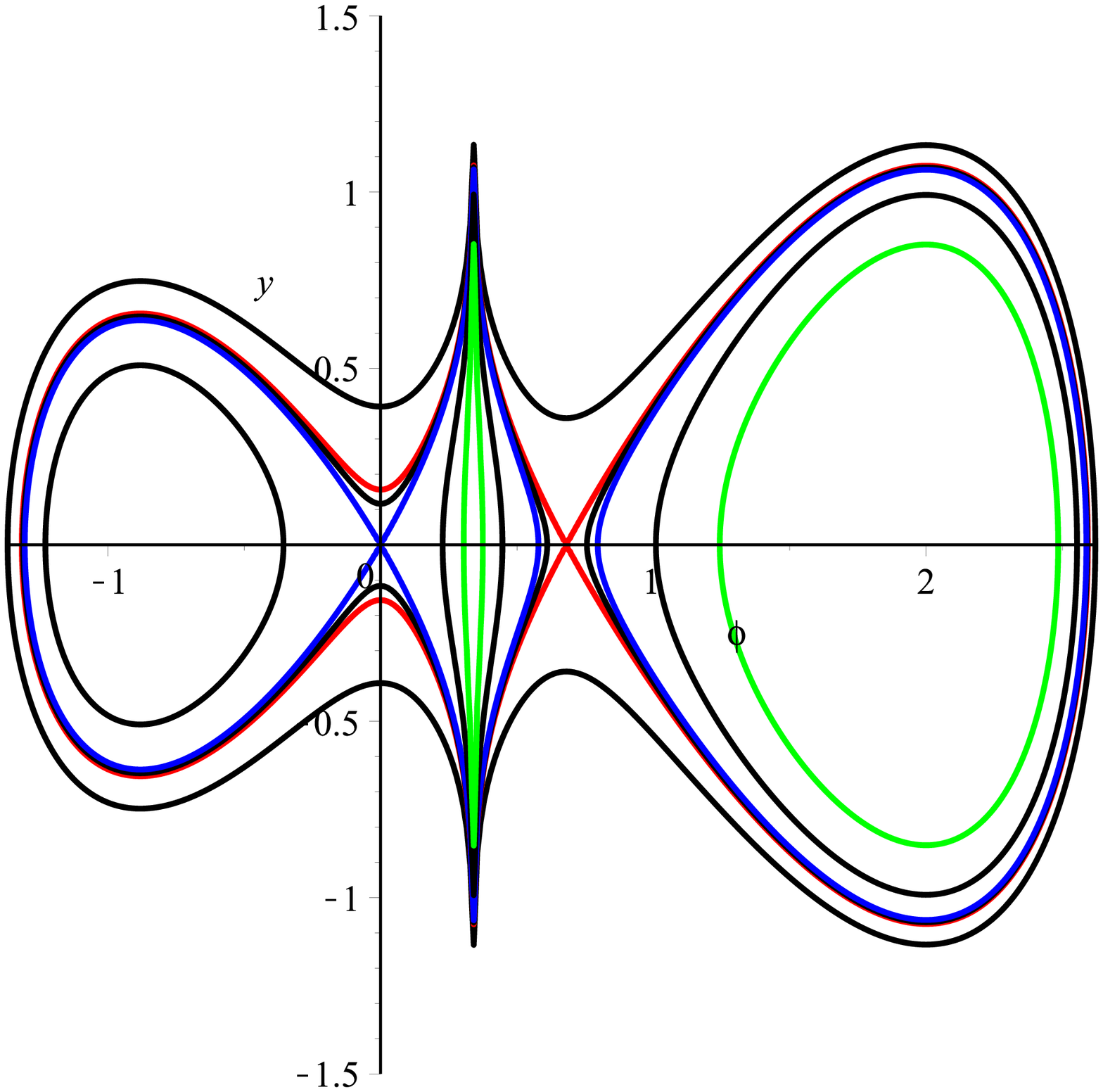}\\
		\footnotesize{ (a) $2C_1>\phi_1$ } & \footnotesize{(b) $\phi_2<2C_1<\phi_1$ }&
		\footnotesize{(c) $0<2C_1<\phi_2$ }
	\end{tabular}
\end{center}

\begin{center}
	\begin{tabular}{cc}
		\epsfxsize=5cm
		\epsfysize=5cm \epsffile{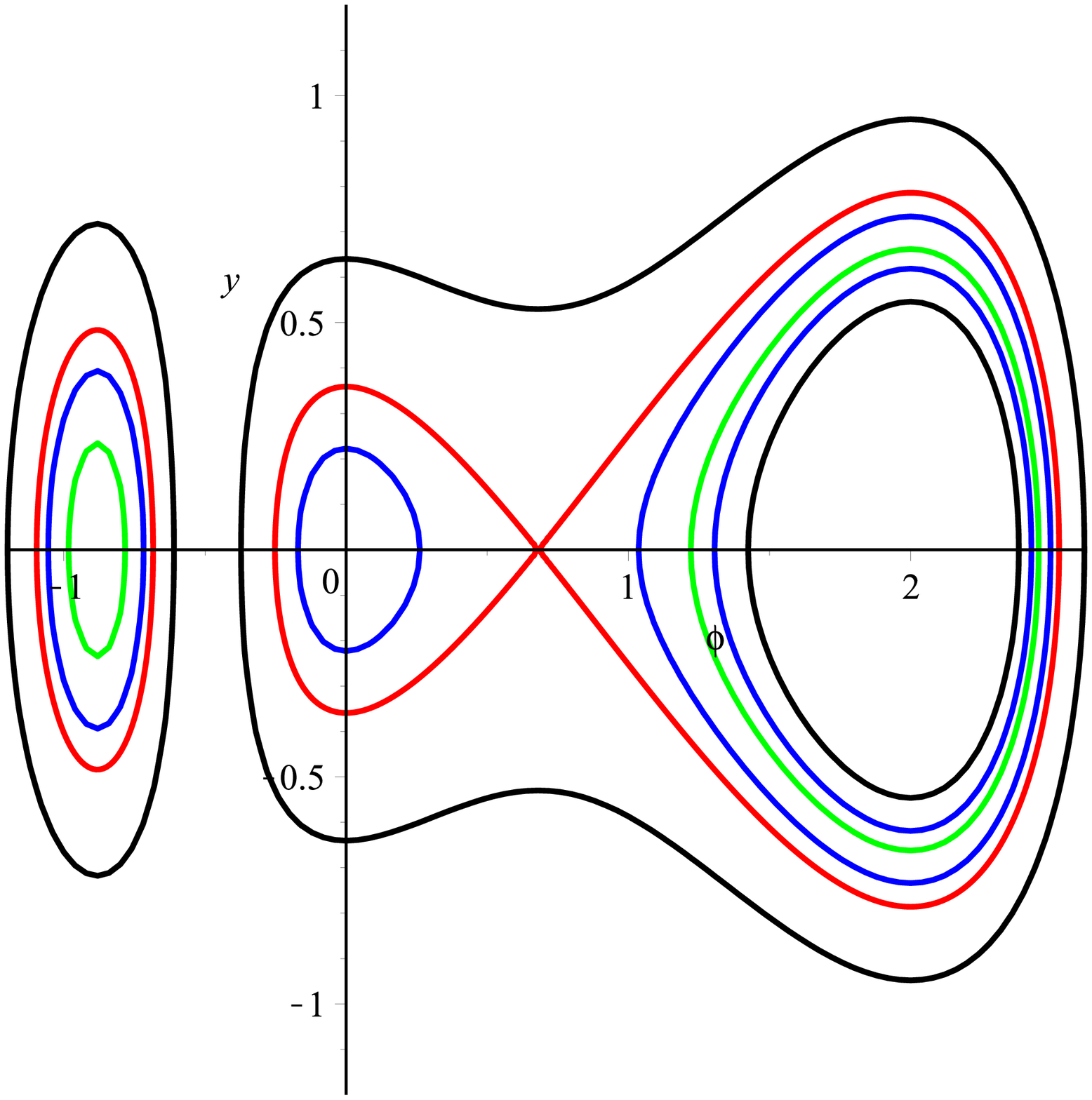}&
		\epsfxsize=5cm \epsfysize=5cm \epsffile{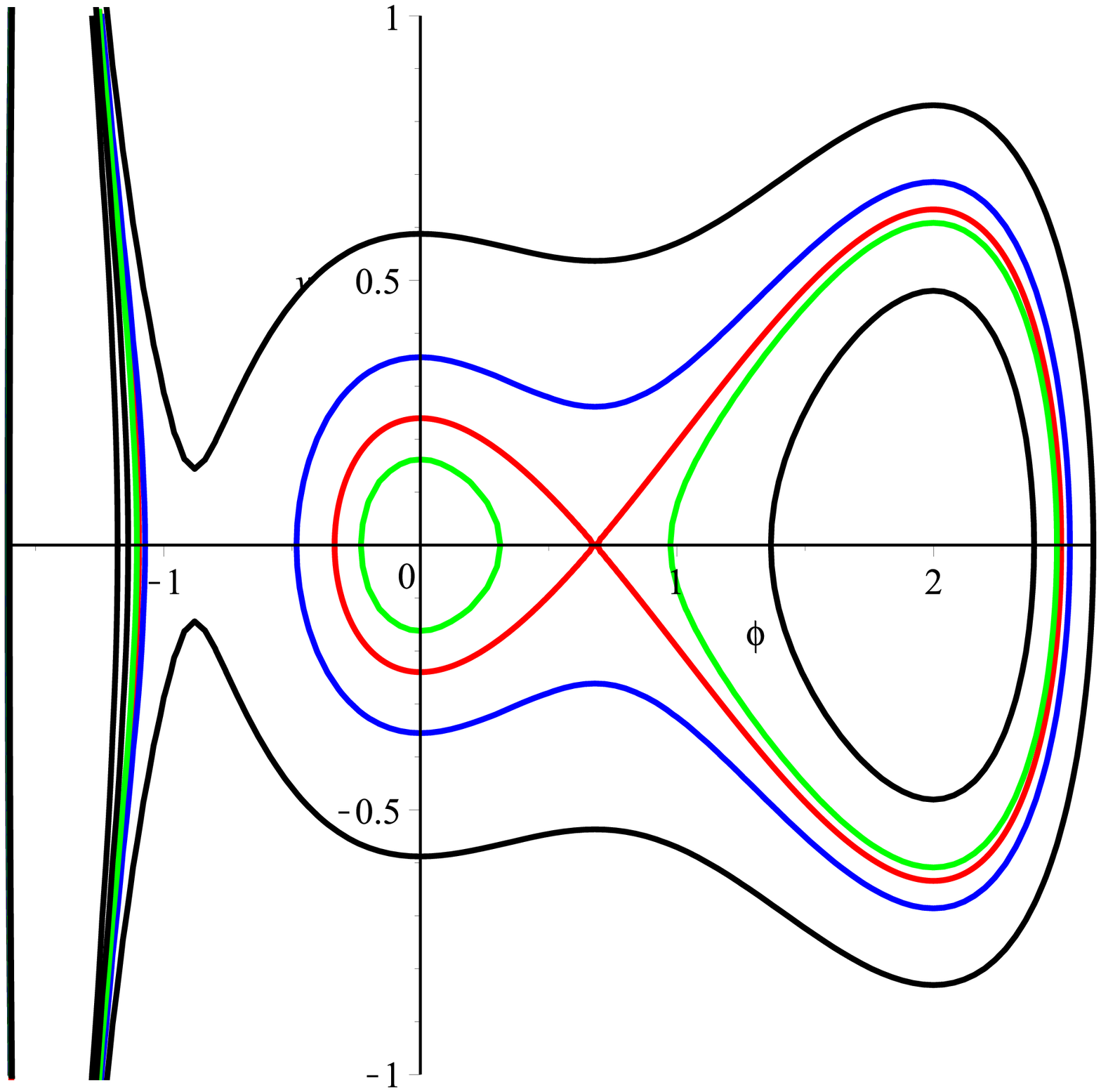}\\
		\footnotesize{ (a) $\phi_3<2C_1<0$ } & \footnotesize{(b) $\phi_3<0$ }
		
	\end{tabular}
\end{center}

\begin{center}
	{\small Fig.11 The function $g(\phi)$ admits three real zeros and $\phi_1<\phi_2<0<\phi_3$.       }
\end{center}

\begin{center}
	\begin{tabular}{ccc}
		\epsfxsize=5cm
		\epsfysize=5cm \epsffile{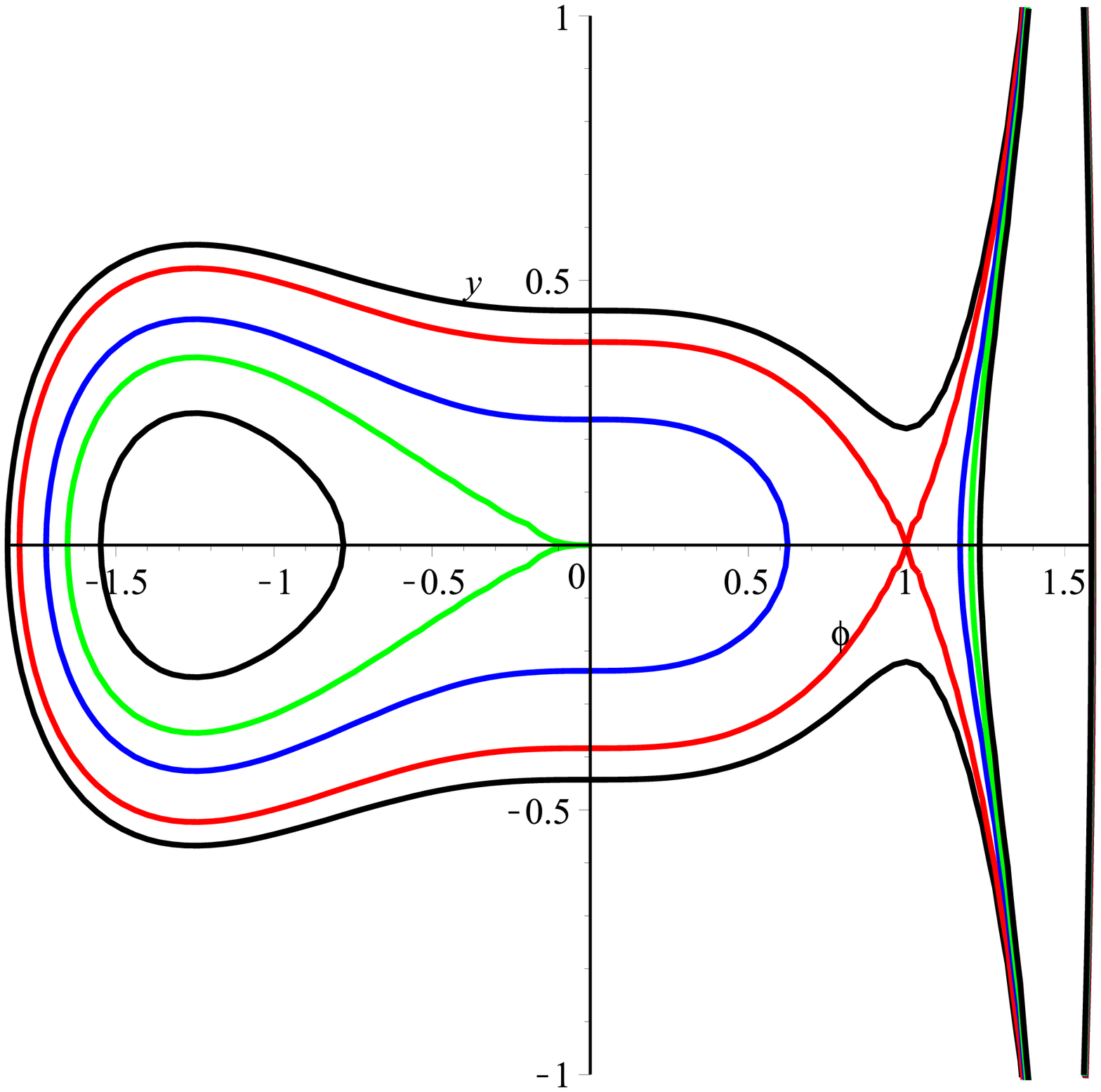}&
		\epsfxsize=5cm \epsfysize=5cm \epsffile{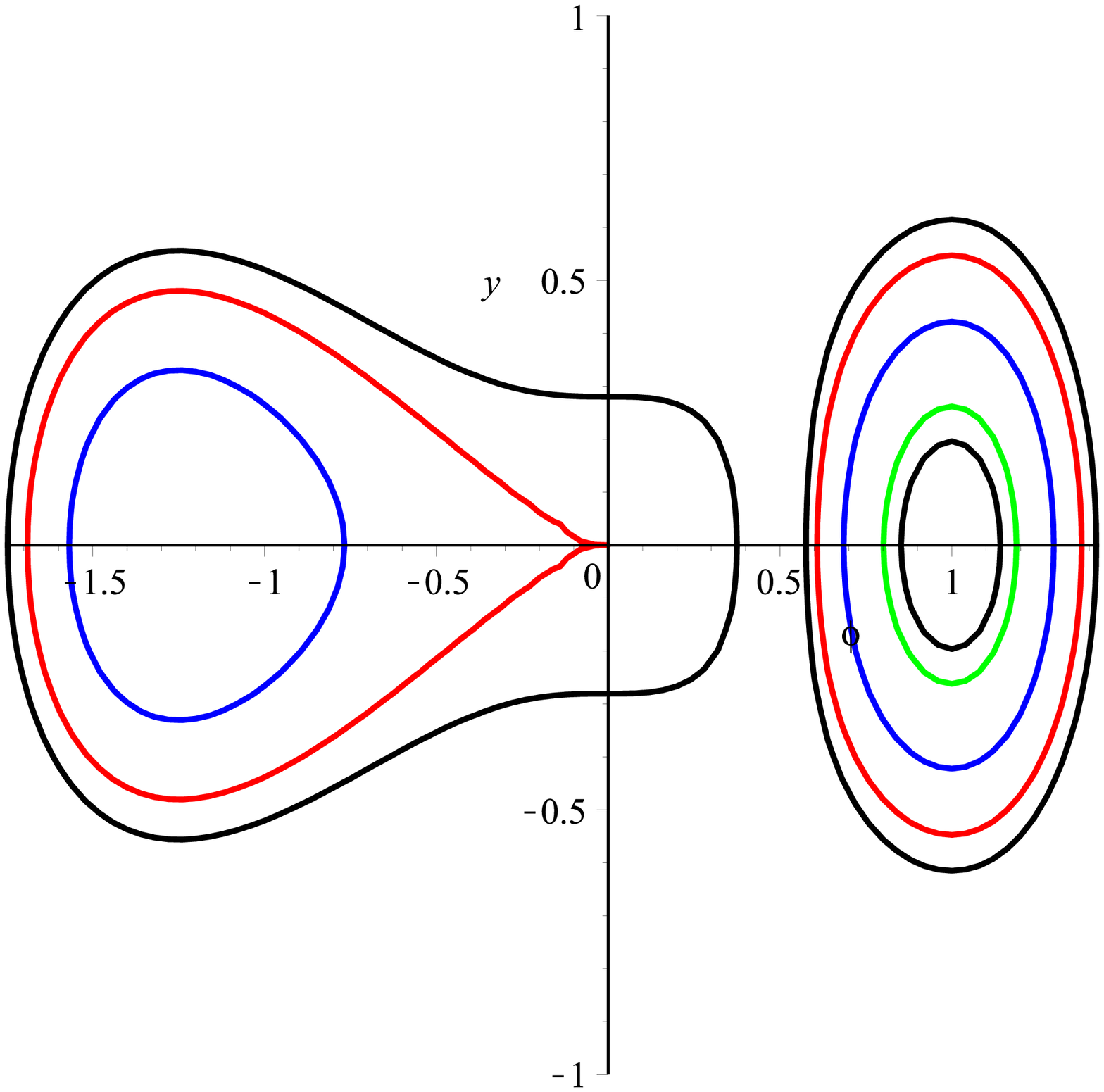}&
		\epsfxsize=5cm \epsfysize=5cm \epsffile{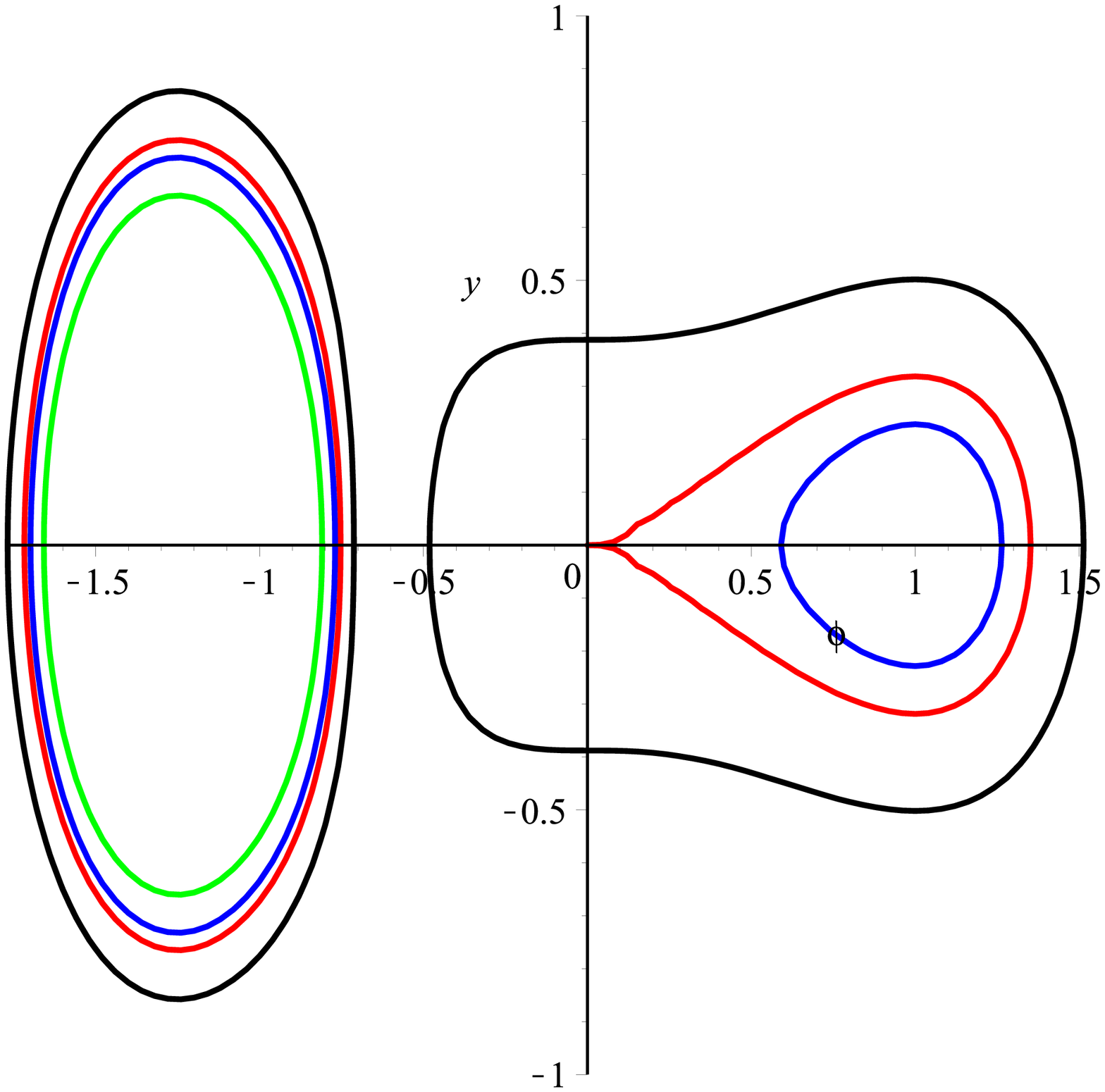}\\
		\footnotesize{ (a) $2C_1>\phi_1$ } & \footnotesize{(b) $0<2C_1<\phi_1$ }&
		\footnotesize{(c) $\phi_2<2C_1<0$ }
	\end{tabular}
\end{center}
\begin{center}
	\begin{tabular}{c}
		\epsfxsize=5cm
		\epsfysize=5cm \epsffile{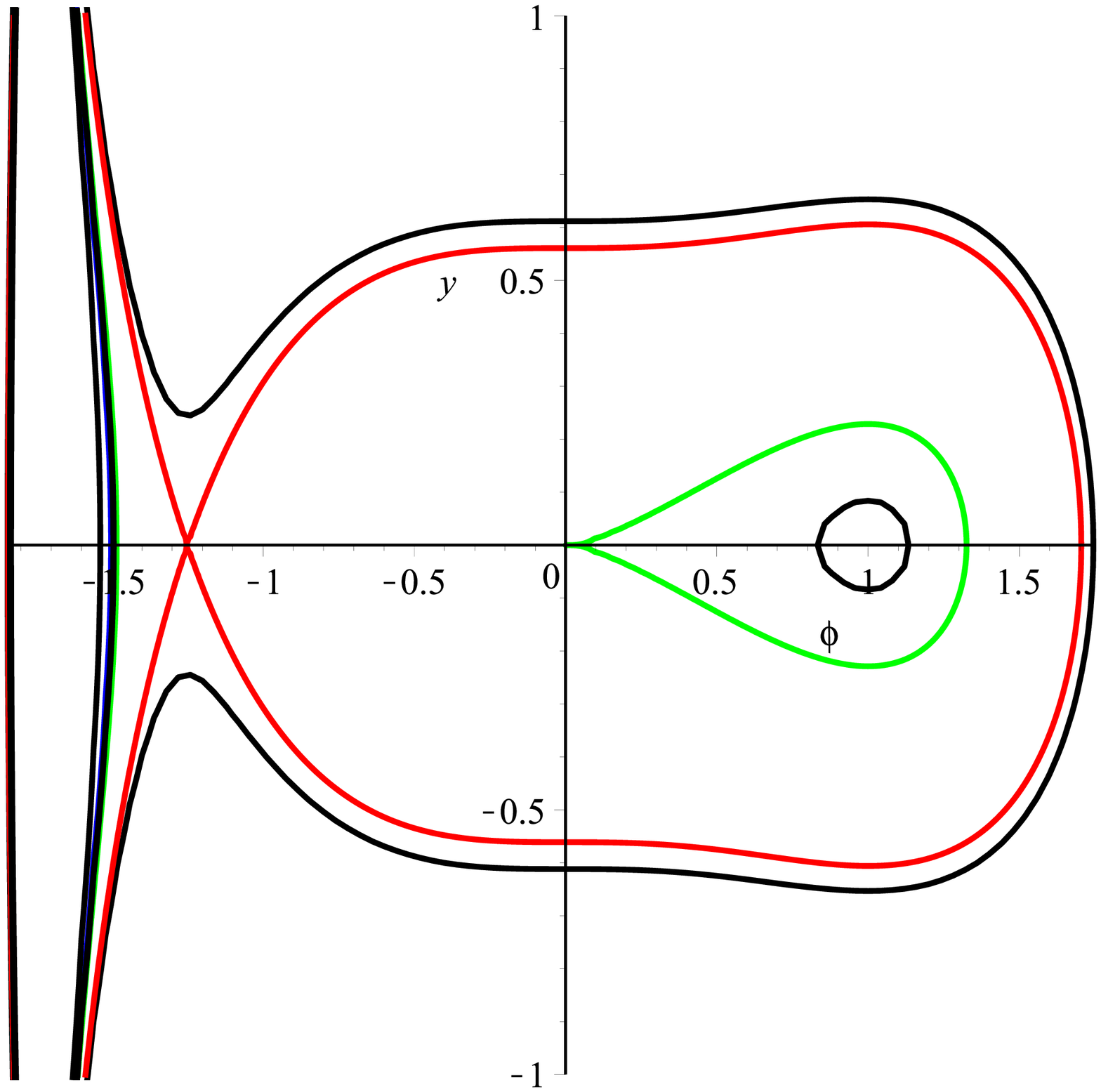}\\
		\footnotesize{(d) $2C_1<\phi_2$}
	\end{tabular}
\end{center}
\begin{center}
	{\small Fig.12  The function $g(\phi)$ admits three real  zeros and $K=0$.}
\end{center}

\begin{center}
	\begin{tabular}{ccc}
		\epsfxsize=5cm
		\epsfysize=5cm \epsffile{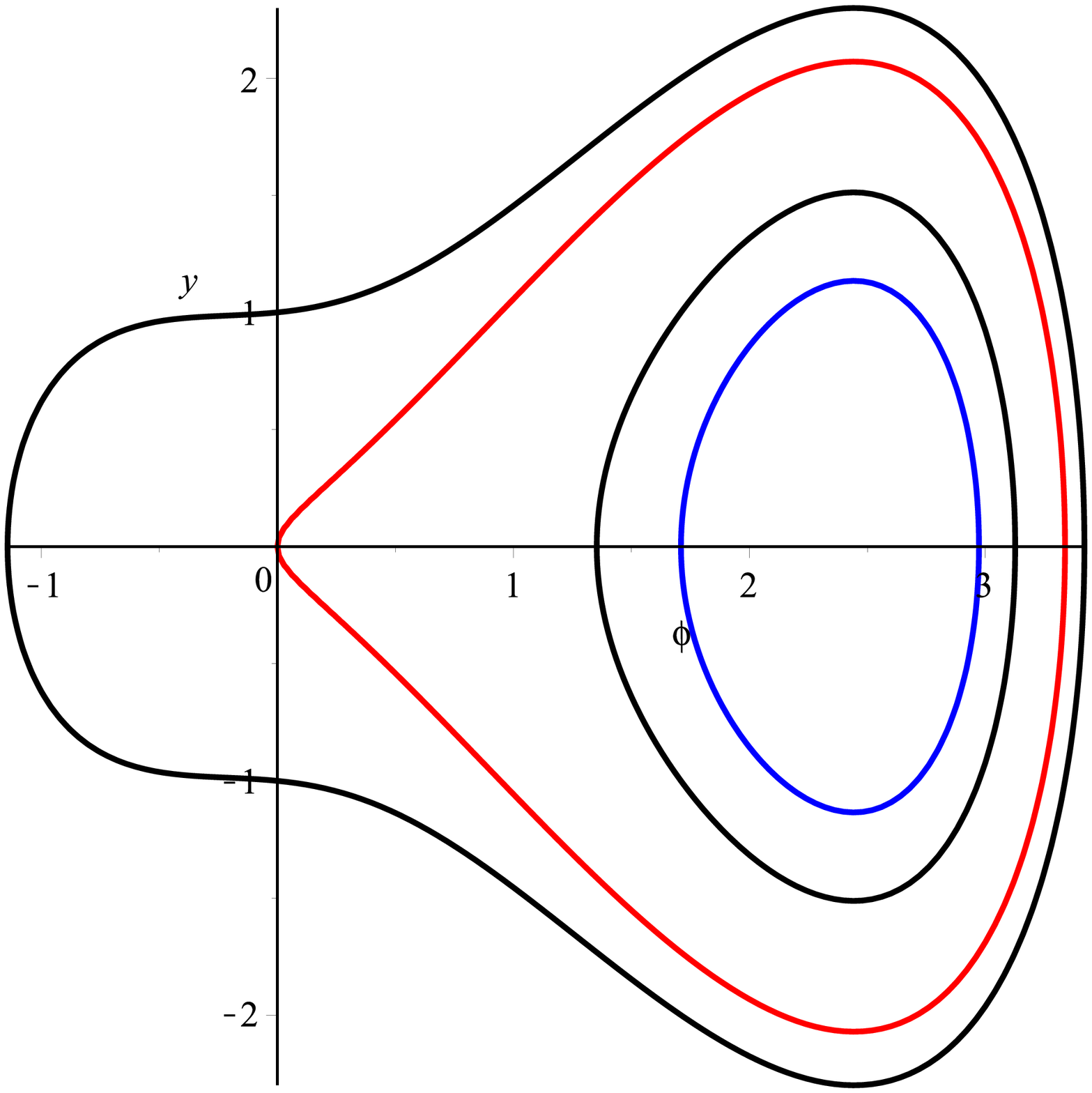}&
		\epsfxsize=5cm \epsfysize=5cm \epsffile{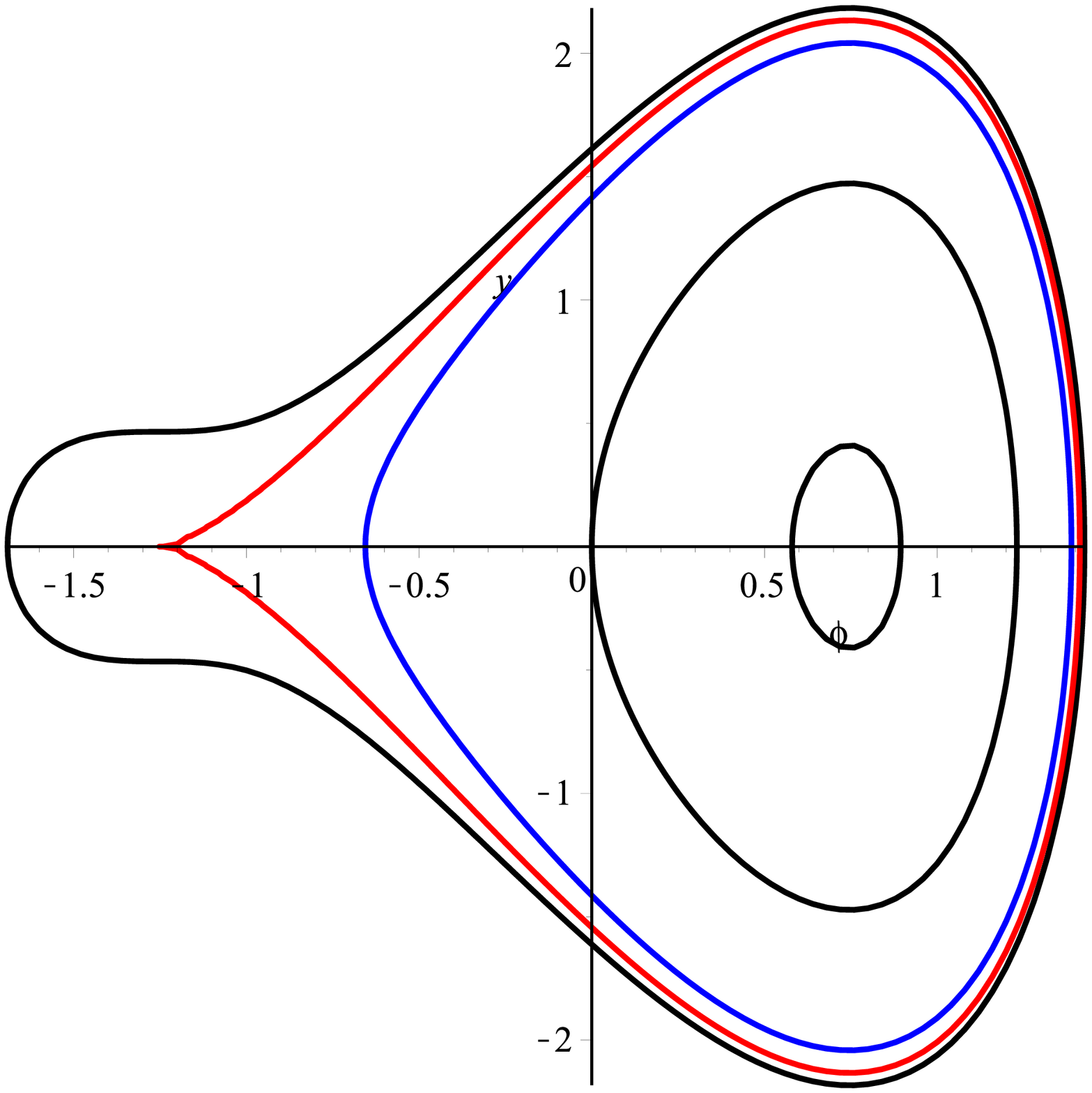}&
		\epsfxsize=5cm \epsfysize=5cm \epsffile{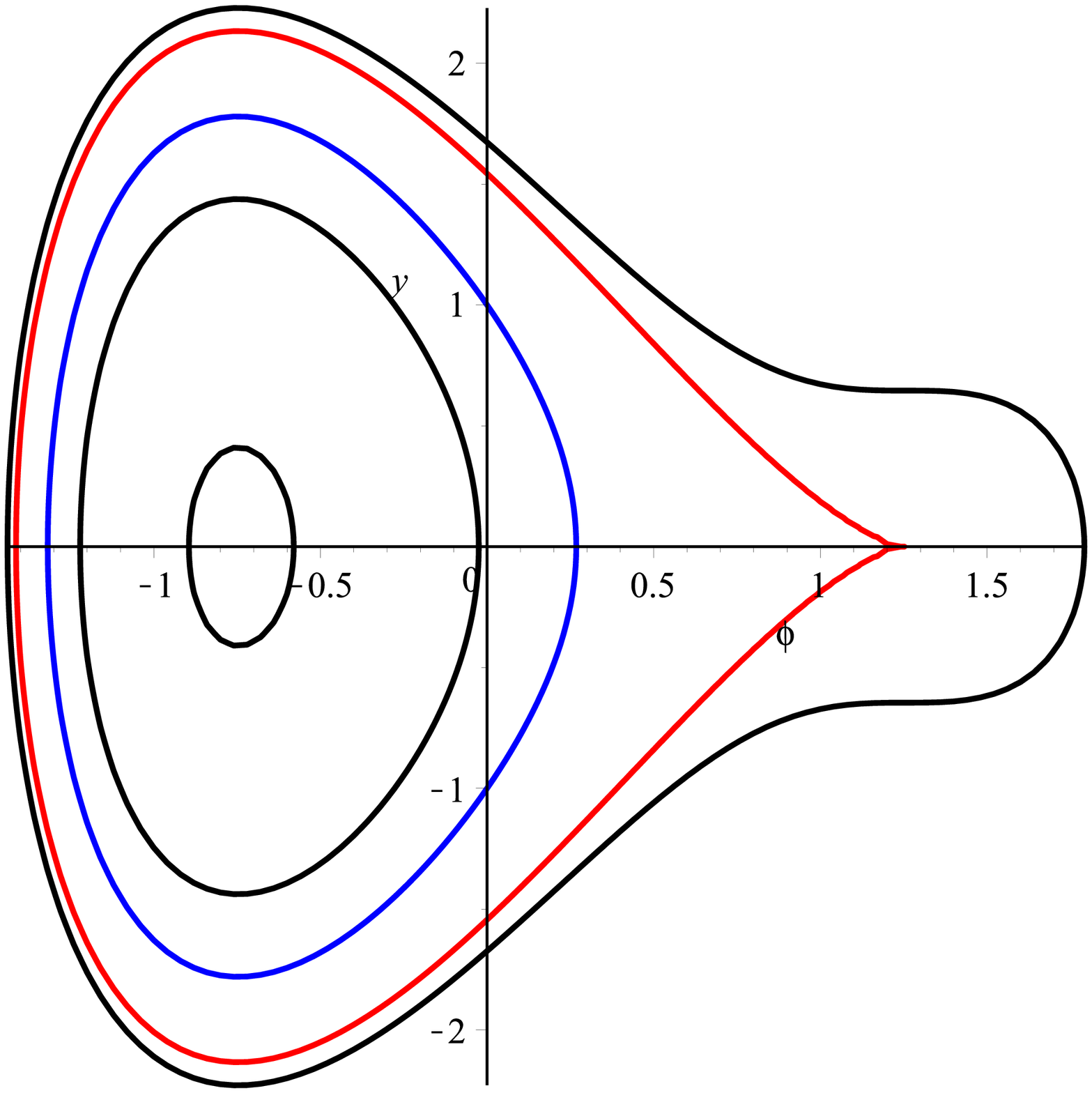}\\
		\footnotesize{ (a) $K\neq0, \Delta_{g'}>0, g(\tilde\phi_{-})>0, C_1=0$. } & \footnotesize{(b) $K\neq0, \Delta_{g'}>0, g(\tilde\phi_{-})=0, C_1=0$. }&
		\footnotesize{(c) $K\neq0, \Delta_{g^{'}}>0, g(\tilde\phi_{+})=0, C_1=0.$ }
	\end{tabular}
\end{center}

\begin{center}
	\begin{tabular}{ccc}
		\epsfxsize=5cm
		\epsfysize=5cm \epsffile{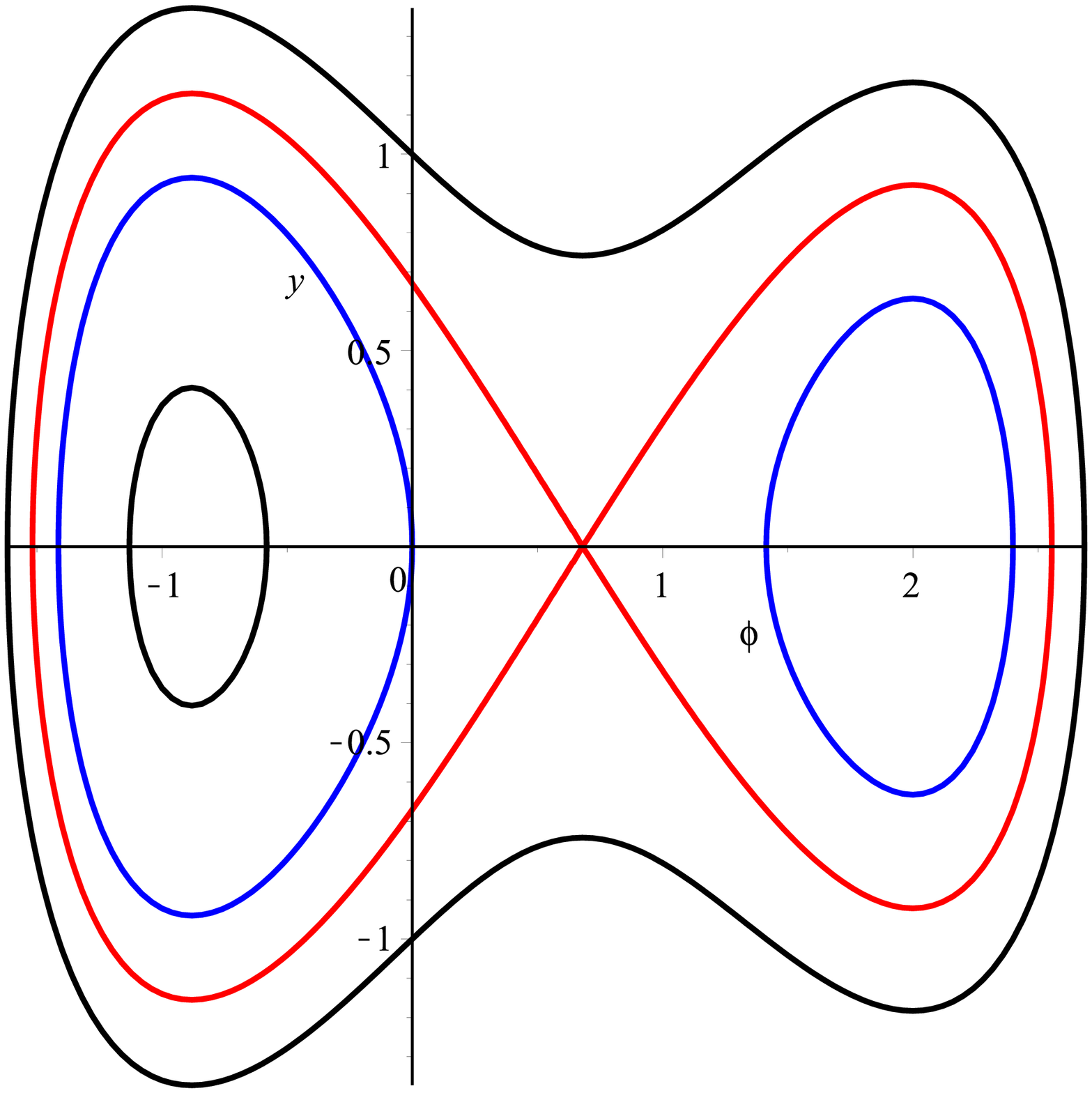}&
		\epsfxsize=5cm \epsfysize=5cm \epsffile{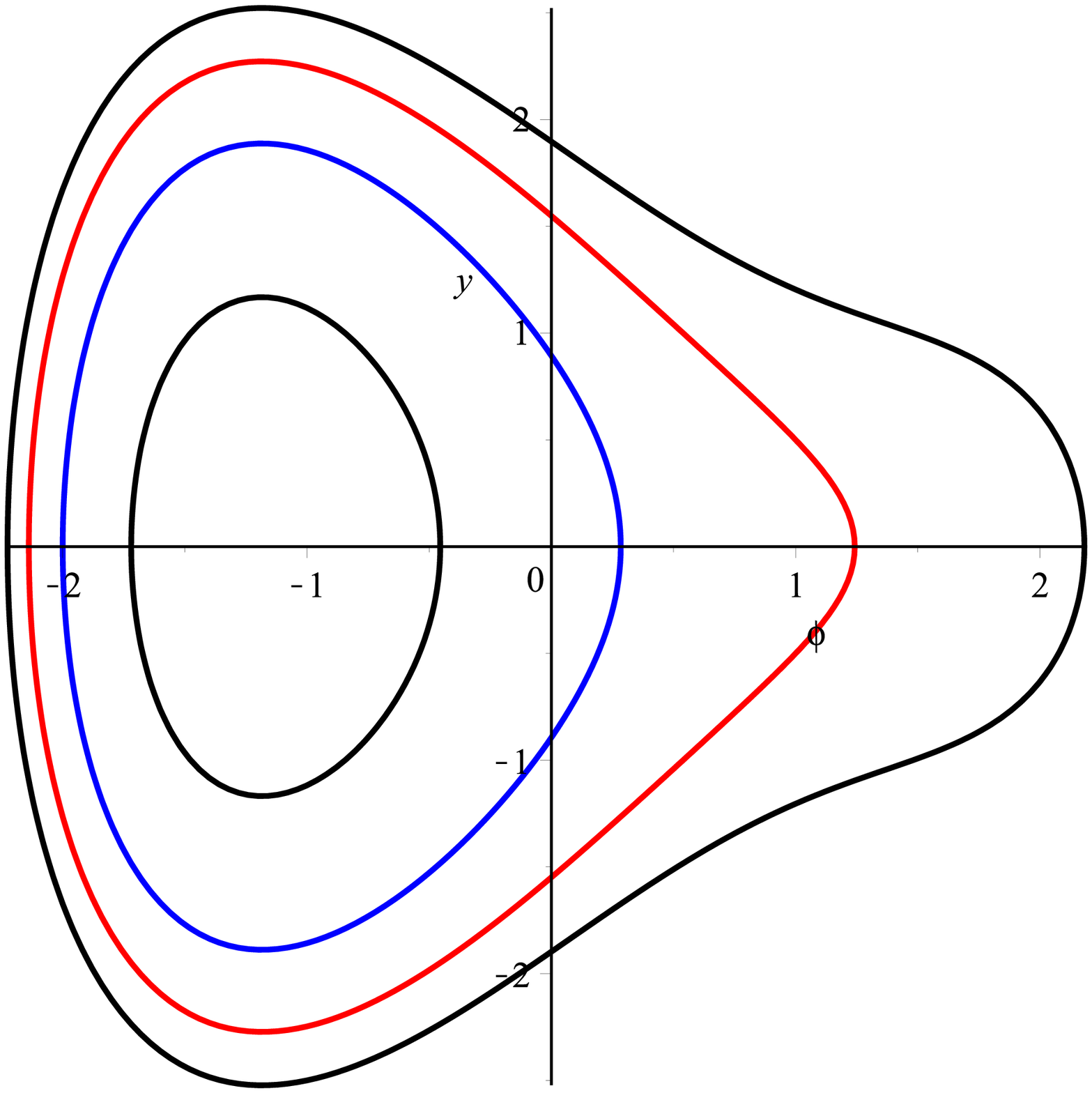}&
		\epsfxsize=5cm \epsfysize=5cm \epsffile{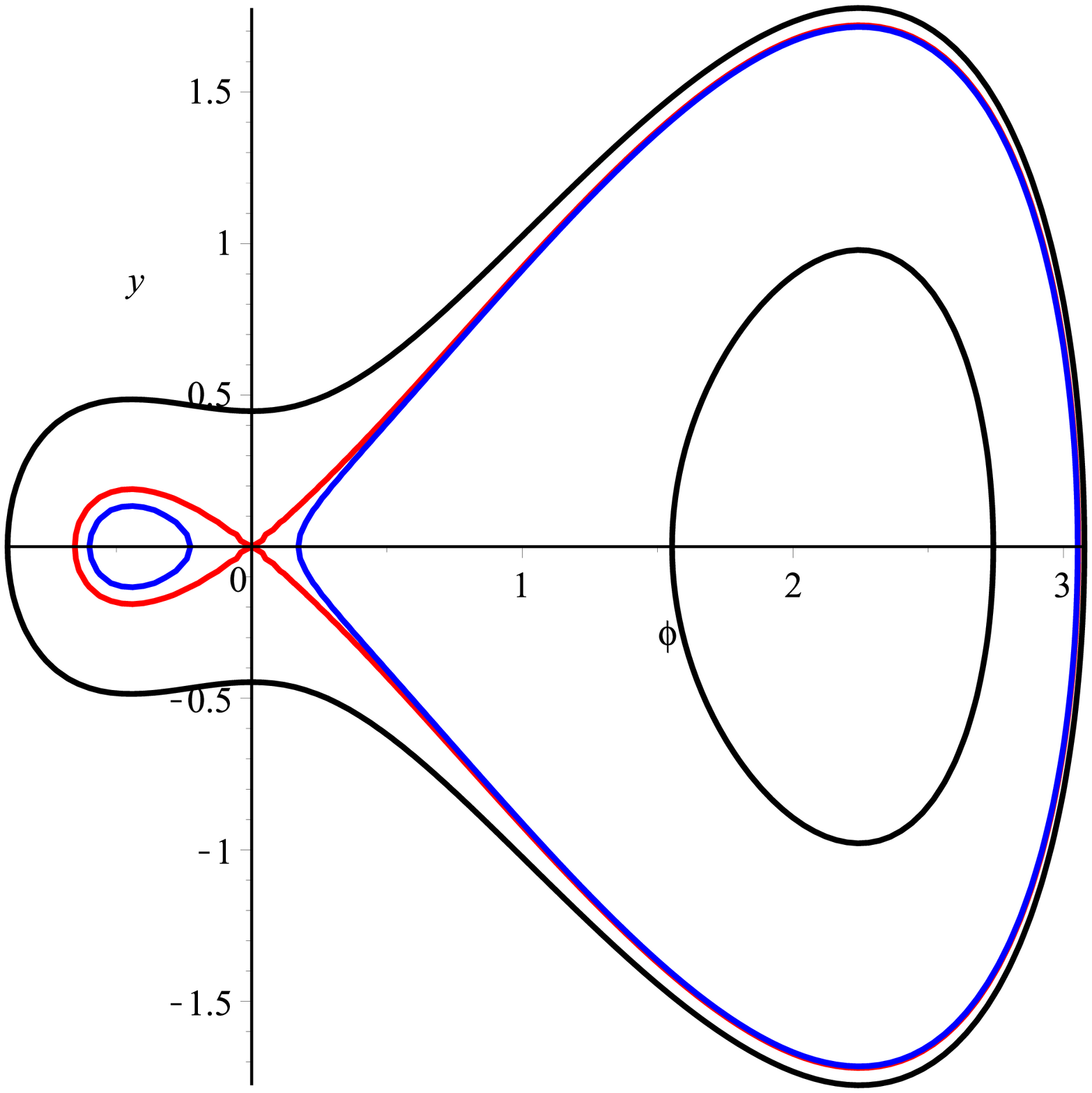}\\
		\footnotesize{ (d)$K\neq0, \Delta_{g'}>0, g(\tilde\phi_{+})>0, C_1=0.$ } & \footnotesize{(e) $K\neq0, \Delta_{g'}>0, g(\tilde\phi_{+})<0, C_1=0.$ }&
		\footnotesize{(f) $K=0, C^2_2>2C_3, C_1=0$. }
	\end{tabular}
\end{center}
\begin{center}
	{\small Fig.13  The special case of $C_1=0$.}
\end{center}

\subsection{Case of \textbf{$\theta=1$}}
If $\theta=1$  then system (1.7) become
\begin{equation}\label{planar system26}
	\begin{cases}
		\frac{d\phi}{d\xi}=y\\
		\frac{dy}{d\xi}=\frac{\frac{1}{2}y^2+C_3\phi^4+C_2\phi^3+\frac{1}{2}\phi^2+K\phi}{\phi-C_1},\\
	\end{cases}
\end{equation}
with its associated regular system
\begin{equation}\label{planar system27}
	\begin{cases}
		\frac{d\phi}{d\tau}=y(\phi-C_1),\\
		\frac{dy}{d\tau}=\frac{1}{2}y^2+C_3\phi^4+C_2\phi^3+\frac{1}{2}\phi^2+K\phi,\\
	\end{cases}
\end{equation}

and Hamiltonian
\begin{equation}\label{Hamiltonian4}
	\begin{aligned}
		H(\phi,y)=y^2(\phi-C_1)+\frac{2}{5}C_3\phi^5+\frac{1}{2}C_2\phi^4+\frac{1}{3}\phi^3+K\phi^2=h.\\
	\end{aligned}
\end{equation}
Note that, for $C_3=-a_3$, $C_2=-a_2$, $a_1=-\frac{1}{4}$ and $K=-a_0$, Eq. (\ref{Hamiltonian4}) is exactly the first integral of the asymptotic Rotation-CH equation considered in \cite{Liang}. Thus, we do not show results.
\section{Dynamical behavior of solutions of system (\ref{planar system})}
In this section, we analyze the dynamical behavior and give some exact traveling wave solutions of system  (\ref{planar system}).

\subsection {Case of \textbf{$\theta=\frac{1}{4}$}}
The expression of the first integral of system (\ref{planar system23}) makes the computation of exact solutions of traveling waves quite a tedious process. In this section, we discuss the dynamical behavior of the solutions, based on the properties of singular points and the obtained phase portraits.
To discuss the traveling wave solutions, we utilize the following two theorems.
\textbf{Theorem A}  (The Rapid-Jump Property of the
Derivative near the Singular Straight Line), and \textbf{Theorem B} (Existence of Finite Time Interval(s)
of Solutions with Respect to Variables in the Positive
or (and) Negative Direction(s))(See \cite{Li-Book}).
The above theorems have been proven in \cite{Lijibin3}.
Consecutively to the above results, we classify the
profiles for the wave function $\phi(\xi)$.

\subsubsection{Smooth solitary and periodic wave solution }

In figures 2a, 2b, 3a, 3b, 5a, 5b, there exists a family of periodic orbits enclosing the equilibrium point $E_0(0,0)$ defined by $H(\phi,y)=h$ with $h \in (h_1,h_3) (or (h_3,h_1))$. In these cases system (\ref{planar system23}) admits a periodic wave solutions(see fig. 7(a)).
Considering figures 2a, 2d, 3a, 3e, 3f, 3g, 4a, 4b, 4g, 5a, 5b, 5c, 5d, 5e, 5f, 5i, 6f, we have homoclinic orbits to the equilibrium points $E_0(0,0)$ or $E_i(\phi_i)$ defined by $H(\phi, y)=h_i$. In these cases, system (\ref{planar system23}) has solitary wave solutions (see fig. 7(b)).

\subsubsection{Peakon, anti-peakon and periodic peakon solution }

The left or right arches connecting the two saddle points on the singular line, in figures 2c, 3c, 4d, 4e, 5c, 5g, are limit solutions for the periodic orbits they enclose (\cite{Liang}). Considering these arches, system (\ref{planar system23}) admits peakon and anti-peakon solutions (see fig.8  (a) and (b)).
Corresponding to the family of triangular periodic orbits enclosed by the singular line $\phi=4C_1$ and the closed orbit in figures 2c, 3c, 4d, 5c, 5g, 6d. They are defined by $H(\phi,y)=h$ with $h \in (h_2,h_3) (or(h_3,h_2))$. The system (\ref{planar system23}) admits periodic peakon solutions (see fig.8 (c)).

\begin{center}
	\begin{tabular}{cc}
		\epsfxsize=5cm
		\epsfysize=5cm \epsffile{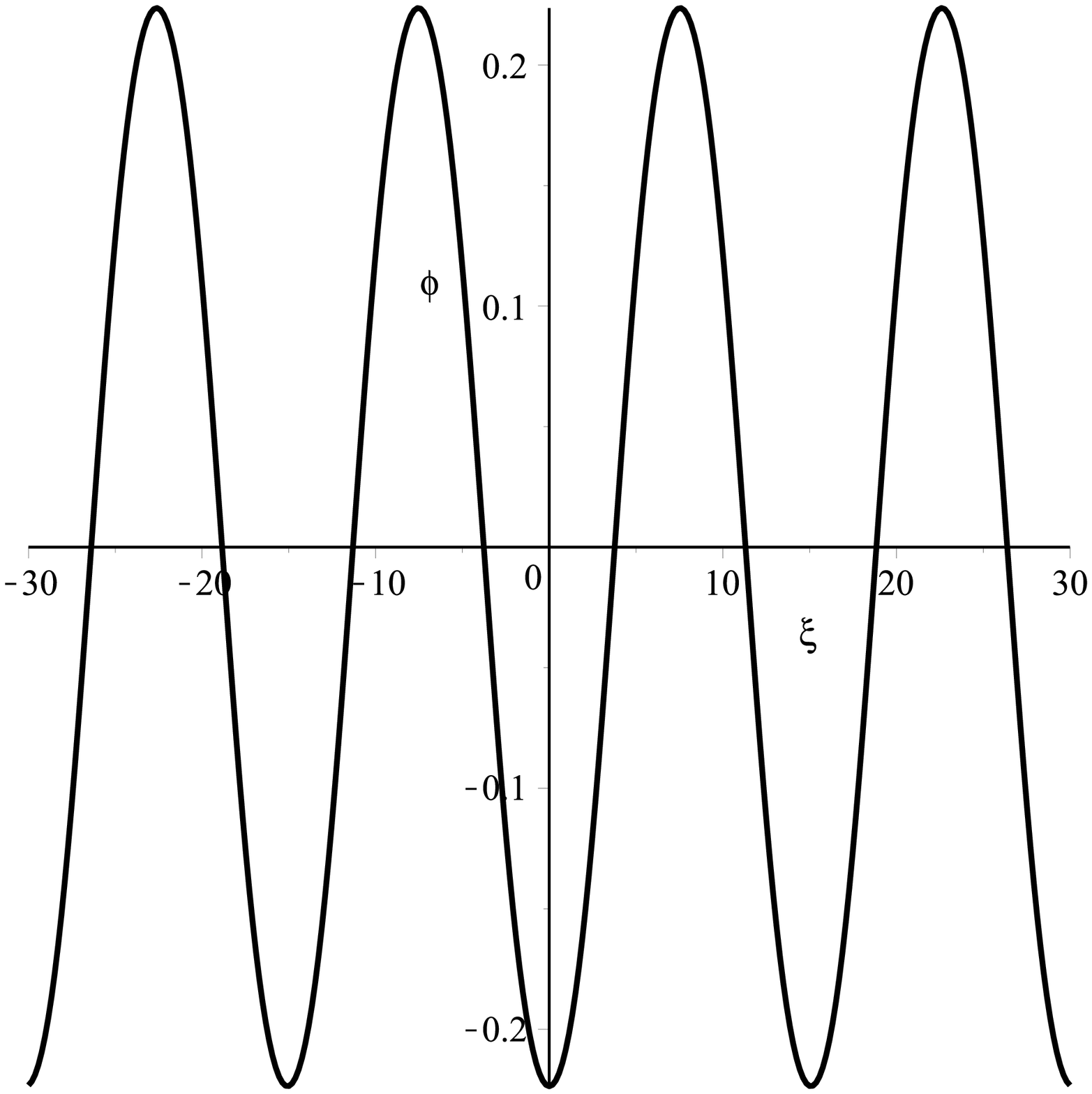}&
		\epsfxsize=5.5cm \epsfysize=5cm \epsffile{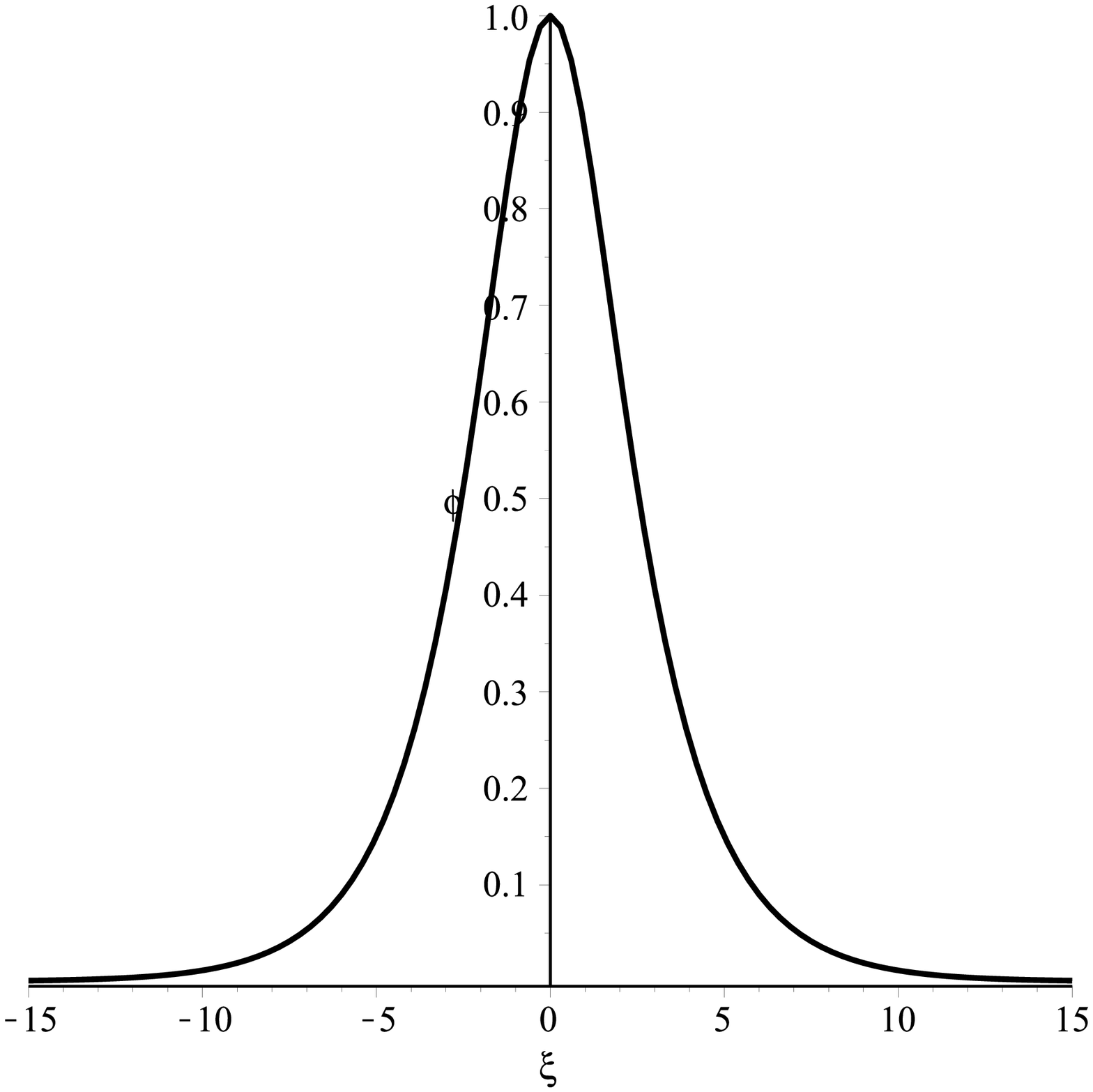}\\
		\footnotesize{ (a)  } & \footnotesize{(b)  }
	\end{tabular}
\end{center}

\begin{center}
	{\small Fig.14  Periodic and solitary wave solution profile of system (\ref{planar system23}).       }
\end{center}

\begin{center}
	\begin{tabular}{cc}
		\epsfxsize=5cm
		\epsfysize=5cm \epsffile{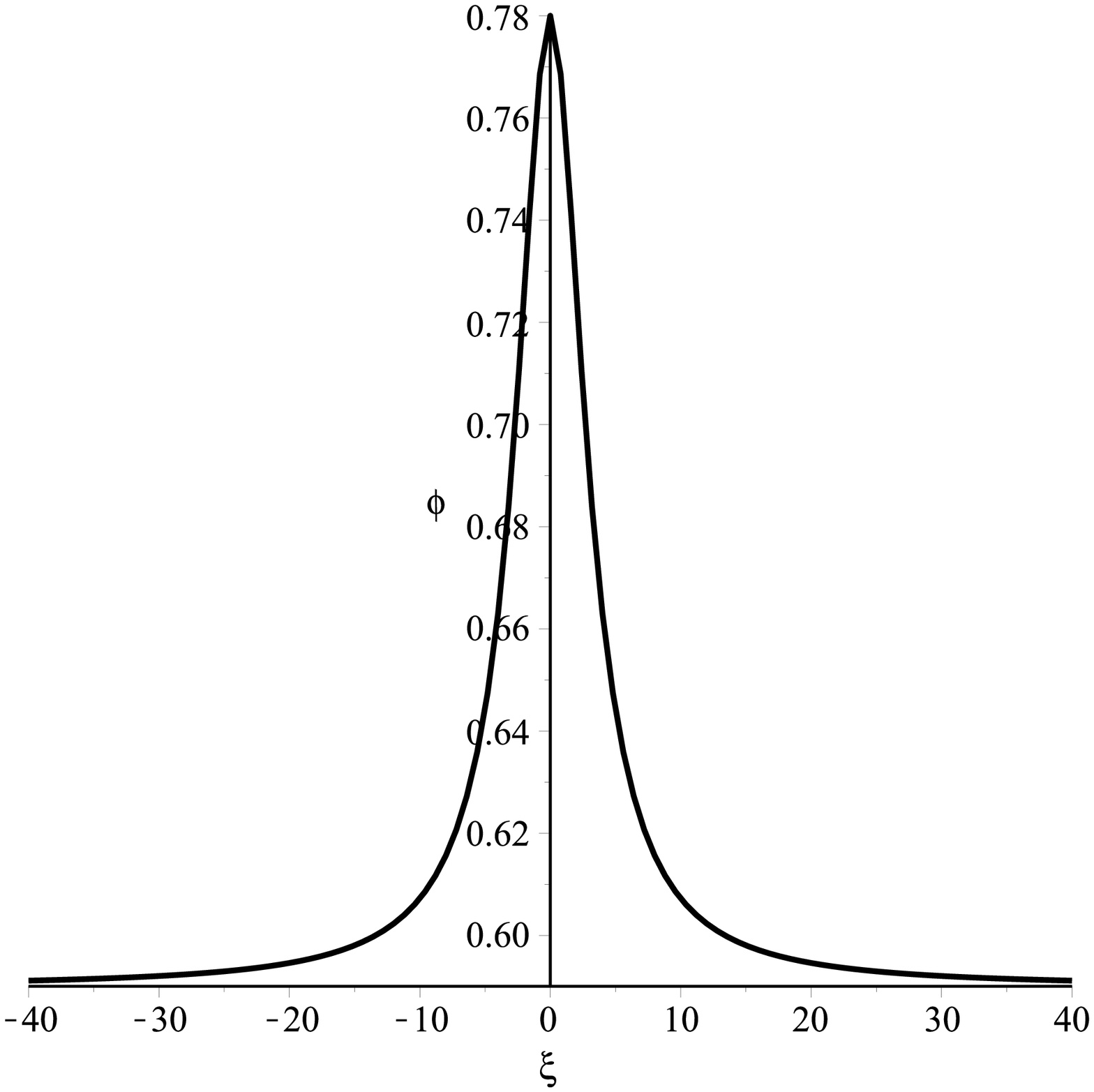}&
		\epsfxsize=5cm \epsfysize=5cm \epsffile{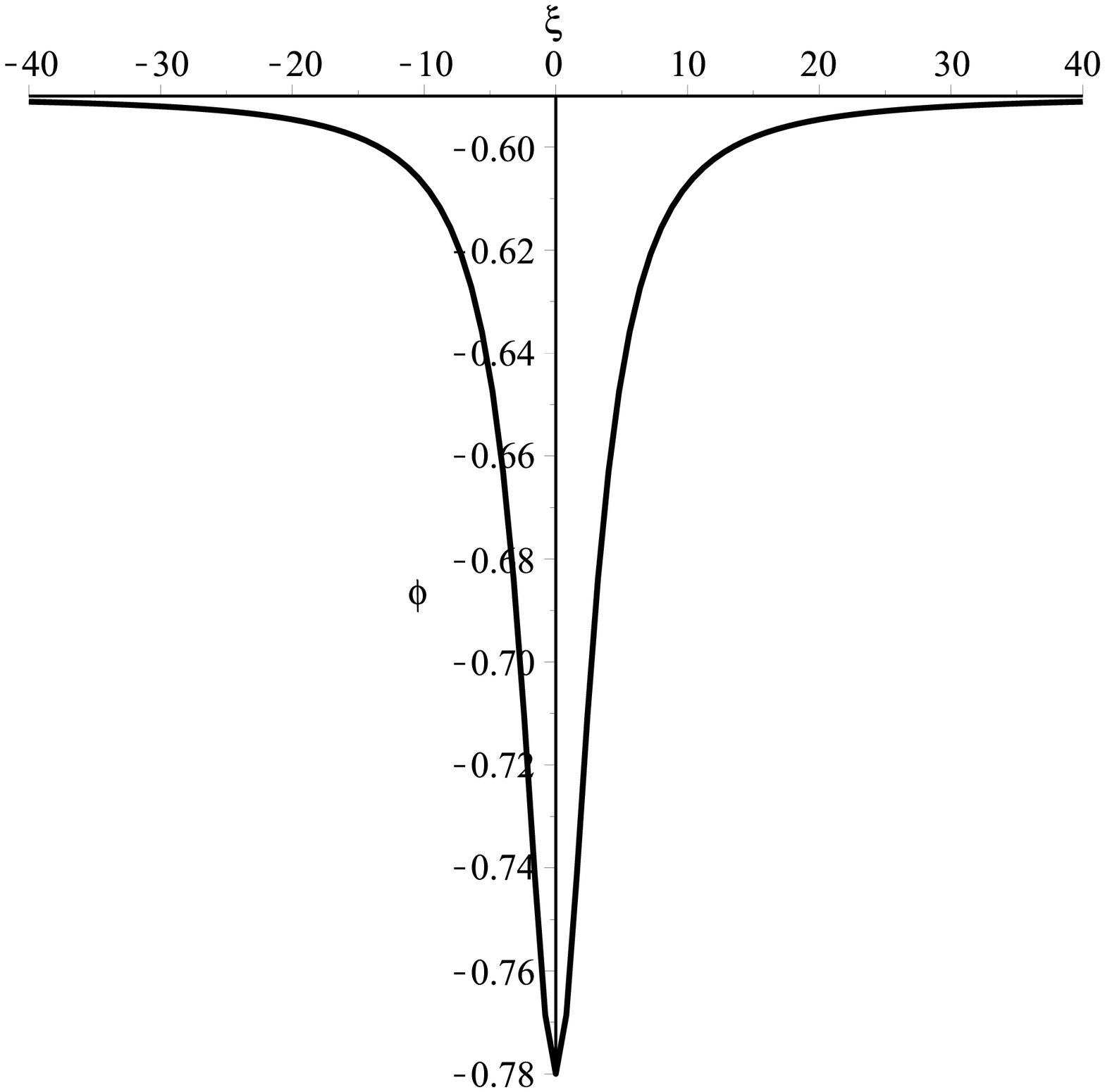}\\
		\footnotesize{ (a)  } & \footnotesize{(b)  }
	\end{tabular}
\end{center}

\begin{center}
	\begin{tabular}{cc}
		\epsfxsize=5cm
		\epsfysize=5cm \epsffile{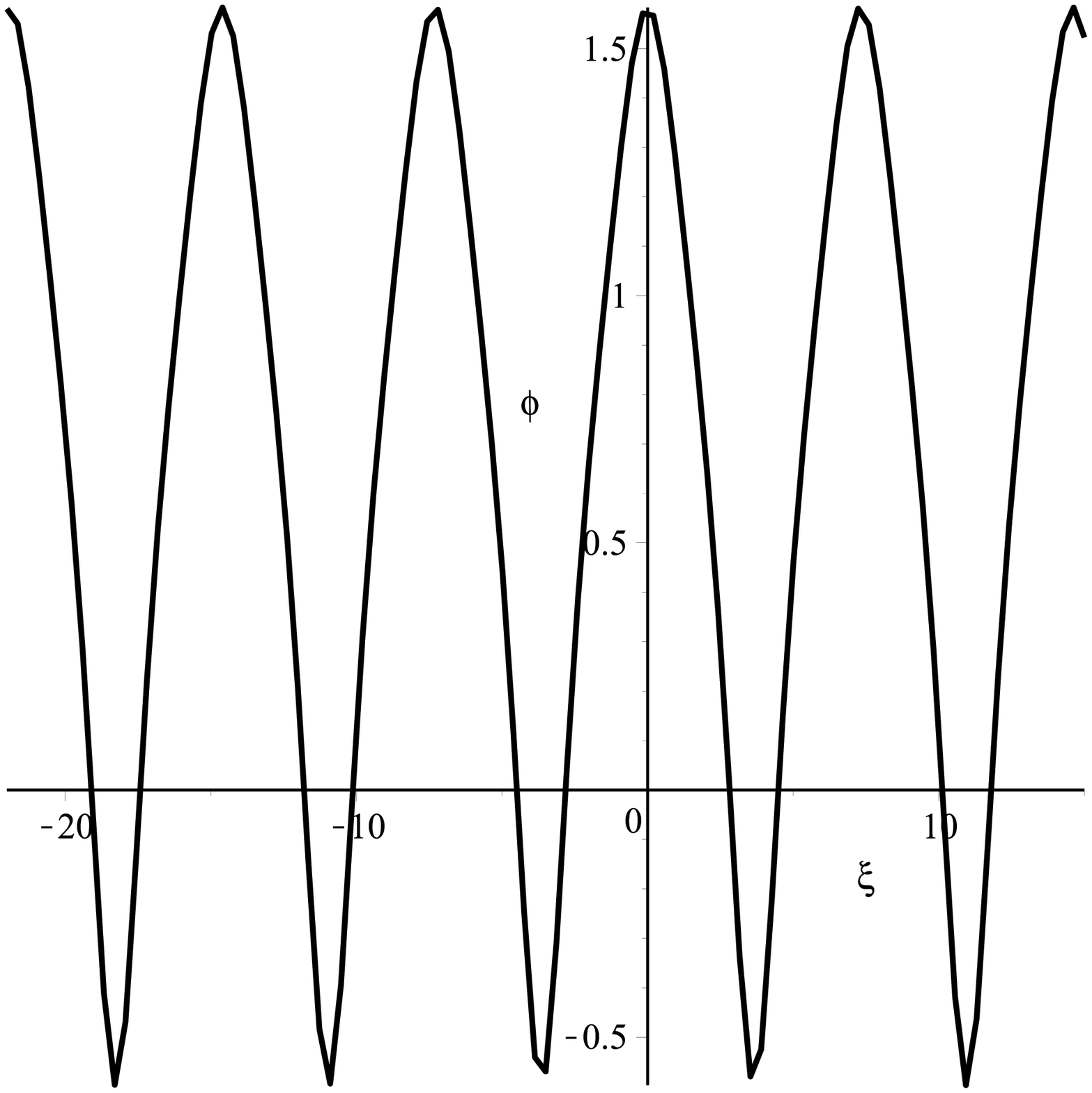}\\
		\footnotesize{(c)  }
	\end{tabular}
\end{center}

\begin{center}

	{\small Fig.15 Peakon, anti-peakon and periodic peakon solution profile of system (\ref{planar system23}).       }
\end{center}

\subsection{Case of \textbf{$\theta=\frac{1}{2}$}}
As previously mentioned, the expression of the first integral of system (\ref{planar system25}), makes explicit expressions of solutions difficult to obtain. Thus, we discuss the dynamical behavior of the solutions, based on the phase portraits and the properties of the singular points. However, for the special case of the parameter $C_1$ being null, we are able to compute some exact traveling wave solutions.

\subsubsection{Dynamical behavior of solutions }

\textbf{(i) Periodic wave solution}

Taking figures 7-12, corresponding to the level curve $H(\phi,y) = h$ with $h \in (h1,h2)(or(h2,h1))$, there exist families of periodic orbits enclosing the center points $E_i(\phi_i,0)$ or (and) $E_0(0,0)$. Thus, system (\ref{planar system25}) admits periodic wave solutions.

\textbf{(ii) Solitary wave solution
}

Considering fig. 7a, d, 8a, c, d, 9a, b, d, 10a, c, 11 and 12a, d. There exist homoclinic orbits corresponding to the level curve $H(\phi,y)=h$ with $h=h_i$. In these cases, system (\ref{planar system25}) admits solitary wave solution.

We notice that for $\theta=\frac{1}{2}$, all the wave solutions appear to be smooth.

\subsubsection{Some exact solutions of system (\ref{planar system25}) for \textbf{$C_1=0$}}
In this subsection, we consider the orbits computed with $C_1=0$.

According to the phase portraits in fig.(13), for a given \textit{h}, $H(\phi,y)=h$ corresponds to level curves of system (\ref{planar system25}). From Eq. (\ref{Hamiltinian2}) we write
\begin{equation}\label{Eq.3.1}
	y^2=\frac{C_3}{2}\phi^4+\frac{2\alpha}{3}\phi^3+\beta\phi^2+2\gamma\phi-2h.
\end{equation}
Furthermore, we know, $\frac{d\phi}{d\xi}=y$. Hence, integrating along the curve $H(\phi,y)=h$, from $\phi(0)=\phi_0$, we derive
\begin{equation}\label{3.2}
	\xi=\int_{\phi_0}^{\phi}\frac{d\phi}{\sqrt{\frac{C_3}{2}\phi^4+\frac{2\alpha}{3}\phi^3+\beta\phi^2+2\gamma\phi-2h}}
\end{equation}

\textbf{i)} Considering figures 13a, 13b, 13d, there is a family of periodic orbits enclosing the equilibrium point $\phi_1$. Let $p_1$ and $p_2$ be the intersections of the curve with the $\phi-axis$. Then, we have
\begin{equation}\label{Eq.3.3}
	y^2=\frac{C_3}{2}(p_1-\phi)(\phi-p_2)(\phi-w)(\phi-\bar{w})
\end{equation}
where $w$ and $\bar{w}$ are imaginary numbers. Thus, from (\cite{Byrd}), we get:
\begin{equation}\label{Eq.3.4}
	\phi(\xi)=\frac{(p_1B_1-p_2A_1)cn(\omega\xi,k)-p_1B_1-P_2A_1}{(B_1-A_1)cn(\omega\xi,k)-(B_1-A_1)},
\end{equation}
where $A_1^2=(p_1-b_1)^2+a_1^2$, $B_1^2=(p_2-b_1)^2+a_1^2$, $a_1^2=\frac{-(w-\bar{w})^2}{4}$, $b_1=\frac{w+\bar{w}}{2}$, $k=\frac{(p_1-p_2)^2-(A_1-B_1)^2}{4A_1B_1}$, $\omega=\sqrt{\frac{AB|C_3|}{2}}$ and $cn(\omega\xi,k)$ is Jacobian elliptic function.

\begin{center}
	\begin{tabular}{c}
		\epsfxsize=5cm
		\epsfysize=5cm \epsffile{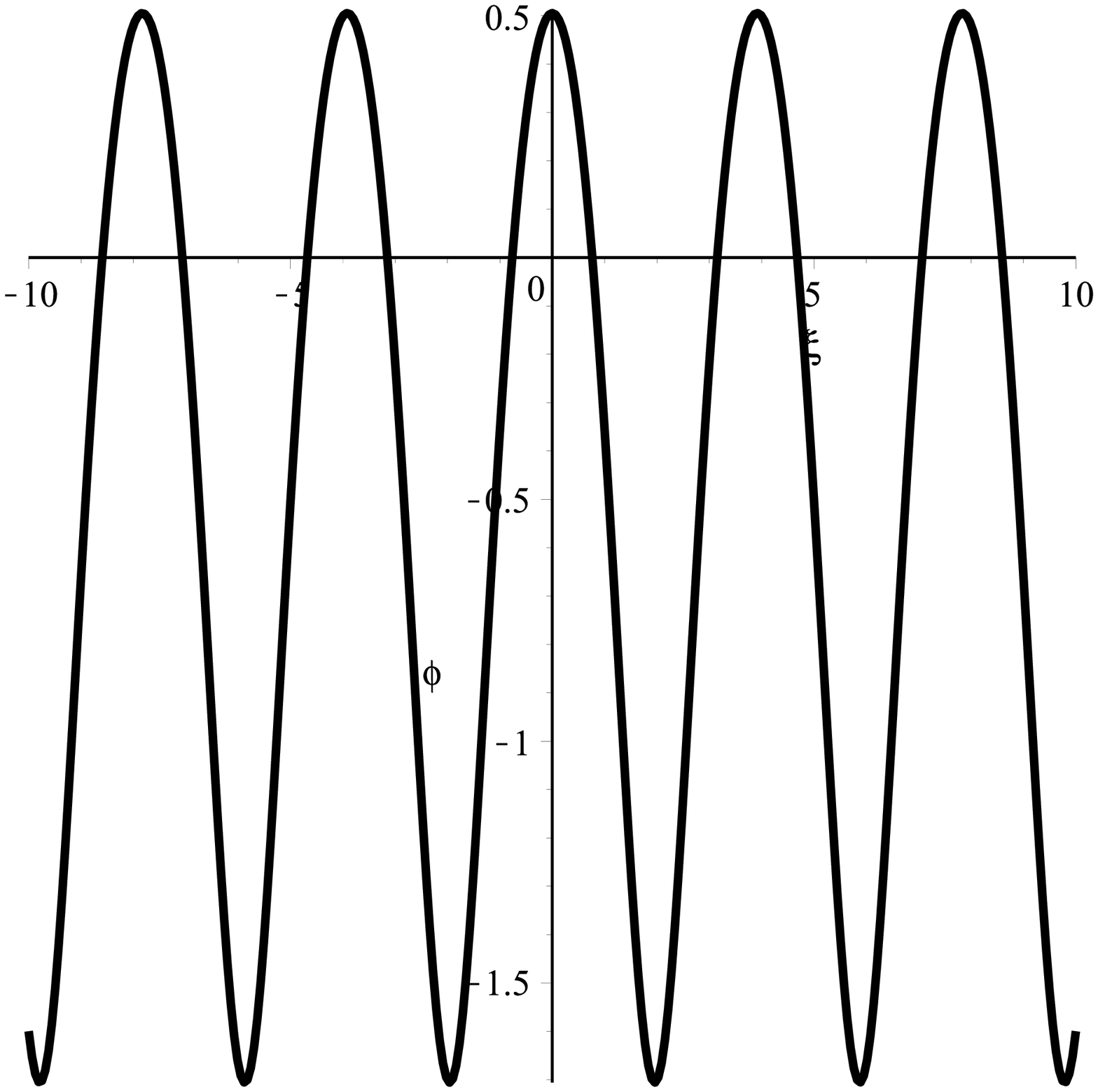}\\
		
	\end{tabular}
\end{center}

\begin{center}

	{\small   Fig.16  Periodic orbit given by (\ref{Eq.3.4})  }
\end{center}

\textbf{ii)} Corresponding to  figures 13d,f, there exists two families of periodic orbits enclosing the equilibrium points $\phi_1$ and $\phi_2$. In term of the right orbit, we have
\begin{equation}\label{Eq.3.5}
	y^2=\frac{C_3}{2}(p_1-\phi)(\phi-p_2)(\phi-p_3)(\phi-p_4).
\end{equation}
Hence,
\begin{equation}\label{Eq.3.6}
	\phi(\xi)=\frac{p_2(p_1-p_3)-p_3(p_1-p_2)sn^2(\omega\xi,k)}{(p_1-p_3)-(p1-p_2)sn^2(\omega\xi,k)},
\end{equation}
where $\omega=\frac{2}{\sqrt{(p_1-p_3)(p_2-p_4)}}\sqrt{\frac{|C_3|}{2}}$, $k^2=\frac{(p_1-p_2)(p_3-p_4)}{(p_1-p_3)(p_2-p_4)}$, and $sn(\omega\xi,k)$ is Jacobian elliptic function.

In term of the left orbit, we have
\begin{equation}\label{Eq.3.7}
	y^2=\frac{C_3}{2}(p_1-\phi)(\phi-p_2)(p_3-\phi)(\phi-p_4),
\end{equation}
which gives us
\begin{equation}\label{Eq.3.8}
	\phi(\xi)=\frac{p_4(p_1-p_3)-p_1(p_3-p_4)sn^2(\omega\xi,k)}{(p_1-p_3)-(p3-p_4)sn^2(\omega\xi,k)},
\end{equation}
where $\omega=\frac{2}{\sqrt{(p_1-p_3)(p_2-p_4)}}\sqrt{\frac{|C_3|}{2}}$, $k^2=\frac{(p_1-p_2)(p_3-p_4)}{(p_1-p_3)(p_2-p_4)}$, and $sn(\omega\xi,k)$ is Jacobian elliptic function.

\begin{center}
	\begin{tabular}{cc}
		\epsfxsize=5cm
		\epsfysize=5cm \epsffile{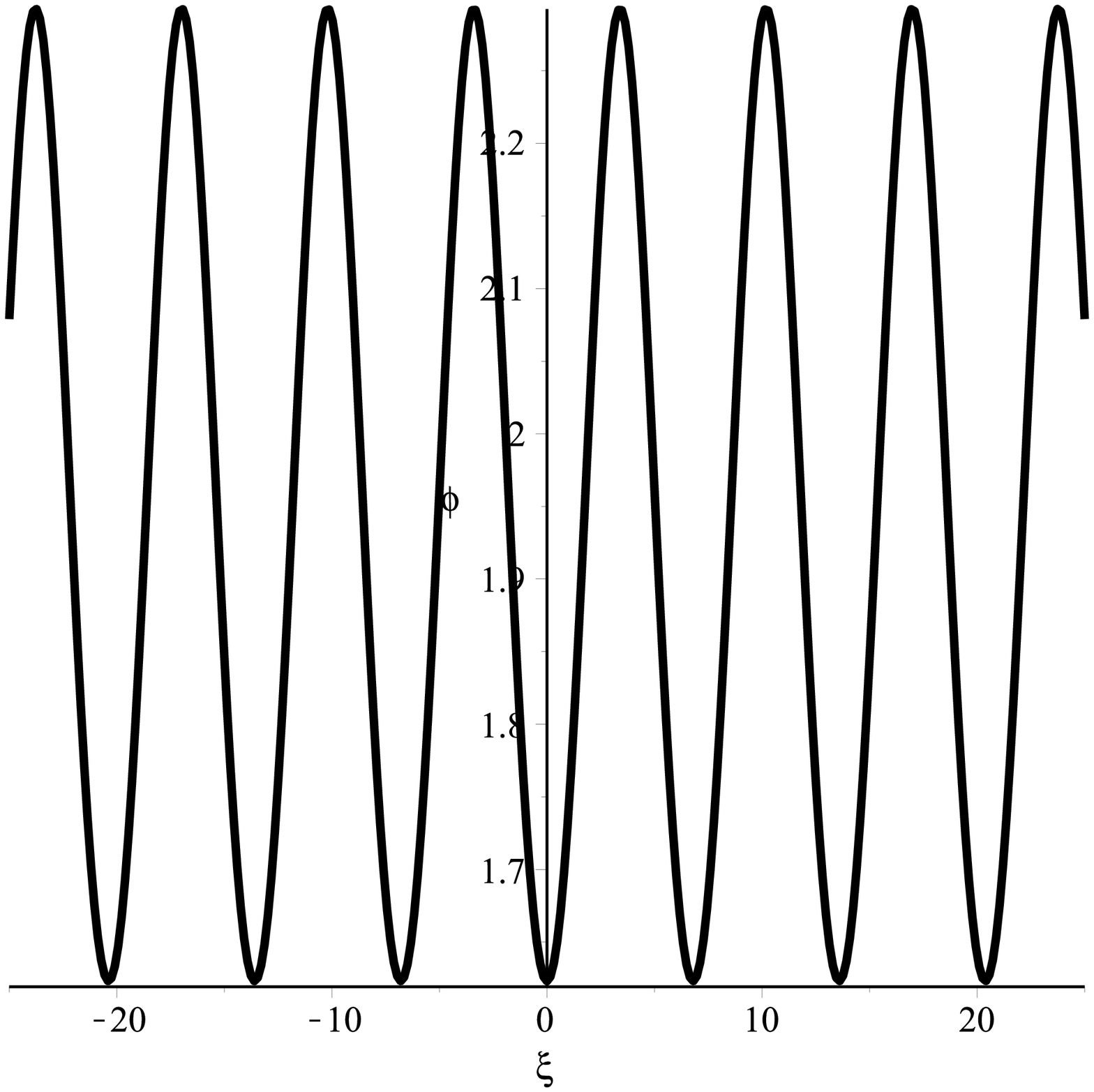} &
		
		\epsfxsize=5cm
		\epsfysize=5cm \epsffile{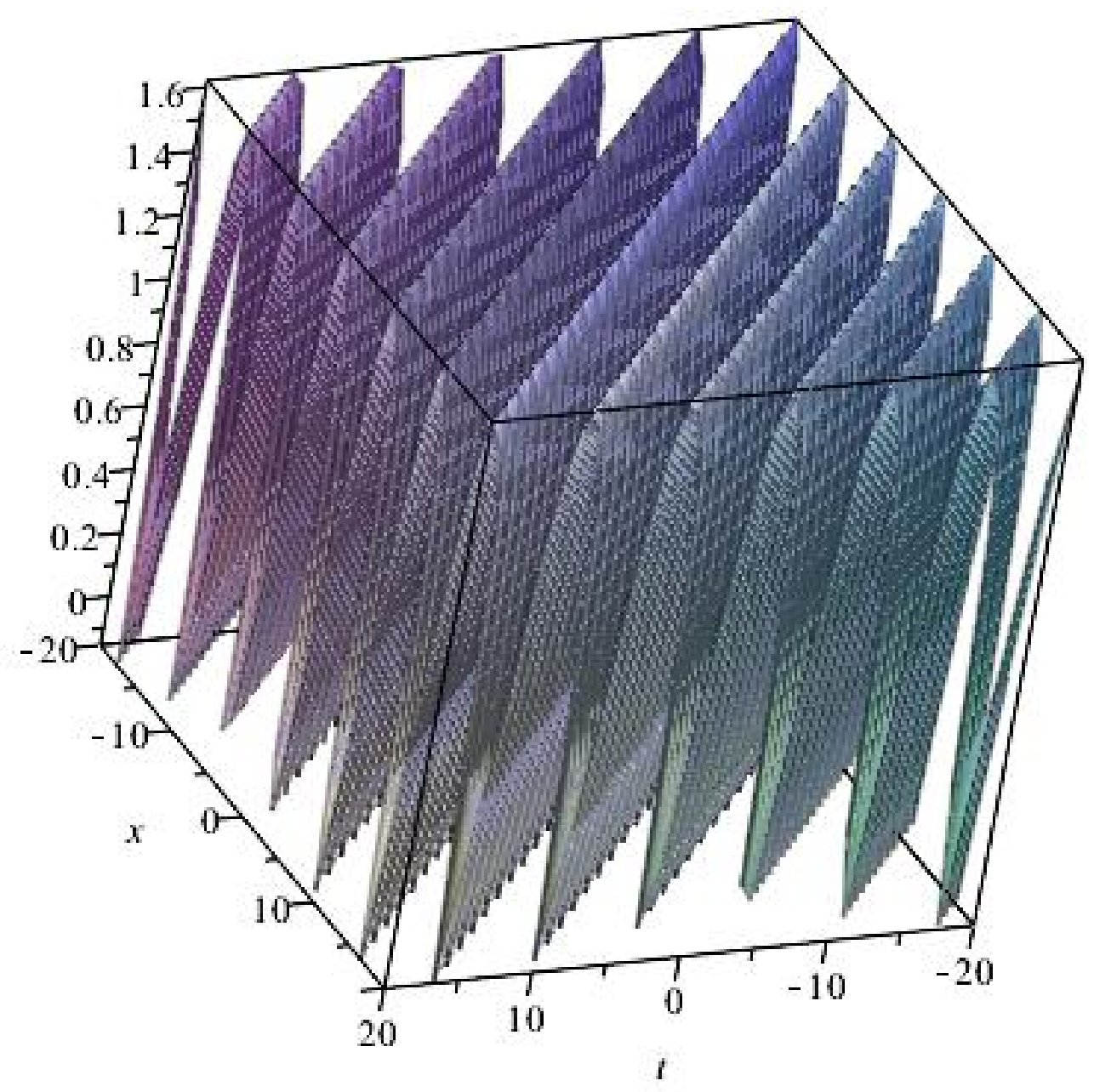}\\
		\footnotesize{ (a) Right orbit given by (\ref{Eq.3.6})  } & \footnotesize{(b) Wave solution given by (\ref{Eq.3.6}) }
	\end{tabular}
\end{center}

\begin{center}

	{\small   Fig.17    }
\end{center}

\begin{center}
	\begin{tabular}{c}
		\epsfxsize=5cm
		\epsfysize=5cm \epsffile{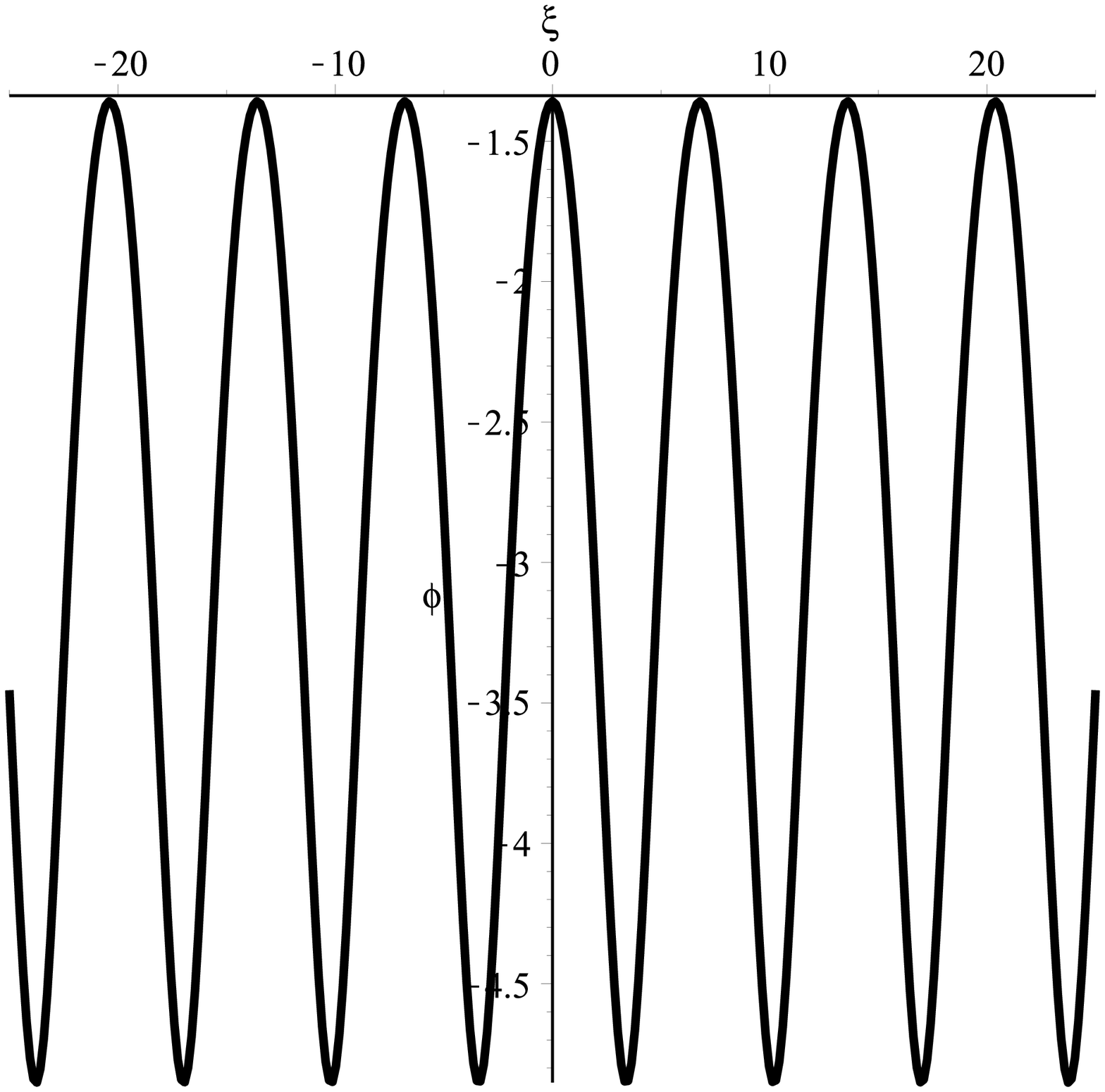}\\
		
	\end{tabular}
\end{center}

\begin{center}

	{\small   Fig.18  Right orbit given by (\ref{Eq.3.6})  }
\end{center}

\textbf{iii)} Considering figures 13d,f, we observe two homoclinic orbits to the equilibrium points $\phi_2$ $O(0,0)$,respectively.
In term of the right orbit, we write:
\begin{equation}\label{Eq.3.9}
	y^2=\frac{C_3}{2}(p_1-\phi)(\phi-p_2)^2(\phi-p_3).
\end{equation}
Hence, we get
\begin{equation}\label{Eq.3.10}
	\phi=p_2+\frac{2a}{(p_3-p_1)cosh(\omega\xi)-b}.
\end{equation}

In term of the right orbit, we have
\begin{equation}\label{Eq.3.11}
	\phi=p_2-\frac{2a}{(p_3-p_1)cosh(\omega\xi)-b} ,
\end{equation}
where $a=p_1p_2-p_1p_3+p_2p_3-p_2^2$, $b=p_1-2p_2+p_3$ and $\omega=\sqrt{\frac{a|C_3|}{2}}$.
\begin{center}
	\begin{tabular}{cc}
		\epsfxsize=5cm
		\epsfysize=5cm \epsffile{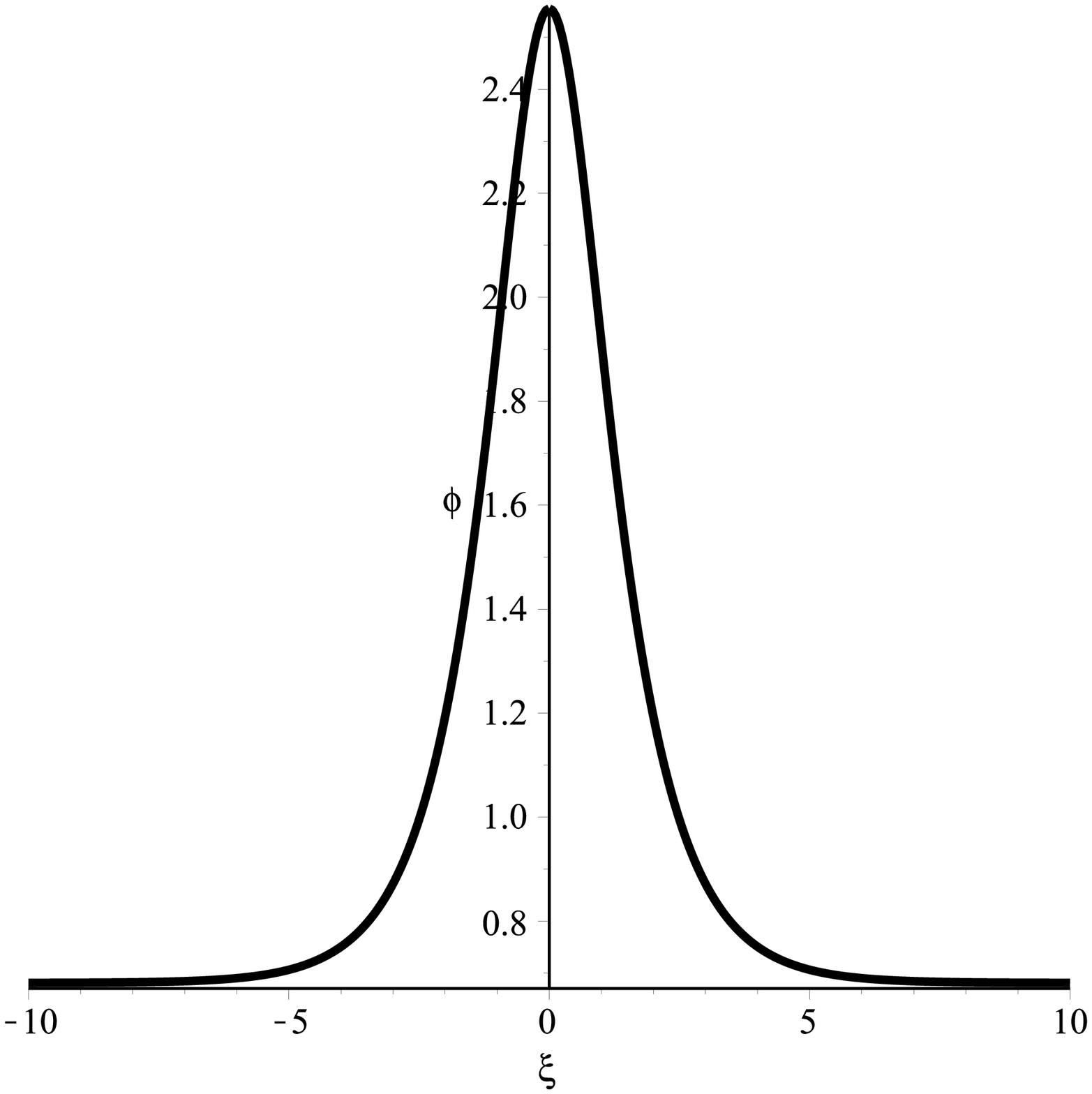} &
		
		\epsfxsize=5cm
		\epsfysize=5cm \epsffile{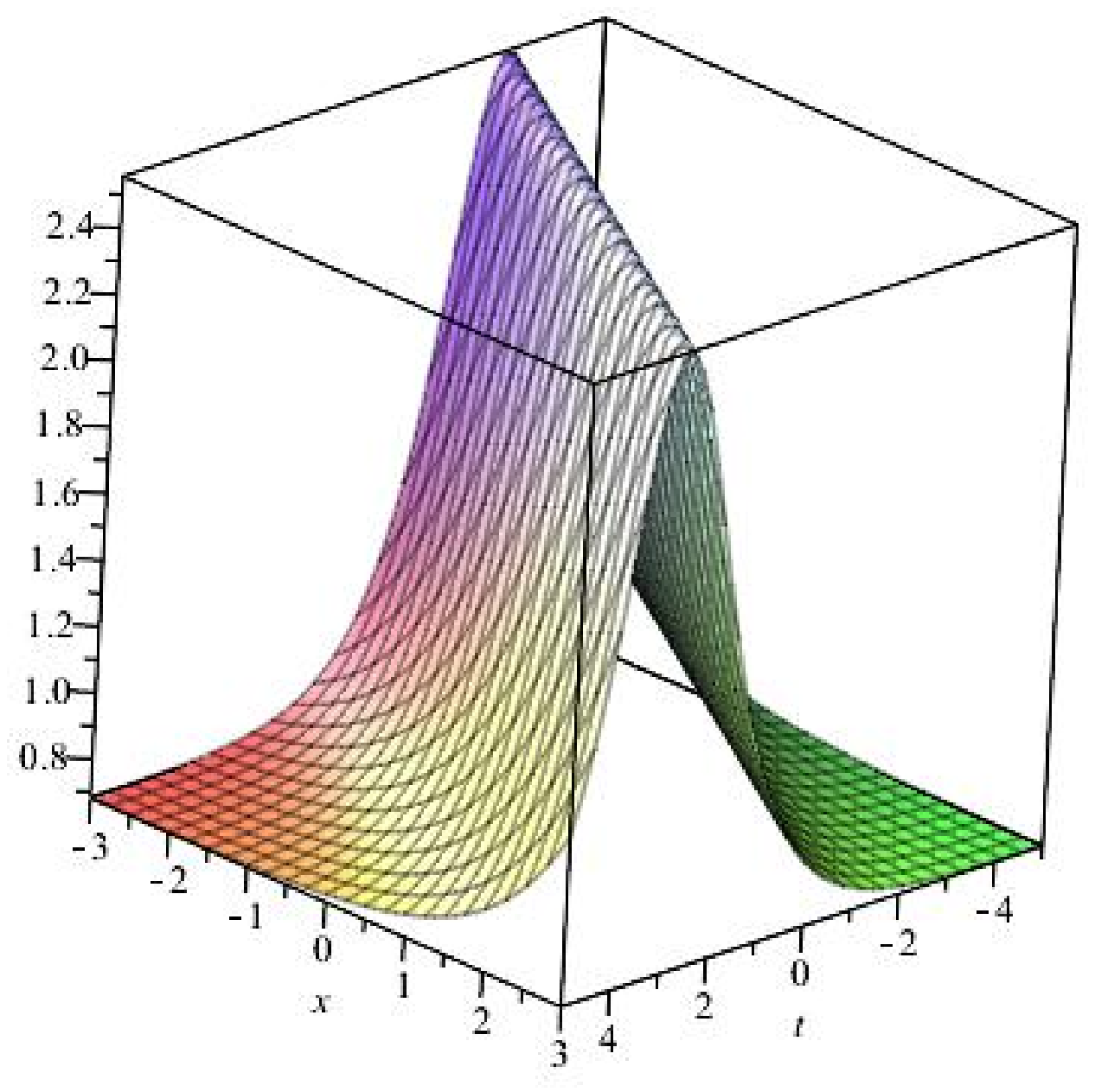}\\
		\footnotesize{ (a) Right homoclinic orbit given by (\ref{Eq.3.10})  } & \footnotesize{(b) Wave solution given by (\ref{Eq.3.10}) }
	\end{tabular}
\end{center}

\begin{center}

	{\small   Fig.19    }
\end{center}

\begin{center}
	\begin{tabular}{cc}
		\epsfxsize=5cm
		\epsfysize=5cm \epsffile{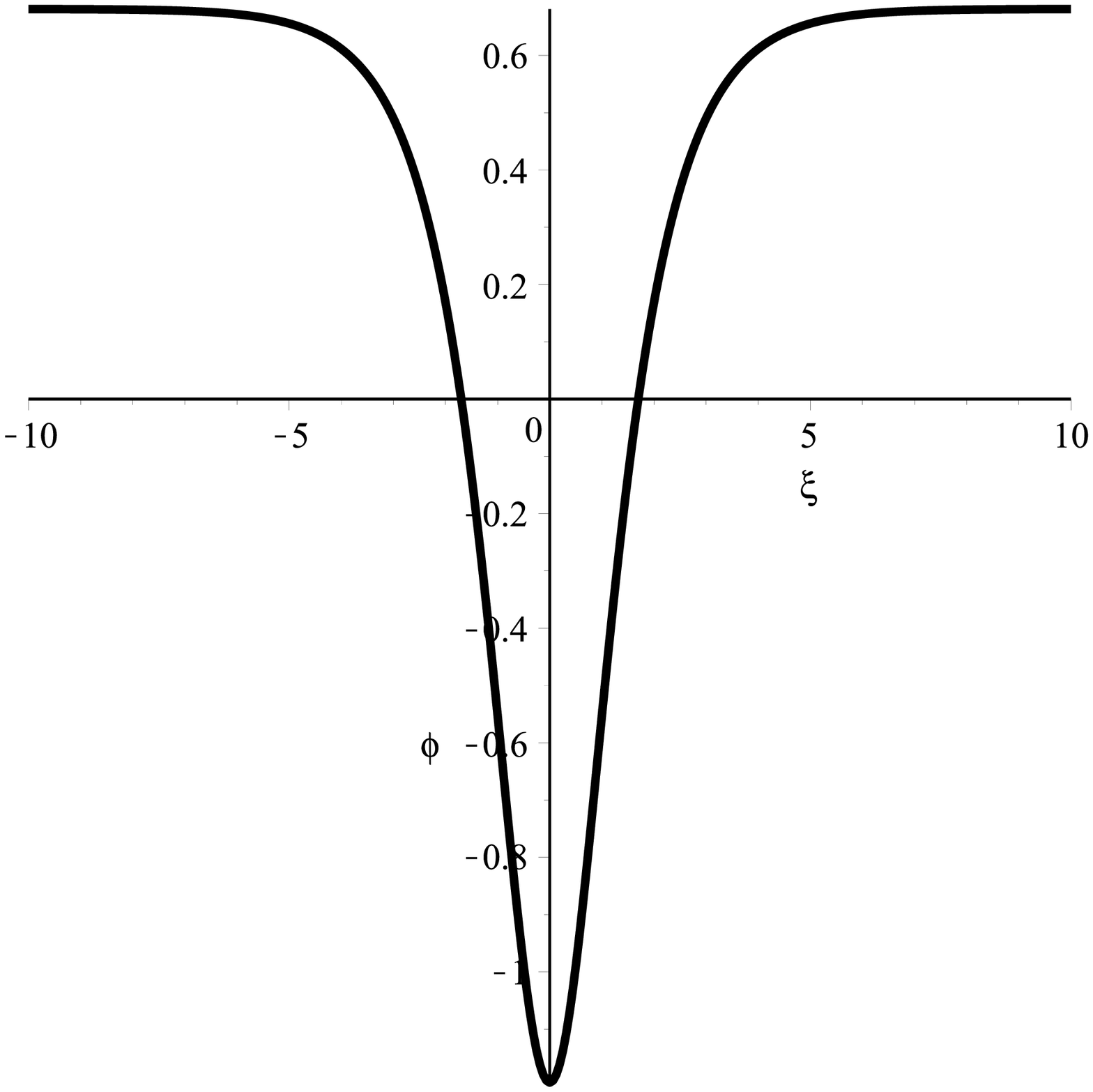} &
		
		\epsfxsize=5cm
		\epsfysize=5cm \epsffile{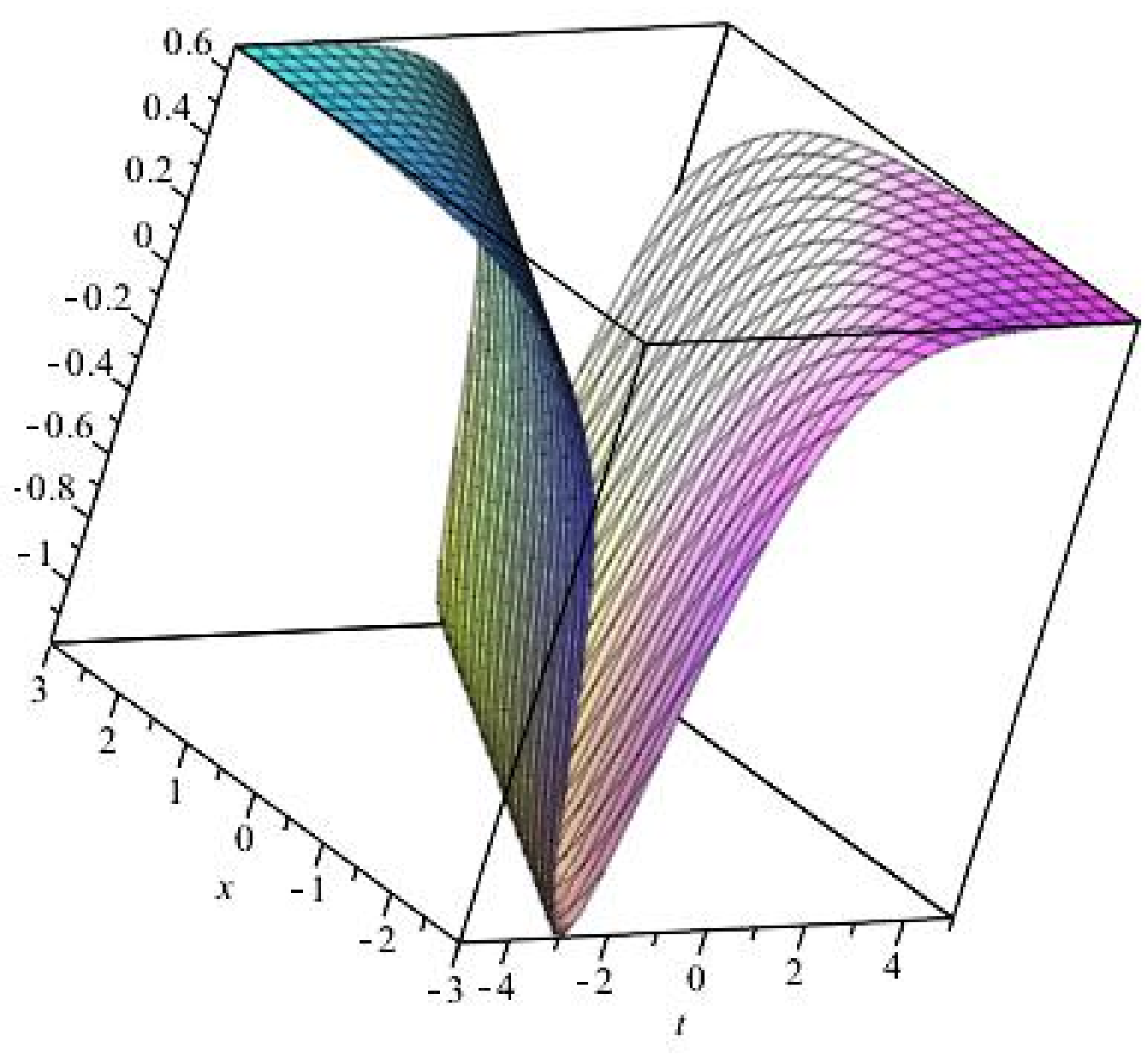}\\
		\footnotesize{ (a) Left homoclinic orbit given by (\ref{Eq.3.11})  } & \footnotesize{(b) Wave solution given by (\ref{Eq.3.11}) }
	\end{tabular}
\end{center}

\begin{center}

	{\small   Fig.20    }
\end{center}

\section{Main results}

We summarize the bifurcations and explicit expressions of waves solutions in the following theorems.

\textbf{Theorem 1. } When $\theta=\frac{1}{4}$,

\textbf{(B1)} In the domain $D_1$=\{$4C_2^2>6C_3$, $g(\tilde\phi_-)>0$\}, moving the singular line $4C_1$ from right to left, there are solitary wave solutions and (or) periodic wave solutions. Additionally, when  $0<4C_1<\phi_1$ there are a peakon solution and two periodic peakon solutions.

\textbf{(B2)} In the domain $D_2$=\{$4C_2^2>6C_3$, $g(\tilde\phi_-)=0$\}, moving the singular line $4C_1$ from right to left, there are solitary wave solutions and (or) periodic wave solutions. Additionally, when $0<4C_1<\phi_1$ there are  two periodic peakon solutions.

\textbf{(B3)} In the domain $D_3$=\{$4C_2^2>6C_3$, $g(\tilde\phi_+)=0$\}, moving the singular line $4C_1$ from right to left, there are solitary wave solutions and (or) periodic wave solutions. Espacially, when $0<4C_1<\phi_1$ there are a peakon and two periodic peakon solutions.

\textbf{(B4)} In the domain $D_4$=\{$4C_2^2>6C_3$, $g(\tilde\phi_+)>0, g(\tilde\phi_-)<0$\}, moving the singular line $4C_1$ from right to left, there are solitary wave solutions and (or) periodic wave solutions. Additionally, when $\phi_3<4C_1<\phi_2$ and $0<4C_1<\phi_1$ there are a peakon solution and two periodic peakon solutions.

\textbf{(B5)} In the domain $D_5$=\{$K=0, 4C_2^2>6C_3$\}, moving the singular line $4C_1$ from right to left, there are solitary wave solutions and (or) periodic wave solutions. Additionally, when $\phi_3<4C_1<\phi_2$ and $0<4C_1<\phi_1$ there are a peakon solution and two periodic peakon solutions.

\textbf{Theorem 2. } When $\theta=\frac{1}{2}$,

\textbf{(B1)} In the domain $D_1$=\{$ 4C_2^2>6C_3,g(\tilde\phi_-)>0$\}, moving the singular line $2C_1$ from right to left, there are solitary wave solutions and (or) periodic wave solutions.

\textbf{(B2)} In the domain $D_2$=\{$4C_2^2>6C_3$, $g(\tilde\phi_-)=0$\}, moving the singular line $2C_1$ from right to left, there are solitary wave solutions and (or) periodic wave solutions.

\textbf{(B3)} In the domain $D_3$=\{$4C_2^2>6C_3$, $g(\tilde\phi_+)=0$\}, moving the singular line $2C_1$ from right to left, there are solitary wave solutions and (or) periodic wave solutions.

\textbf{(B4)} In the domain $D_42$=\{$4C_2^2>6C_3$, $g(\tilde\phi_+)<0$\}, moving the singular line $2C_1$ from right to left, there are solitary wave solutions and (or) periodic wave solutions.

\textbf{(B5)} In the domain $D_5$=\{$4C_2^2>6C_3$, $g(\tilde\phi_+)>0, g(\tilde\phi_-)<0$\}, moving the singular line $2C_1$ from right to left, there are solitary wave solutions and (or) periodic wave solutions.

\textbf{(B6)} In the domain $D_5$=\{$K=0, 4C_2^2>6C_3$\}, moving the singular line $4C_1$ from right to left, there are solitary wave solutions and (or) periodic wave solutions.

\textbf{Theorem 3. } When $\theta=\frac{1}{2}$ and $C_1=0$,

\textbf{(B1)} In the domain $D_1$=\{$ 4C_2^2>6C_3,g(\tilde\phi_-)\geq0$\}, there is a periodic solution given by Eq. (\ref{Eq.3.4}).

\textbf{(B2)} In the domain $D_2$=\{$4C_2^2>6C_3$, $g(\tilde\phi_-)\leq0$\}, there is a periodic solution with the same parametric representation as Eq. (\ref{Eq.3.6}).

\textbf{(B3)} In the domain $D_2$=\{$4C_2^2>6C_3$, $g(\tilde\phi_-)<0, g(\tilde\phi_+>0$\}, there are two periodic wave solutions given by Eqs. (\ref{Eq.3.6}), (\ref{Eq.3.8}) and two solitary wave solutions given by Eqs. (\ref{Eq.3.10}), (\ref{Eq.3.11}).

	\section{Conclusion}
	In this paper, we studied the Rotation-$\theta$-equation (the $\theta$-equation with the Coriolis force effect), by mean of the bifurcation method and qualitative theory of dynamical system. Using traveling wave solution transformation, we obtained the first integral for the model we built, which to our knowledge has not been brought to investigation. After qualitative analysis, we were able to obtain phase portraits for certain values $\theta$ and the parameters. Smooth periodic and solitary waves, as well as, peakons and periodic peakons solutions were observed. Interestingly, we noticed that the proximity of the singular line alters the nature of an equilibrium point.
	One would notice that for $\theta=\frac{1}{4}$, the Rotation-$\theta$ equation is reduced to the Rotation Degasperis-Procesi (R-DP) equation. Furthermore, for $\theta=1$, the Rotation-$\theta$ equation is similar to the asymptotic R-CH equation in (\cite{Liang}), only differing by the values of certain parameters.
	
	The chosen values of $\theta$, allow us to assess the difference in behavior between the classical $\theta$-equation and the Rotation-$\theta$ equation.
	We can conclude that the coriolis effect does affect the traveling wave solutions.
	Due to the form of the model and the high power of the hamiltonian, we were unable to compute some of the   explicit expressions of the solutions. To the best of our knowledge, there are no mathematical tools to solve the elliptic integrals  whose the power is higher than 5 (\cite{Byrd}).  It is an interesting question and remains open.
	
	\section{Acknowledgement}
	This manuscript was supported from the Natural
	Science Foundation of Zhejiang Province under Grant (No. LY20A010016), National Natural
	Science Foundation of China under Grant (No. 11671176, 11931016).

\end{document}